%% file: main.tex
\pdfoutput=1
\documentclass[a4paper]{ar-1col}
\usepackage{preview}
\usepackage{amsmath}
\usepackage{graphicx,tabularx,float}
\usepackage{url}
\usepackage{footmisc}
\usepackage{amsfonts}
\usepackage{cancel}
\usepackage{color}
\usepackage{multirow} 
\usepackage{amssymb}
\usepackage{pifont}
\usepackage{epstopdf}
\usepackage{slashed}
\usepackage{comment}
\usepackage{booktabs} 
\usepackage[numbers]{natbib}
\usepackage{array}
\usepackage{mathrsfs}
\usepackage{caption}
\usepackage{subcaption}
\usepackage[toc,page]{appendix}
\usepackage{mathtools}
\usepackage{romannum}
\usepackage{ulem}
\usepackage{bbold}

\usepackage{atbegshi,picture}

\allowdisplaybreaks

\newcommand{\met}{\ensuremath{{\not\mathrel{E}}_T}}

\def\met{\ensuremath{\slashed{E}_T}}
\def\n{\ensuremath{\tilde\chi^0_1}}
\def\nn{\ensuremath{\tilde\chi^0_2}}

\def\cha{\ensuremath{\tilde\chi^\pm_1}}
\def\champ{\ensuremath{\tilde\chi^\mp_1}}

\def\chai{\ensuremath{\tilde\chi^\pm_i}}
\def\nj{\ensuremath{\tilde\chi^0_j}}

\def\pt{\ensuremath{\slashed{p}_T}}

\def\ifb{~fb$^{-1}$}

\def\beq{\begin{equation}}
\def\eeq{\end{equation}}
\def\bea{\begin{eqnarray}}
\def\eea{\end{eqnarray}}

\jname{Annu. Rev. Nucl. Part. Sci.}
\jyear{2020}
\doi{10.1146/annurev-nucl-031020-121031}
\AtBeginShipoutNext{\AtBeginShipoutUpperLeft{%
  \put(\dimexpr\paperwidth-4.cm\relax,-2.5cm){FERMILAB-PUB-16-565-T}%
  \put(\dimexpr\paperwidth-4.cm\relax,-3.cm){PITT-PACC-2001}%
}}
\begin{document}

\markboth{A. Canepa et al.}{The Search for Electroweakinos}

\title{The Search for Electroweakinos}


\author{Anadi Canepa,$^1$ Tao Han,$^2$ and Xing Wang$^3$
\affil{$^1$Fermi National Accelerator Laboratory, P.O. Box 500, Batavia, Illinois 60510, USA; email: acanepa@fnal.gov}
\affil{$^2$Department of Physics and Astronomy, University of Pittsburgh, Pittsburgh, PA 15260, USA; email: than@pitt.edu}
\affil{$^3$Department of Physics, University of California, San Diego, 9500 Gilman Drive, La Jolla, CA  92093, USA; email: xiw006@physics.ucsd.edu}}

\begin{abstract}
In this review, we consider a general theoretical framework for fermionic color-singlet states, including a singlet, a doublet and a triplet under the standard model SU(2)$_{\rm L}$ gauge symmetry, corresponding to the Bino, Higgsino and Wino in Supersymmetric theories, generically dubbed as ``electroweakinos" for their mass eigenstates. Depending on the relations among their three mass parameters and the mixings after the electroweak symmetry breaking, this sector leads to rich phenomenology potentially accessible at the current and near-future experiments. We discuss the decay patterns of the electroweakinos and their observable signatures at colliders. We review the existing bounds on the model parameters. We summarize the current status for the comprehensive searches from the ATLAS and CMS experiments at the LHC. We comment on the prospects for future colliders. An important feature of the theory is that the lightest neutral electroweakino can be identified as a WIMP cold dark matter candidate. We take into account the existing bounds on the parameters from the dark matter direct detection experiments and discuss the complementarity for the electroweakino searches at colliders.
\end{abstract}

\maketitle

\tableofcontents


\newpage

\input{1_intro.tex}

\input{2_model.tex}

\input{3_dm.tex}

\input{4_collider.tex}

\input{5_sum.tex}

\vskip 0.1cm
\noindent {\bf Acknowledgment} \\
{We would like to thank Tathagata Ghosh, Zhen Liu, Xerxes Tata and Lian-Tao Wang for comments on the manuscript. TH was supported in part by the U.S.~Department of Energy under grant No.~DE-FG02-95ER40896 and in part by the PITT PACC. XW was supported by the National Science Foundation under Grant No.~PHY-1915147. We would also like to thank the Aspen Center for Physics for hospitality, where part of the work was completed. The Aspen Center for Physics is supported by the NSF under Grant No.~PHYS-1066293.
We thank the ATLAS and CMS collaborations for their contributions to the search for Supersymmetry at the LHC. This manuscript has been authored by Fermi Research Alliance, LLC under Contract No. DE-AC02-07CH11359 with the U.S. Department of Energy, Office of Science, Office of High Energy Physics.
}

\bibliographystyle{ar-style5} 
\bibliography{ref.bib}

\end{document}

%% file: 1_intro.tex
\section{Introduction}
\label{sec:intro}

The Higgs boson ($h$) discovered at the CERN Large Hadron Collider (LHC) by the ATLAS \cite{Aad:2012tfa} and CMS \cite{Chatrchyan:2012xdj} collaborations completes the particle spectrum of the Standard Model (SM), which can be a self-consistent effective field theory valid up to an exponentially high scale. Yet from the observational point of view, the SM is incomplete. The missing component of  dark matter (DM), the lack of ingredients for generating the baryon-antibaryon asymmetry and a satisfactory account for neutrino masses all imply the existence of physics beyond the Standard Model (BSM). On the other hand, theoretical considerations, such as the hierarchy puzzle between the electroweak (EW) scale and the Planck scale \cite{Weinberg:1975gm, Gildener:1976ai, Susskind:1978ms, tHooft:1980xss}, gauge coupling unification \cite{Ellis:1990wk, Amaldi:1991cn, Langacker:1991an, Giunti:1991ta}, new space-time symmetry \cite{Golfand:1971iw, Volkov:1973ix, Wess:1974tw, Wess:1974jb, Ferrara:1974pu, Salam:1974ig}, new strong dynamics \cite{Kaplan:1983fs,Kaplan:1983sm,Georgi:1984af} or warped extra dimensions \cite{Randall:1999ee,Randall:1999vf}, all indicate the need for new physics at a scale not far from the electroweak scale \cite{Nilles:1983ge,Haber:1984rc,Martin:1997ns,Giudice:2008bi,Feng:2013pwa}. Therefore, the search for TeV-scale new physics 
in experiments at the energy frontier continues to be of high priority for particle physics in the coming decades. 

Current measurements of the Higgs boson properties at the LHC support the interpretation of its being a SM-like, weakly-coupled elementary particle. In this regard, weak-scale Supersymmetry may be arguably the most compelling incarnation for new  physics at the next scale. The introduction of the new space-time symmetry requires the existence of SUSY partners of the SM particles with predictable couplings and will lead to profound theoretical and experimental implications. The pressing question is the unknown mechanism for SUSY breaking and the associated scale that determines the mass spectrum for the SUSY partners, preferably not much heavier than the EW scale. If the weak-scale SUSY is realized in nature, the definitive confirmation will require the discovery of the supersymmetric partners, such as the QCD colored states gluinos $(\tilde{g})$, squarks $(\tilde{q})$ and the electroweak partners, such as the gauginos $(\tilde{B}, \tilde{W})$ and Higgsinos $(\tilde{H})$, or their mass eigenstates the charginos $(\chai)$ and neutralinos $(\nj)$. Here and henceforth we generically refer them as  ``electroweakinos'' (EWkinos). 
If a discrete symmetry, called $R$-parity that classifies the SM particles ($R$-even) and the SUSY partners (sparticles, $R$-odd), is conserved, then the SUSY particles and their  antiparticle will be produced in pair, and the lightest Supersymmetric particle (LSP), most commonly the lightest neutralino, will be practically stable. Such a stable neutral particle will escape from the direct detection and thus lead to a missing momentum signature in collider experiments.
It is particularly interesting to note that such a weakly-interacting massive particle (WIMP) will be a natural cold dark matter candidate \cite{Jungman:1995df}. 
The search for SUSY at colliders thus becomes especially important because of the connection with the DM detection. 

Given an underlying theory for SUSY breaking and a mechanism for mediating the breaking effects to the SM sector, SUSY partner masses may be calculable in terms of the SUSY breaking scale. 
The null results from SUSY searches performed at the LHC so far\footnote{We refer the readers to the comprehensive programs for ATLAS \cite{ATLASWeb} and CMS \cite{CMSWeb}.
Also see, {\it e.g.}, \cite{Sirunyan:2019ctn, ATLAS-CONF-2019-040}.} 
especially in final states with substantial missing transverse momenta plus large hadronic activities implies that the colored supersymmetric particles under QCD strong interaction may not have been copiously produced. With some simple assumptions, the interpretation of the current LHC data leads to the multi-TeV mass bound for the gluinos and light-generation squarks, making their direct discovery at the LHC increasingly difficult due to the kinematic limitation. 
On the other hand, it is quite conceivable that the charginos and neutralinos in the EW sector could be significantly lighter than the colored SUSY partners, as argued in the scenarios of ``natural SUSY'' \cite{Feng:1999zg,Hall:2011aa,Baer:2012up,Baer:2020kwz}. The direct production of electroweak supersymmetric particles at the LHC is of lower rate \cite{Baer:1994nr} and the current direct search bounds are thus rather weak \cite{Athron:2018vxy}. 
In addition, some DM considerations favor a situation for nearly degenerate EWkinos \cite{ArkaniHamed:2006mb}, making their identification more challenging \cite{Giudice:2010wb} owing to the lack of substantial missing transverse momenta. 
It is thus strongly motivated to target EWkinos in the hope to extend the SUSY search coverage. In this review, we focus on the electroweakinos and decouple the SUSY color and the scalar states.
We present a status summary for the EWkino searches at the LHC, and outline the near-future prospects. We also make connection with the DM direct detections. 

It is interesting to note that, although throughout the paper we work in a framework of the Minimal Supersymmetric extension of the Standard Model (MSSM) because of its clarity and predictability, our analyses and conclusions will be equally applicable to other color-singlet fermionic states (such as BSM heavy leptons) of ${\rm SU}(2)_{\rm L}$ singlet/doublet/triplet with a conserved global quantum number to assure the existence of a stable light neutral particle as the WIMP DM candidate. 

The rest of the article is organized as follows. We first present the model setup in Sec.~\ref{sec:model} by specifying the EWkino states and the model parameters of their masses and mixing. This also sets the tone for the parameter coverage in the searches. In Sec.~\ref{sec:DM}, we consider the DM direct detection and present the current bounds on the model parameters, that will serve as qualitative guidance and target for the future searches. The main body of this review is presented in Sec.~\ref{sec:Collider}, where we first show the predicted production cross sections for the EWkinos at hadron colliders and their decay modes in various theoretical scenarios, and then summarize the current bounds from LEP and LHC, and finally comment on the expectations for future colliders. We summarize the presentation and discuss some future prospects in Sec.~\ref{sec:sum}.

%% file: 2_model.tex
\section{Model Setup}
\label{sec:model} 

We start with the general BSM formulation with the new fermionic states of the $\rm SU(2)_L$ multiplets: a singlet $\tilde{B}$ (Bino), a triplet $\tilde{W}$ (Wino), and two doublets $\tilde{H}_d$ and $\tilde{H}_d$ (Higgsinos), as in the gaugino and Higgsino sectors in the MSSM, with soft-SUSY breaking masses as\footnote{If without any specification, $M_1, M_2$ and $\mu$ refer to their absolute values.}
\begin{equation}
   M_1,\quad M_2,\quad {\rm and}\quad \mu.
   \label{eq:masses}
 \end{equation}
The mass matrix for the neutral components in the gauge-eigenstate basis of  
$\psi^0=(\tilde{B}, \tilde{W}^0, \tilde{H}_d^0, \tilde{H}_u^0)$ is
\begin{equation}
M_{\tilde{N}}=
\left(
\begin{array}{cccc}
M_1&0&-c_\beta s_W m_Z&s_\beta s_W m_Z \\
0&M_2&c_\beta c_W m_Z&-s_\beta c_W m_Z \\
-c_\beta s_W m_Z&c_\beta c_W m_Z&0&-\mu\\
s_\beta s_W m_Z&-s_\beta c_W m_Z&-\mu&0
\end{array}
\right),
\label{eq:mN}
\end{equation}
where we have used the abbreviations $s_W=\sin\theta_W,\ c_W=\cos\theta_W$ with $\theta_W$ being the weak mixing angle, and $s_\beta=\sin\beta$ and $c_\beta=\cos\beta$ with 
$\tan\beta=\langle \tilde{H}_u^0 \rangle /\langle \tilde{H}_d^0 \rangle$.
Similarly, the mass matrix of the charged components 
in the basis of $\psi^\pm=(\tilde{W}^+, \tilde{H}_u^+, \tilde{W}^-, \tilde{H}_d^-)$ is
\begin{equation}
M_{\tilde{C}}=
\left(
\begin{array}{cc}
0_{2 \times 2}&X^T_{2 \times 2} \\
X_{2 \times 2} &0_{2 \times 2}
\end{array}
\right),\ \ \ 
{\rm with} \ \ \ 
X_{2 \times 2}=
\left(
\begin{array}{cc}
M_2&\sqrt{2}s_\beta m_W \\
\sqrt{2} c_\beta m_W &\mu
\end{array}
\right).
\label{eq:mC}
\end{equation}
After the diagonalization, we arrive at the neutral and charged mass eigenstates: the neutralinos $\tilde{\chi}_i^0\ (i=1,2,3,4)$ and the charginos $\tilde{\chi}_i^\pm\ (i=1,2)$, respectively, with increasing mass for a higher label $i$. We generically refer them as ``electroweakinos" (EWkinos). 

As such, $\tilde{\chi}_1^0$ is the lightest electroweakino and we will refer it as the ``lightest supersymmetric partner'' (LSP). If an electroweakino carries a dominant component of a gaugino or Higgsino with an approximate mass given by $M_1, M_2$ or $\mu$, we will call the state ``Bino-like'', ``Wino-like'' or ``Higgsino-like'', respectively. 
Furthermore, if one of the three mass scales is significantly lower than the other two, the LSP could be essentially a pure Bino, a pure Wino or a pure Higgsino. In this case, it has become customary to liberally label the nearly degenerate multiplets as Wino LSPs or Higgsino LSPs. Obviously, the LSP $\tilde{\chi}_1^0$ is most characteristic since it  can produce missing momentum in collider experiments if R-parity is conserved and serves as the WIMP DM candidate. 
However, the ``next lightest supersymmetric partners'' (NLSPs) can be of special importance as well, since they may govern the collider signatures by the production and subsequent decays to the LSP. 
In the rest of this section, we categorize the parameter configurations into several characteristic cases according to the nature of the LSPs and NLSPs, and discuss their mass spectra.

\subsection{Scenario 1: Bino LSP 
\label{sec:Bino}}

First we consider the scenario where $M_1$ is lower than the other two $M_2$, and $\mu$. This is a quite generic scenario and most common example is the minimal Super-Gravity Model (mSUGRA) with  universal gaugino masses \cite{Chamseddine:1982jx, Barbieri:1982eh, Ibanez:1982ee, Hall:1983iz, Ohta:1982wn, Ellis:1982wr, AlvarezGaume:1983gj}.
The Bino LSP is a gauge singlet Majorana state whose annihilation in the early universe occurs through squark and slepton exchange. In the scope of this review, we assume the scalar sector is heavy and thus decoupled. Therefore, a pure Bino as the dark matter would lead to an over-closure of the universe, and we will consider its mixing with the Wino and Higgsino for the two cases
\begin{eqnarray}
&{\rm Scenario~1a~~ }& M_1 < M_2< \mu:~~\tilde{\chi}^0_1~\text{Bino-like LSP;}~\tilde{\chi}^\pm_1,\tilde{\chi}^0_{2}~\text{Wino-like NLSPs.}\\
&{\rm Scenario~1b~~ }& M_1 <\mu< M_2:~~\tilde{\chi}^0_1~\text{Bino-like LSP;}~\tilde{\chi}^\pm_1,\tilde{\chi}^0_{2,3}~\text{Higgsino-like NLSPs.}
\end{eqnarray}
For Scenario 1a, we focus on the Bino-Wino mixing and the Higgsino can be decoupled by taking $|\mu| \gg M_1, M_2$. The effective neutralino mass matrix can be expressed as
\begin{equation}
M =
\begin{pmatrix} 
M_1 & 0  \\
0   & M_2
\end{pmatrix} - s_{2\beta}\frac{M_Z^2}{\mu}
\begin{pmatrix}
s_W^2 & -s_Wc_W \\
-s_Wc_W & c_W^2
\end{pmatrix}
+ \mathcal{O}\left(\frac{M_Z^3}{\mu^2} \right)
\end{equation}
The mixing only occurs through the mixture of Higgsino states at the order of ${\cal O}(M_Z^2/\mu)$. The mass splitting between Wino-like NLSPs $\tilde{\chi}_1^\pm$ and $\tilde{\chi}_2^0$ 
is generated at the order of ${\cal O}(M_Z^3/\mu^2)$ or at one-loop level. 
For Scenario 1b, we focus on the Bino-Higgsino mixing and the Wino states can be decoupled by taking 
$M_2 \gg M_1, \mu$. The effective neutralino mass matrix in the basis $\tilde{B}, \tilde{H}^0_{1,2} \equiv (\tilde{H}^0_u \mp \tilde{H}^0_d)/\sqrt{2}$ is
\begin{equation}
\aligned
M = &
\begin{pmatrix} 
M_1 & -\frac{s_\beta+c_\beta}{\sqrt{2}}s_WM_Z & \frac{s_\beta-c_\beta}{\sqrt{2}}s_WM_Z \\
-\frac{s_\beta+c_\beta}{\sqrt{2}}s_WM_Z & \mu & 0\\
\frac{s_\beta-c_\beta}{\sqrt{2}}s_WM_Z & 0 & -\mu
\end{pmatrix} \\
&  - \frac{M_W^2}{2M_2}
\begin{pmatrix}
0 & 0 & 0\\
0 & 1+s_{2\beta} & c_{2\beta}\\
0 & c_{2\beta} & 1-s_{2\beta}
\end{pmatrix}
+ \mathcal{O}(\frac{M_W^3}{M_2^2})
\endaligned
\end{equation}

\begin{figure}
\centering
\begin{subfigure}{.4\textwidth}
\centering
  \includegraphics[width=\textwidth]{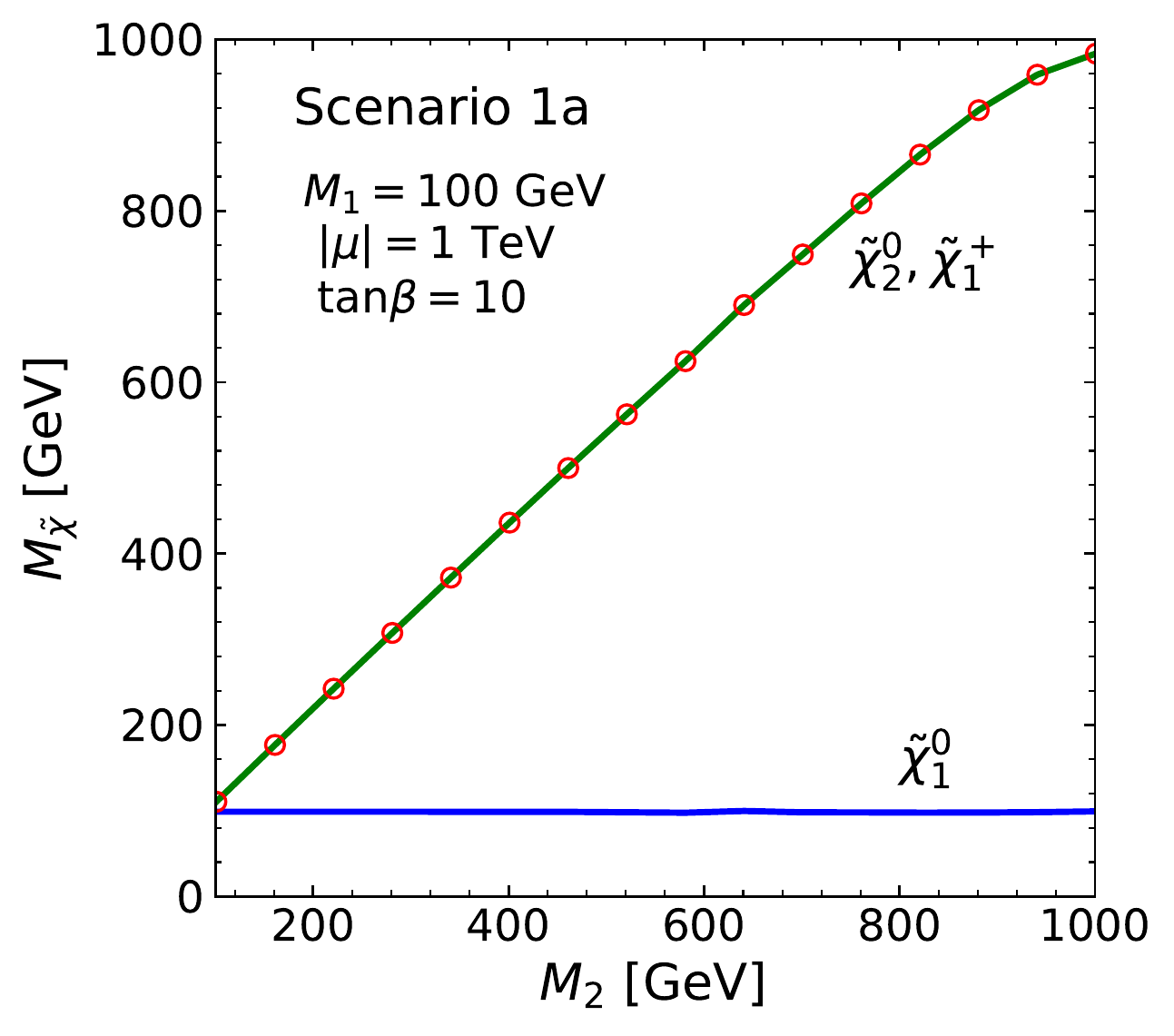}
\vspace{-0.7cm}
\caption{}
\vspace{0.7cm}
\end{subfigure}
\begin{subfigure}{.4\textwidth}
\centering
\includegraphics[width=\textwidth]{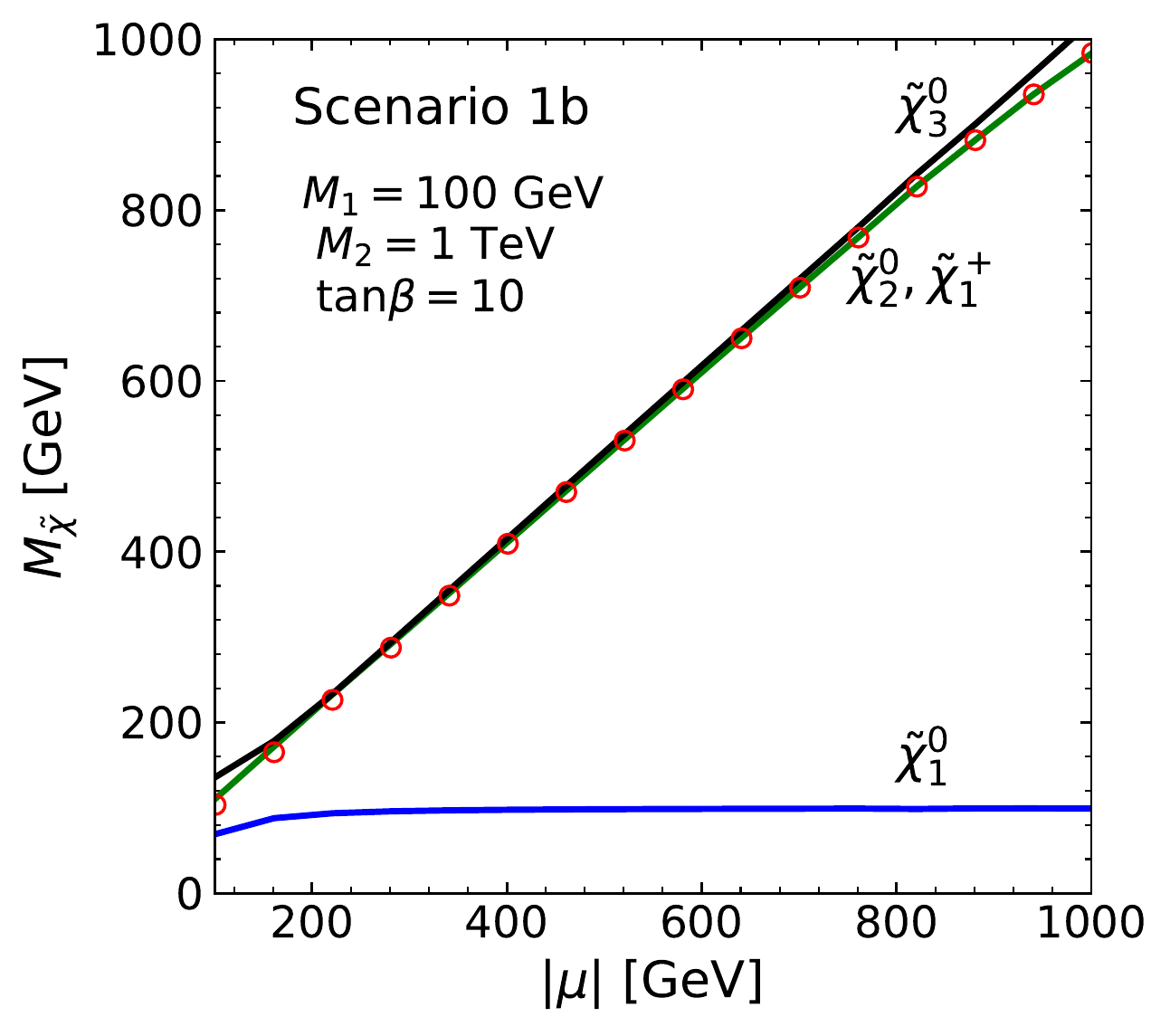}
\vspace{-0.7cm}
\caption{}
\vspace{0.7cm}
\end{subfigure} \\
\begin{subfigure}{.4\textwidth}
\centering
\includegraphics[width=\textwidth]{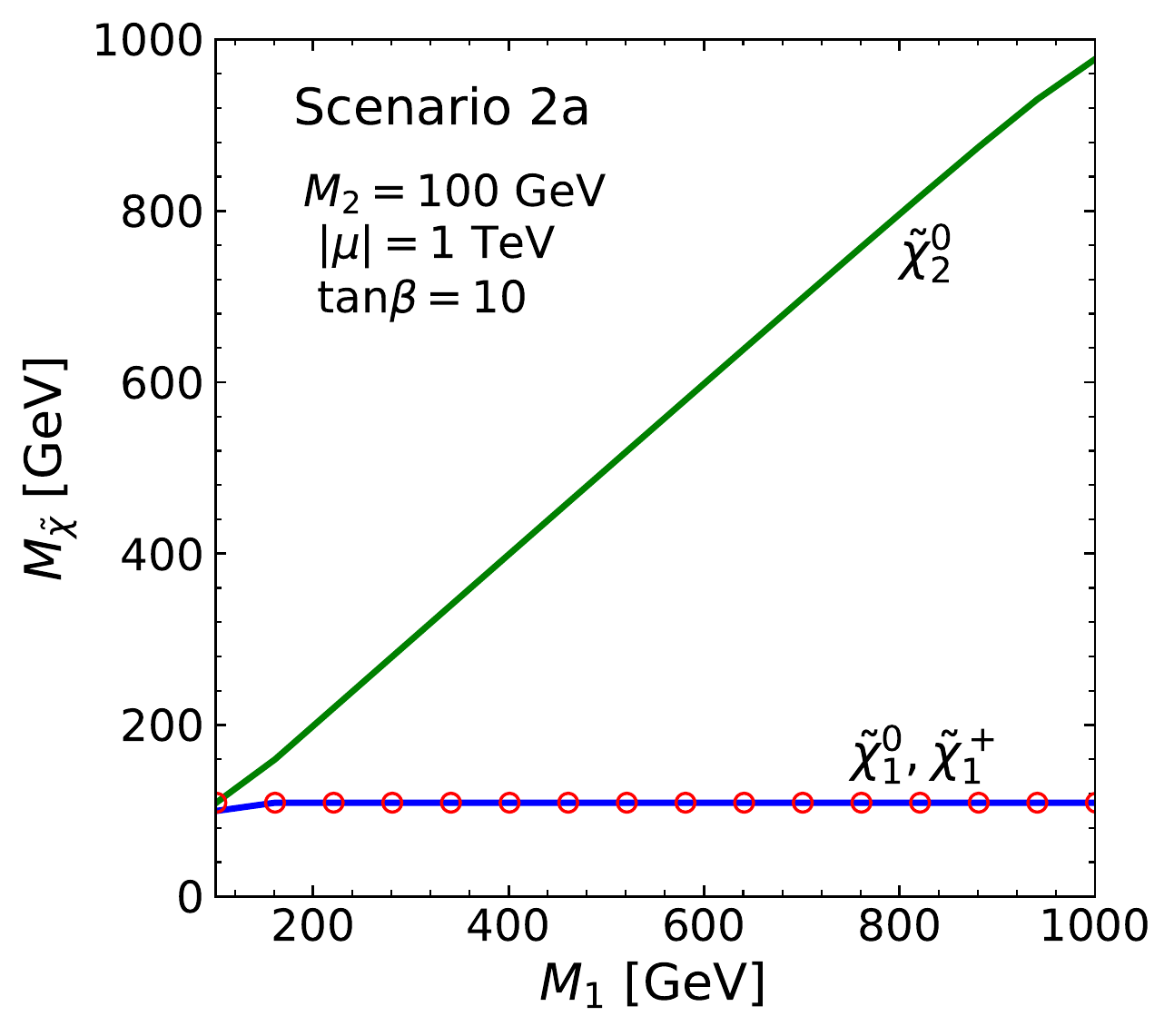}
\vspace{-0.7cm}
\caption{}
\vspace{0.7cm}
\end{subfigure}
\begin{subfigure}{.4\textwidth}
\centering
\includegraphics[width=\textwidth]{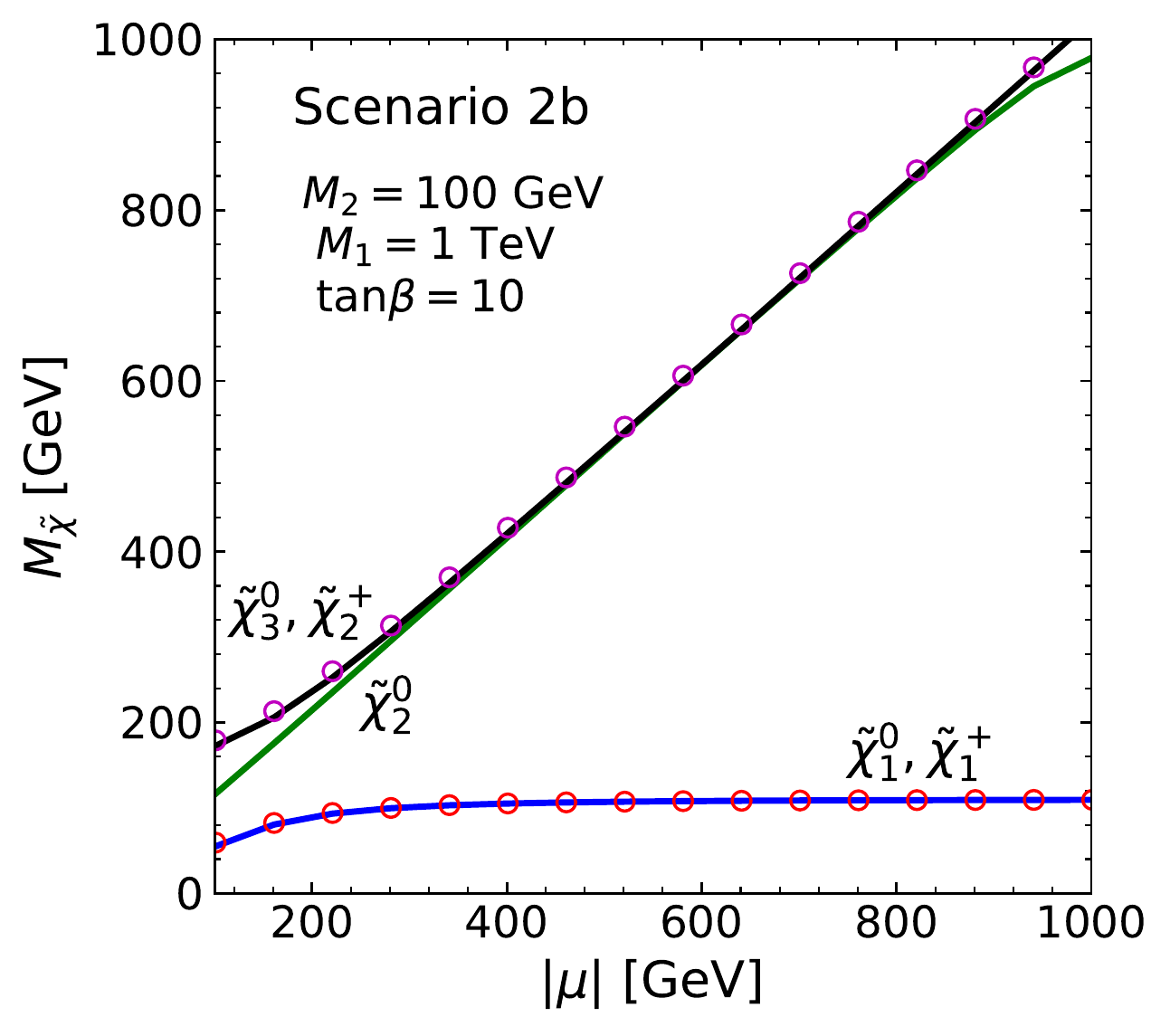}
\vspace{-0.7cm}
\caption{}
\vspace{0.7cm}
\end{subfigure} \\
\begin{subfigure}{.4\textwidth}
\centering
\includegraphics[width=\textwidth]{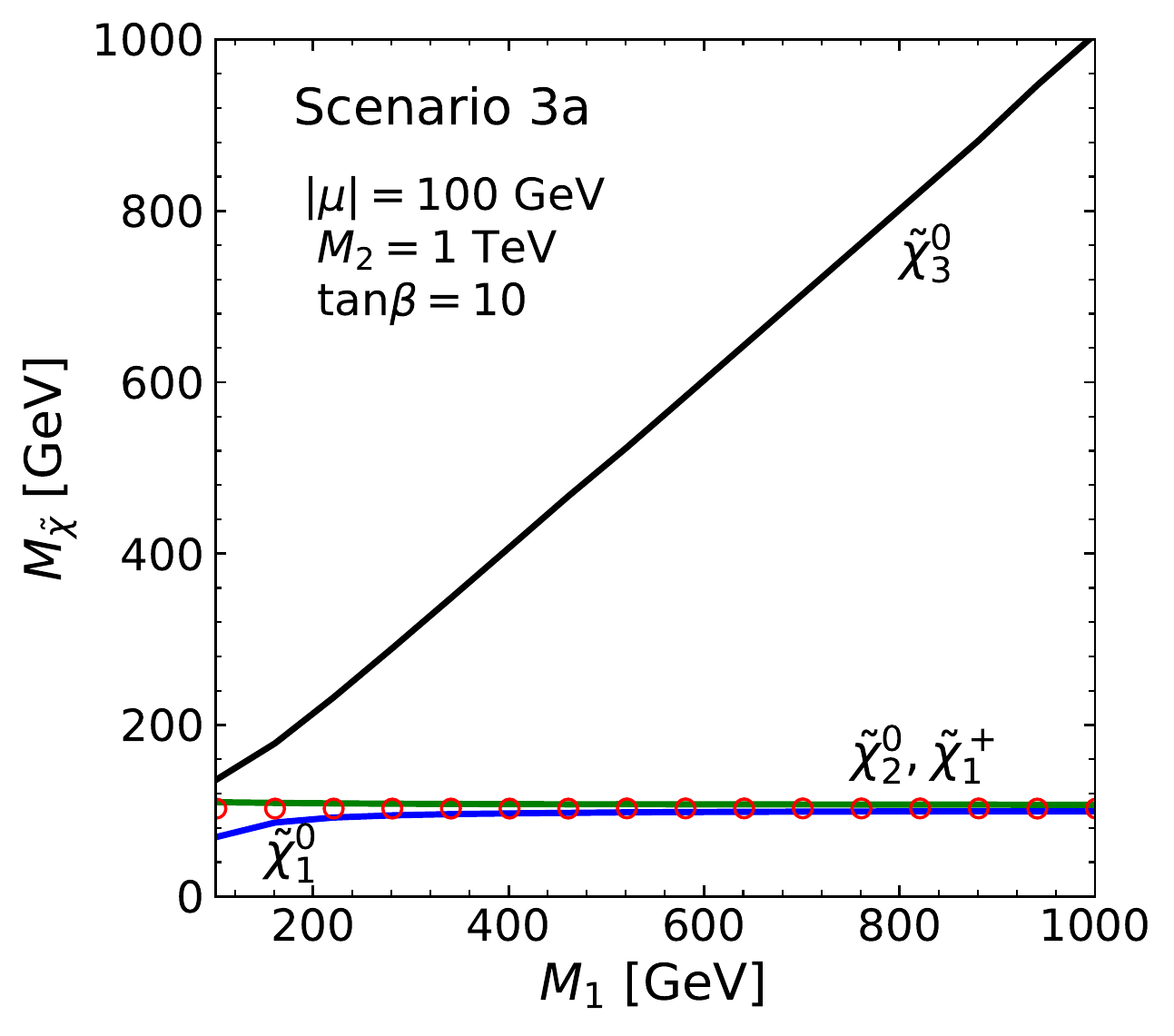}
\vspace{-0.7cm}
\caption{}
\vspace{0.7cm}
\end{subfigure}
\begin{subfigure}{.4\textwidth}
\centering
\includegraphics[width=\textwidth]{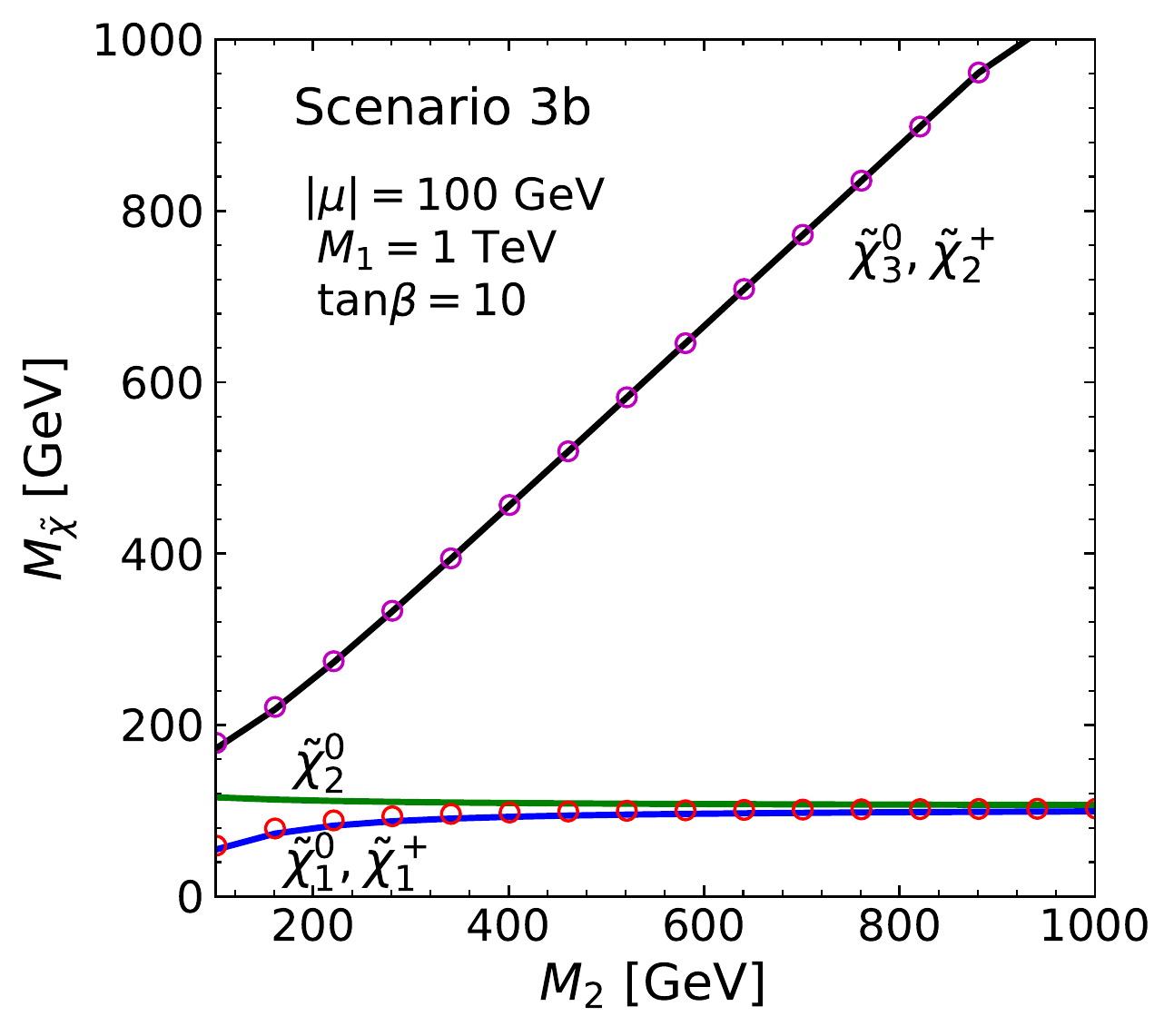}
\vspace{-0.7cm}
\caption{}
\vspace{0.7cm}
\end{subfigure}
\vspace{-0.5cm}
\caption{Electroweakino masses (vertical) of the LSP and NLSP states versus the NLSP mass parameter (horizontal) for the three scenarios described in the text. Solid curves are for neutralino states and circles for chargino states.  The LSP mass parameter is set to be 100 GeV, the heaviest mass parameter is set to be 1 TeV, and $\tan\beta =10$.} 
\label{fig:s1-mass}
\end{figure}

In Fig.~\ref{fig:s1-mass}, we illustrate the EWkino masses of the LSP states (nearly horizontal lines) and NLSP states (nearly diagonal lines) versus the NLSP mass parameter.
Solid curves are for neutralino states and circles for chargino states. Without losing much generality, for illustrative purposes, we fix the LSP mass parameter to be 100 GeV, the heaviest mass parameter to be 1 TeV, and $\tan\beta =10$.
Figures \ref{fig:s1-mass}(a) and \ref{fig:s1-mass}(b) are for Scenario~1a versus the mass parameters $M_{2}$, and for Scenario~1b versus $\mu$, respectively. We see, in Scenario~1b, {\it e.g.}, that a mass splitting among the Higgsino multiplet is only appreciable when $|\mu| \sim M_1$ or $|\mu| \sim M_2$. 

\subsection{Scenario 2: Wino LSPs\label{sec:modelWinos}}
We next consider the scenario where $M_2$ is lower than the other two, $M_1$ and $\mu$. This is a  scenario with Wino-like LSP favored by the anomaly-mediation of SUSY breaking (AMSB) \cite{Randall:1998uk, Giudice:1998xp, Gherghetta:1999sw}.
The dimension-4 effective  Lagrangian describing the interaction of  the Wino triplet ($\tilde W$) with the SM electroweak gauge bosons  is given by 
\begin{equation}
\mathcal{ L}_{V\tilde W\tilde W} \supseteq -g \left(\overline{\tilde{W}^0} \gamma^\mu \tilde{W}^{+} W^{-}_\mu +{\rm h.c.} \right) + g \overline{\tilde{W}^-} \gamma^\mu \tilde{W}^{-} \left(\cos \theta_W Z_\mu +\sin \theta_W A_\mu \right),
\end{equation}
where $g$  is the ${\rm SU}(2)_{\rm L}$ gauge coupling.
In the absence of large corrections from couplings with the fermion and sfermion sectors of the MSSM, these gauge interactions induce a mass splitting between the charged and neutral Winos ($\delta m_{\tilde{W}}$), which, at the two-loop order can be parametrized as follows~\cite{Ibe:2012sx}
\begin{align}
\frac{\delta m_{\tilde{W}}}{1 {~\rm MeV}} & = -413.315 + 305.383 \left(\log \frac{m_{\tilde{\chi}_0}}{1 {~\rm GeV}}\right) - 60.8831 \left(\log \frac{m_{\tilde{\chi}_0}}{1 {~\rm GeV}}\right)^2  \\ \nonumber
 &  + 5.41948 \left(\log \frac{m_{\tilde{\chi}_0}}{1 {~\rm GeV}}\right)^3- 0.181509 \left(\log \frac{m_{\tilde{\chi}_0}}{1 {~\rm GeV}}\right)^4,
\end{align}
where $m_{\tilde{\chi}_0}$ is the neutral Wino mass. The $m_{\tilde{\chi}_0}$-dependence of the mass difference is rather weak and it is approximately 160 MeV. 
The corresponding decay lifetime of the charged Wino to a neutral Wino and a charged pion is given in terms of the $c\tau$-value by Ref.~\cite{Ibe:2012sx}. 
\begin{equation}
c\tau \simeq 3.1 {~\rm cm} \left[\left(\frac{\delta m_{\tilde{W}}}{164 {~\rm MeV}}\right)^{3} \sqrt{1- \frac{m_\pi^2}{\delta m_{\tilde{W}}^2}}\ \right]^{-1},
\label{eq:Wino_split}
\end{equation}
with $m_\pi$ being the charged pion mass. We have normalized the mass difference to 164 MeV, which is the mass splitting in the limit $m_{\tilde{\chi}_0} \gg M_W$. 

Beyond the pure Wino situation, we consider two distinctive scenarios for the lower-lying state mixing
\begin{eqnarray}
&{\rm Scenario~2a}& M_2 < M_1< \mu:~\tilde{\chi}^\pm_1,\tilde{\chi}^0_{1}~\text{Wino-like LSPs;}~\tilde{\chi}^0_{2}~\text{Bino-like NLSP.}\\
&{\rm Scenario~2b}& M_2 < \mu < M_1:\tilde{\chi}^\pm_1,\tilde{\chi}^0_{1}~\text{Wino-like LSPs;}~\tilde{\chi}^\pm_2,\tilde{\chi}^0_{2,3}~\text{Higgsino-like NLSPs.}~~~~~
\end{eqnarray}

As for the Wino-Higgsino mixing in Scenario 2b, the Bino can be decoupled by taking $M_1 \gg M_2\;{\rm and}\;\mu$, and the effective neutralino mass matrix can be effectively described by
\begin{equation}
\aligned
M = &
\begin{pmatrix} 
M_1 & \frac{s_\beta+c_\beta}{\sqrt{2}}c_WM_Z & -\frac{s_\beta-c_\beta}{\sqrt{2}}c_WM_Z \\
\frac{s_\beta+c_\beta}{\sqrt{2}}c_WM_Z & \mu & 0\\
-\frac{s_\beta-c_\beta}{\sqrt{2}}c_WM_Z & 0 & -\mu
\end{pmatrix} \\
&  - \frac{M_Z^2s_W^2}{2M_1}
\begin{pmatrix}
0 & 0 & 0\\
0 & 1+s_{2\beta} & c_{2\beta}\\
0 & c_{2\beta} & 1-s_{2\beta}
\end{pmatrix}
+ \mathcal{O}(\frac{M_Z^3}{M_1^2})
\endaligned
\end{equation}

Figures \ref{fig:s1-mass}(c) and \ref{fig:s1-mass}(d) show the physical LSP/NLSP masses for Scenario~2a versus the mass parameters $M_{1}$, and for Scenario~2b versus $|\mu|$, respectively. 

\subsection{Scenario 3: Higgsino LSPs\label{sec:modelHiggsinos}}
For $\mu$ to be lower than the other two, $M_1$ and $M_2$, the Higgsino multiplet is essentially the LSPs. This scenario is favored by the argument of the ``natural SUSY'' \cite{Feng:1999zg,Hall:2011aa,Baer:2012up}. 
The effective interaction Lagrangian at dimension-4 for charged ($\tilde H^\pm$) and neutral Dirac 
($\tilde H^0$) Higgsinos with the SM electroweak gauge bosons is given by
\begin{align} 
  {\cal L}_{V\chi H \chi H} & \supseteq  -\frac{g}{\sqrt{2}} \left( \overline{\tilde{H}^0}\gamma^\mu {\tilde H}^-\, W^+_\mu +{\rm h.c.} \right)
+g \overline{{\tilde H}^-}\gamma^\mu \tilde H^-\,\left(\frac{1/2-s^2_W}{c_W}\,Z_\mu \, + s_W A_\mu \right) \nonumber \\
                                    & -  \frac{g}{2 c_W}\, \overline{{\tilde H}^0}\gamma^\mu {\tilde H}^0\, Z_\mu,
\end{align}
with $s_W=\sin\theta_W$ and $c_W=\cos\theta_W$.
The above interactions induce a one-loop mass splitting between the charged and neutral states ($\delta m_{\tilde{H}}$) which can be written as
\begin{eqnarray}
\delta m_{\tilde{H}} &=& \frac{g^2}{16\pi^2}m_{\tilde{H}}\sin^2\theta_W f\left(\frac{M_Z}{m_{\tilde{H}}}\right),\\
f(r) &=& r^4\ln r - r^2 - r\sqrt{r^2 - 4}(r^2 + 2)\ln\frac{\sqrt{r^2 - 4} + r}{2}.
\nonumber
\end{eqnarray}
The corresponding decay lifetime of the charged Higgsino to a charged pion
can be parametrized in terms of the $c\tau$-value as~\cite{Fukuda:2017jmk} 
\begin{equation}
c\tau \simeq 0.7 {~\rm cm} \times \left [\left(\frac{\delta m_{\tilde{H}}}{340 {~\rm MeV}}  \right)^3 \sqrt{1-\frac{m_\pi^2}{\delta m^2_{\tilde{H}}}} \   \right]^{-1}.
\label{eq:Higgsino_split}
\end{equation}
As we can observe from Eqs.~(\ref{eq:Wino_split}) and (\ref{eq:Higgsino_split}), for typical values of the mass splitting between the charged and neutral states, the charged Wino has a considerably larger decay length compared to the charged Higgsino.  This makes  the searches for long-lived particles potentially more favorable for Winos than for Higgsinos. 

Depending on which one is lighter between $M_1$ and $M_2$, there are two scenarios for the lower-lying state mixing
\begin{eqnarray}
&{\rm Scenario~3a}& \mu < M_1< M_2:~\tilde{\chi}^\pm_1,\tilde{\chi}^0_{1,2}~\text{Higgsino-like LSPs;}
~~\tilde{\chi}^0_{3}~\text{Bino-like NLSP.}~\\
&{\rm Scenario~3b}& \mu <M_2 < M_1:\tilde{\chi}^\pm_1,\tilde{\chi}^0_{1,2}~\text{Higgsino-like LSPs;}
~\tilde{\chi}^\pm_2,\tilde{\chi}^0_{3}~\text{Wino-like NLSPs.}~~~~~
\end{eqnarray}

The physical masses of the LSPs/NLSPs are shown in Fig.~\ref{fig:s1-mass}(e) for Scenario 3a versus the mass parameters $M_1$ and Fig.~\ref{fig:s1-mass}(f) for Scenario 3b versus $M_2$ with $\mu=100$ GeV. Relatively large mixing occurs for smaller values $M_1,M_2<300$ GeV, when being close to $\mu$.

\subsection{Simplified model and phenomenological MSSM}

The SUSY partner mass spectrum crucially depends on the SUSY breaking scale and the mechanism to mediate the effects to the SM sector \cite{Chung:2003fi}. Well-formulated scenarios include the mSUGRA \cite{Chamseddine:1982jx, Barbieri:1982eh, Ibanez:1982ee, Hall:1983iz, Ohta:1982wn, Ellis:1982wr, AlvarezGaume:1983gj} that predicts a Bino-like LSP with  $M_1:M_2:M_3\approx 1:2:7$; the minimal gauge-mediation (GMSB) that yields a very light gravitino LSP \cite{Dine:1981gu, Nappi:1982hm, AlvarezGaume:1981wy, Dine:1993yw, Dine:1994vc, Dine:1995ag, Giudice:1998bp}, anomaly-mediation (AMSB) \cite{Randall:1998uk, Giudice:1998xp, Gherghetta:1999sw} that prefers a Wino-like LSP with $M_2:M_1:M_3\approx 1:3:8$; and the ``natural SUSY'' that argues for a Higgsino LSP 
with $\mu\sim {\cal O}(M_Z)$ \cite{Feng:1999zg,Hall:2011aa,Baer:2012up}. However, those minimal and predictive scenarios are too restrictive and highly constrained by the current experimental observations, such as the direct searches at the LHC and 125 GeV SM-like Higgs boson for mSUGRA and GMSB \cite{Arbey:2011ab, Ajaib:2012vc, Kang:2012ra, Craig:2012xp, Albaid:2012qk}, 
and by astronomical constraints for AMSB \cite{Cohen:2013ama}.
It is therefore prudent to consider the less restrictive situation where the soft-SUSY breaking masses are treated as independent free parameters as outlined in the previous sections, in accordance with the ``simplified model'', defined by an effective Lagrangian~\cite{0810.3921,1105.2838}. In the simplified models under the current consideration,  the nature of the sparticles is set to pure states, while the masses and decay branching fractions are set to chosen values. In the phenomenological MSSM, or pMSSM~\cite{0812.0980}, the masses, cross-sections, and branching fractions are instead derived from the $\mu$, $M_1$ and $M_2$ values, assumed to be free parameters. The pMSSM therefore captures the complex pattern of the EWkinos masses and decay channels realized when the electroweakinos  have sizable mixings among the Bino, Winos and Higgsinos.

%% file: 3_dm.tex
\section{Dark Matter Relic Density and Direct Detection Constraints}
\label{sec:DM} 

The nature of Dark Matter  is one of the most outstanding puzzles in contemporary physics. While there is stunning evidence for its existence in the Universe in the form of cold non-baryonic matter, and it provides a clear argument for physics beyond the Standard Model, there is no particular indication on what form it actually takes. This is due to the fact that, so far, it only manifests itself through gravitational interactions. 
There is, however, a strong theoretical preference for DM to be weakly interacting massive particles (WIMPs) near the EW scale, because of the coincidence to yield the correct ballpark of the relic abundance and the possible connection to the next scale of  BSM physics. Among the options of viable cold DM candidates, the lightest EWkino (LSP) in $R$-parity conserving SUSY theories, provides a natural candidate for DM \cite{Jungman:1995df}. In this section, we discuss the DM connection of the EWkinos.

\subsection{Relic density}
The paradigm of thermal decoupling, based upon applications to cosmology of statistical mechanics, particle and nuclear physics, is enormously successful at making detailed predictions for observables in the early universe, including the abundances of light elements and the cosmic microwave background. It is somewhat natural to invoke a similar paradigm to infer the abundance of DM as a thermal relic from the early universe uniquely from the underlying DM particle properties.
The relic abundance of dark matter particles is set by their annihilation cross section $\sigma \propto g_{\rm eff}^4/M_{\rm DM}^2$ in the early universe ~\cite{Lee:1977ua,Goldberg:1983nd,Steigman:2012nb}  
\begin{equation}
\Omega h^2 = 0.11 
\times \left(\frac{2.2 \times 10^{-26}~{\rm cm}^3/{\rm s}}{\langle \sigma v \rangle_{{\rm freeze}}}\right) ,
\end{equation}
To avoid over-closure of the universe, today's relic abundance $\Omega h^2 \sim 0.11$ translates to a bound on the dark matter mass as 
\begin{equation}
M_{\rm DM} < 1.8~{\rm TeV} \left(\frac{g^2_{\rm eff}}{0.3}\right) .
\end{equation}
The natural presence of the TeV scale and the EW coupling strength leads to the notion of ``WIMP Miracle'' \cite{Jungman:1995df}. 
Owing to the efficient annihilation to the SM particles in the early universe, the Wino-like and Higgsino-like DM will typically be under-abundant. A heavier Wino (Higgsino) DM with a mass of 3.1 TeV (1.1 TeV), however, could fully account for the thermal relic density \cite{Cohen:2013ama,Bramante:2015una}, that provides a well-motivated target for collider searches.

Beyond the generic consideration above, acceptable WIMP DM relic density may be achievable 
by tuning the mass parameters. Widely explored examples include the co-annihilation mechanisms \cite{Griest:1990kh,Mizuta:1992qp,Edsjo:1997bg,Baer:2002fv}, in which the LSP mass is close to that of another sparticle so that they effectively annihilate into SM particles to reach a desirable relic abundance, such as squark co-annihilation \cite{Boehm:1999bj,Ellis:2001nx,Arnowitt:2001yh}, slepton co-annihilation \cite{Ellis:1998kh,Ellis:1999mm,Gomez:1999dk,Nihei:2002sc} and Bino-Wino co-annihilation \cite{Baer:2005jq}. They all lead to rich and characteristic phenomenology at colliders because of the co-existence of light SUSY states. $A$-funnel annihilation is another example \cite{Berlin:2015wwa,Freese:2015ysa}, in which the mass of the CP-odd Higgs boson is tuned to be $m_{A}\approx 2m_{\tilde \chi^0_1}$ for effective LSP annihilation. In this case, it is possible to make the EWkinos as heavy as ${\cal O}(10$ TeV), still consistent with the bound of thermal relic abundance \cite{Gilmore:2007aq}.  
For such a heavy WIMP DM mass, indirect detections of the relic DM annihilation by astro-physical observations may achieve better sensitivities \cite{Cohen:2013ama,Kar:2019mcq}. 

\subsection{Direct detection}
If the halo of the Milky Way consists of WIMPs, then a WIMP flux of about $10^2-10^3$cm$^{-2}$s$^{-1}$ must pass through the Earth's surface.
A convincing proof of the WIMP hypothesis would be the direct detection of these particles, by, for example, observation of nuclear recoil after WIMP-nucleus elastic scattering on a nuclear target in the underground experiments. 

For EWkinos as the DM candidate, the neutralino LSP couples to the spin of the nucleus 
via the axial-vector interaction $Z\tilde{\chi}_1^0\tilde{\chi}_1^0$ (spin-dependent, SD), and is independent of the nucleus spin via the scalar interaction $h\tilde{\chi}_1^0\tilde{\chi}_1^0$ (spin-independent, SI). 
The scattering cross section on a heavy nuclear target with atomic number $A$ will be proportional to $A^2$ in SI interactions  due to the coherent effect of the nucleons. 
DM direct detections are thus more sensitive to the SI interactions. On the other hand, the SD interactions may still be significant because of the stronger gauge interactions via the $Z$-exchange.

\subsubsection{Current bounds on WIMP-nucleon cross-sections from direct detection}

\begin{figure}
\centering
\includegraphics[width=\textwidth]{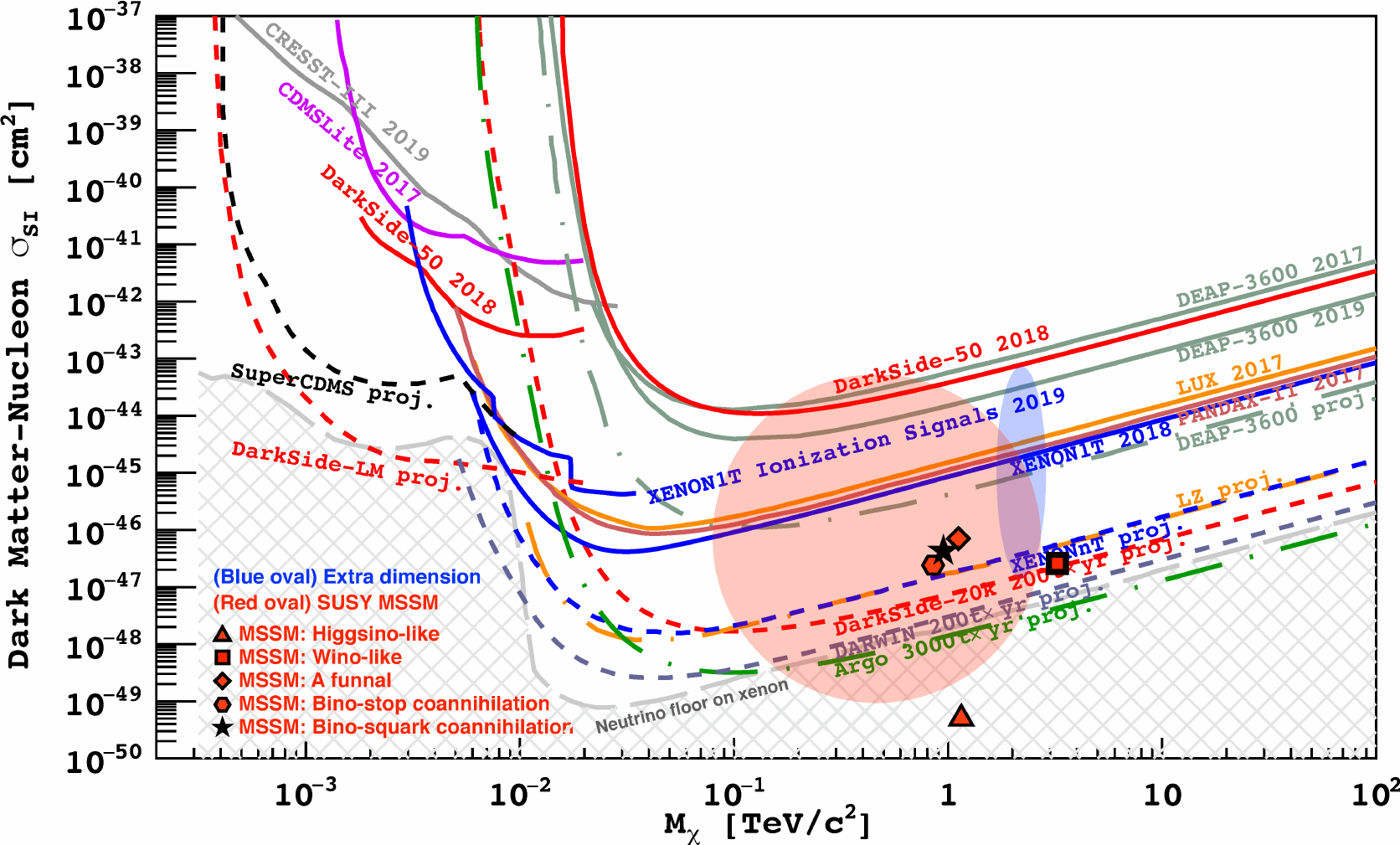}
\caption{90\% CL exclusion limits from DM direct detection \cite{Strategy:2019vxc} on the SI cross section versus the DM mass (solid and dashed lines) and some representative predictions in SUSY models (circle, oval and red dotted symbols) \cite{Cushman:2013zza, Hisano:2015rsa, Chen:2019gtm}.}
\label{fig:dm_dd}
\end{figure}

At present, direct detection searches \cite{Strategy:2019vxc} have excluded spin-independent dark matter-nucleon cross sections as low as $10^{-46}{\rm cm}^2$, shown as solid curves in Fig.~\ref{fig:dm_dd}, and spin-dependent cross sections as low as $10^{-41}{\rm cm}^2$. In Fig.~\ref{fig:dm_dd}, the leading results in the 5 GeV range and below come from the DarkSide-50 LAr TPC low-mass search and from cryogenic solid-state detectors, while at higher masses from cryogenic noble liquids, led for the past decade by the pioneering XENON program at LNGS. Projected sensitivities of near-future direct detection dark matter searches are shown in Fig.~\ref{fig:dm_dd} as dashed curves. Three mid-term searches using Xe TPCs—LZ, PANDA, and XENON-nT, all aim to reach $10^{-48}{\rm cm}^2$ scale sensitivity at 30 GeV dark matter mass. The DarkSide-20k experiment expects to reach the $10^{-47}{\rm cm}^2$ scale at 1 TeV. Long-term future searches using Xe (DARWIN) and Ar (ARGO) project reaching beyond $10^{-48}{\rm cm}^2$ in the next decade. For spin-dependent interactions, near-term future experiments using Xe and ${\rm CF}_3$ targets project to reach sensitivity to $10^{-42}{\rm cm}^2$ WIMP-neutron and WIMP-proton cross sections, at 50 GeV. At low mass (around 1 to 10 GeV), solid state experiments, {\it e.g.}, SuperCDMS, expect to achieve $10^{-42}{\rm cm}^2$ cross section reach on a 5-year time scale.

\subsubsection{Theory parameter space and complementarity of direct detection-collider searches}

The null results from the DM direct detection have put stringent limits on the dark matter-nucleon scattering cross sections, excluding much of the parameter region for many WIMP dark matter models and thus challenging the WIMP miracle paradigm. Yet, caution needs to be taken when interpreting the current DM direct detection results since the DM interactions with the SM particles may be rather subtle. 
In Fig.~\ref{fig:dm_dd}, we include the theoretical predictions for the general MSSM (large red circle) and the Kaluza-Klein universal extra-dimensional model (blue oval). We also show the special cases of loop-suppressed Wino-like (red square, Scenario 2) and Higgsino-like (red triangle, Scenario 3) DM. Of particular interest are the cases to yield the correct relic abundance via Bino-stop co-annihilation (red hexagon), Bino-squark co-annihilation (black star), and via the CP-odd Higgs boson funnel (red diamond). 

It has been realized that there exist ``blind spots'' in the SUSY neutralino parameter space where the direct detection cross section is highly suppressed due to subtle cancellation of the couplings \cite{Cheung:2012qy}. The direct detection rate of the neutralino dark matter in the underground laboratories sensitively depends on the couplings of $h\tilde{\chi}^0_1\tilde{\chi}^0_1$ and $Z\tilde{\chi}^0_1\tilde{\chi}^0_1$, which are governed by the components of the $\tilde{\chi}^0_1$ admixture. If the theory parameters satisfy the following condition
\begin{equation}
(m_{\tilde{\chi}^0_1}+\mu\sin2\beta)\left(m_{\tilde{\chi}^0_1} -\frac{1}{2}(M_1+M_2+(M_1-M_2)\cos\theta_W)\right) = 0,
\end{equation}
the $h\tilde{\chi}^0_1\tilde{\chi}^0_1$ coupling vanishes, and thus leads to a vanishing SI cross section. Analogously, the $Z\tilde{\chi}^0_1\tilde{\chi}^0_1$ coupling vanishes if 
\begin{equation}
M_1 = M_2,\quad |\mu| > \left|\frac{M_{1,2}}{\sin2\beta}\right|,\quad \text{sign}(\frac{M_{1,2}}{\mu}) = -1,
\end{equation}
or
\begin{equation}
\tan\beta = 1,
\end{equation}
which would lead to a vanishing SD cross section \cite{Han:2016qtc}. If the heavy CP-even Higgs boson in MSSM is not decoupled, it can also destructively interfere with the scattering via the light CP-even Higgs boson, leading to a new SI blind spot \cite{Arnowitt:2001yh, Huang:2014xua, Baum:2017enm}. The condition can be approximately written as 
\begin{equation}
2(m_\chi + \mu\sin2\beta)\frac{1}{m_h^2} \simeq -\mu\tan\beta\frac{1}{m_H^2},
\end{equation}
for moderate or large values of $\tan\beta$.
It has been shown~\cite{Han:2018gej} that the blind spots still exist after the one-loop corrections are included, with their exact locations slightly shifted at an order of $\mathcal{O}(1\%)$. In some regions, the one-loop corrections to the SI cross section can reach values up to a few times $10^{-47}~{\rm cm}^2$, which will be detectable at future multi-ton scale liquid Xenon experiments. 

While the above arguments clearly indicate the need to improve the detection sensitivity for discovery, yet it calls for complementary searches at colliders. 
Indeed, SUSY searches at the LHC will substantially extend the coverage of the DM direct detections to the TeV mass region, regardless the direct detection blind-spot scenarios \cite{Han:2016qtc}. 
In the optimistic situation where a signal is observed either in the DM direct detection or at the LHC experiments, determining its mass scale and coupling is of ultimate importance. Only with the achievements in both experiments, can one reach a full characterization of SUSY dark matter.


%% file: 4_collider.tex
\section{Collider Searches }
\label{sec:Collider}


\subsection{Production at $e^+e^-$ colliders}

The EWkinos can be pair-produced by electroweak processes at colliders. At $e^+e^-$ colliders, assuming decoupling of the sleptons, the leading production processes are through $s$-channel exchange of $\gamma/Z$ bosons 
\begin{equation}
e^+e^-\rightarrow \gamma^*/Z^* \to \tilde{\chi}^+_i\tilde{\chi}^-_j, \quad \tilde{\chi}^0_i\tilde{\chi}^0_j,
\end{equation}
where $i,j=1\ldots 4$ for neutralinos and $i,j=1\ldots 2$ for charginos. 
The pair production cross sections scale like
\begin{equation}
\sigma \approx {\pi \alpha^2 Q_{ij}^2 \over s}\beta,
\end{equation}
where $s$ is the c.m.~energy squared, $\beta = \sqrt{1- (m_i+m_j)^2/s}$, and $Q_{ij}$ some gauge charges \cite{Choi:1998ut,Choi:2001ww}. The pair production rate can reach 1 fb$-$100 fb at $\sqrt{s} =1000$ GeV \cite{Djouadi:2007ik,Moortgat-Picka:2015yla}. 

The signal observation through their decay products in the SM particles would be straightforward owing to the clean experimental environment in $e^+e^-$ collisions \cite{Djouadi:2007ik}. In the case where the final states contain neither reconstructed tracks nor significant energy deposit from electroweakinos decays, the searches rely on the initial state radiation \cite{Carena:1986jp, Chen:1995yu,Hensel:2002bu,Berggren:2013vfa,Birkedal:2004xn,Dreiner:2012xm} 
\begin{equation}
e^+e^-\rightarrow \tilde{\chi}^+_i\tilde{\chi}^-_j \gamma, \quad \tilde{\chi}^0_i\tilde{\chi}^0_j \gamma,
\end{equation}
to identify an isolated hard photon plus large recoil missing mass $m_{recoil}^2 = (p_{e^+} + p_{e^-} - p_\gamma)^2$. The sensitivity reach is essentially kinematically limited, with $M_1, M_2, \mu \sim \sqrt s/2$. We refer further detailed discussions to some general reports \cite{Lebrun:2012hj}. 


Through the precision measurement of the $Z$ boson  invisible width, 
the LEP experiments placed a lower bound on the mass of \n\ at 45.5~GeV 
under the assumption of a significant \n-$Z$ coupling~\cite{Tanabashi:2018oca}.  
Massless neutralino are however allowed in scenarios with small couplings~\cite{0901.3485}.  By scanning particle production at the threshold, the LEP experiments also probed for the
existence of charginos in a quasi-model independent fashion. Results from the
searches in the LEP data led to the model-independent bound on the chargino mass 
\begin{equation}
m_{\tilde{\chi}^\pm} > 103.5\ {\rm GeV\ if}\ \Delta M(\cha,\n) \geq 3~{\rm GeV}. 
\end{equation}
The bound is reduced to 92.4~GeV for smaller $\Delta M$ values~\cite{LEP1}. We will take 100 GeV as our benchmark LSP mass for future illustrations. 


\subsection{Production at hadron colliders and NLSP decays}
\label{sec:ProDecay}

Assuming decoupling of the squarks, the leading contribution at hadron colliders are the $s$-channel Drell-Yan (DY) processes  with $\gamma/W/Z$ exchanges
\begin{equation}
p p\rightarrow {\ensuremath{\tilde{\chi}^+_i}}{\ensuremath{\tilde{\chi}^-_j}}X,\quad{\ensuremath{\tilde{\chi}^\pm_i}}{\ensuremath{\tilde{\chi}^0_j}}X, \quad{\ensuremath{\tilde{\chi}^0_i}}{\ensuremath{\tilde{\chi}^0_j}}X
\end{equation}
where $X$ generically denotes the hadronic remnants associated with the protons. Dominant processes are typically those that involves two Wino-like or two Higgsino-like states, since their couplings to 
$W$, $Z$ and $\gamma$ are unsuppressed. The EWkino pair production via $W$-exchange has the largest cross section due to the 
large SU(2)$_{\rm L}$ coupling. 
\begin{figure}
\centering
\begin{subfigure}{.4\textwidth}
\centering
  \includegraphics[width=\textwidth]{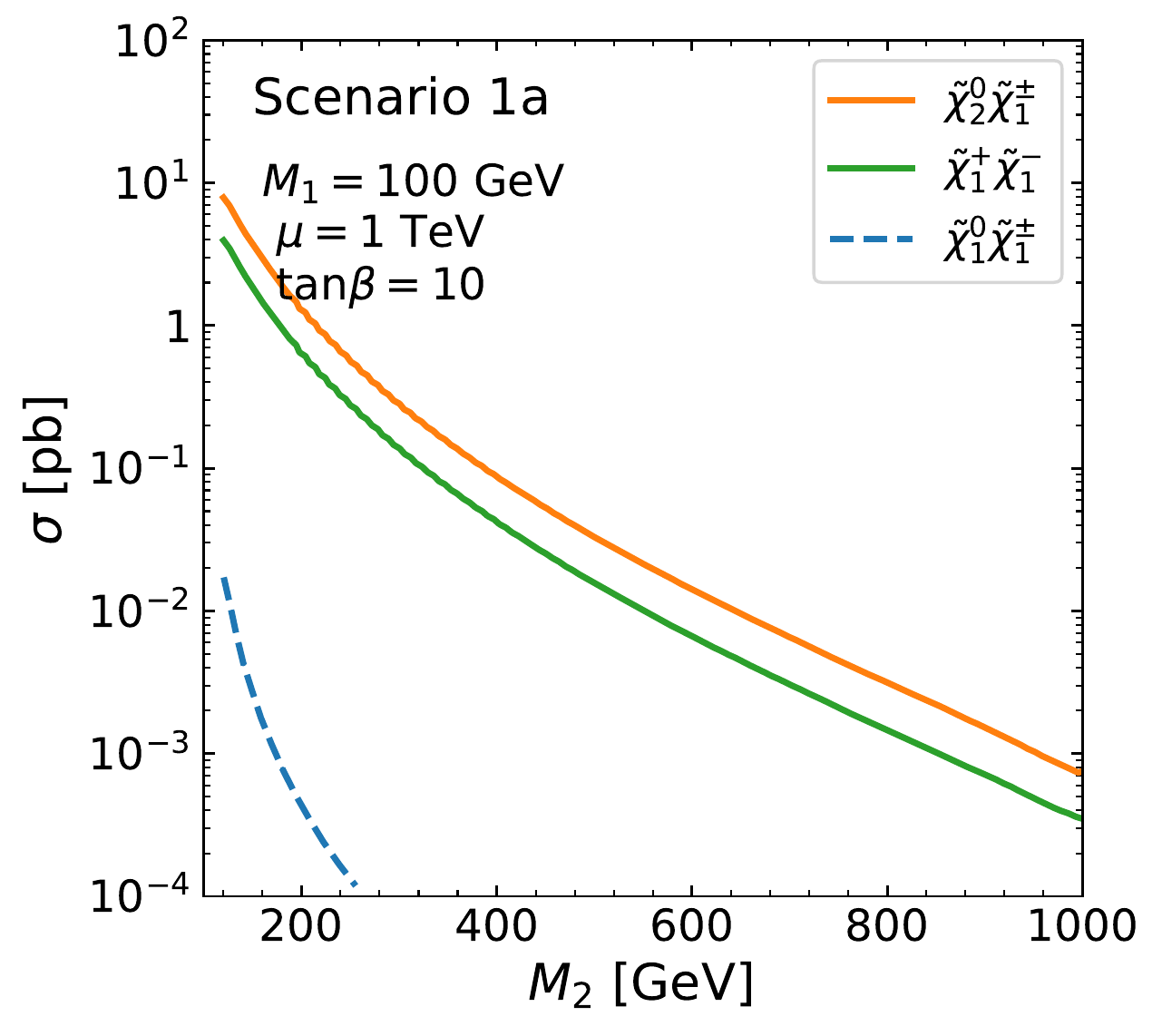}
\vspace{-0.7cm}
\caption{}
\vspace{0.7cm}
\end{subfigure}
\begin{subfigure}{.4\textwidth}
\centering
\includegraphics[width=\textwidth]{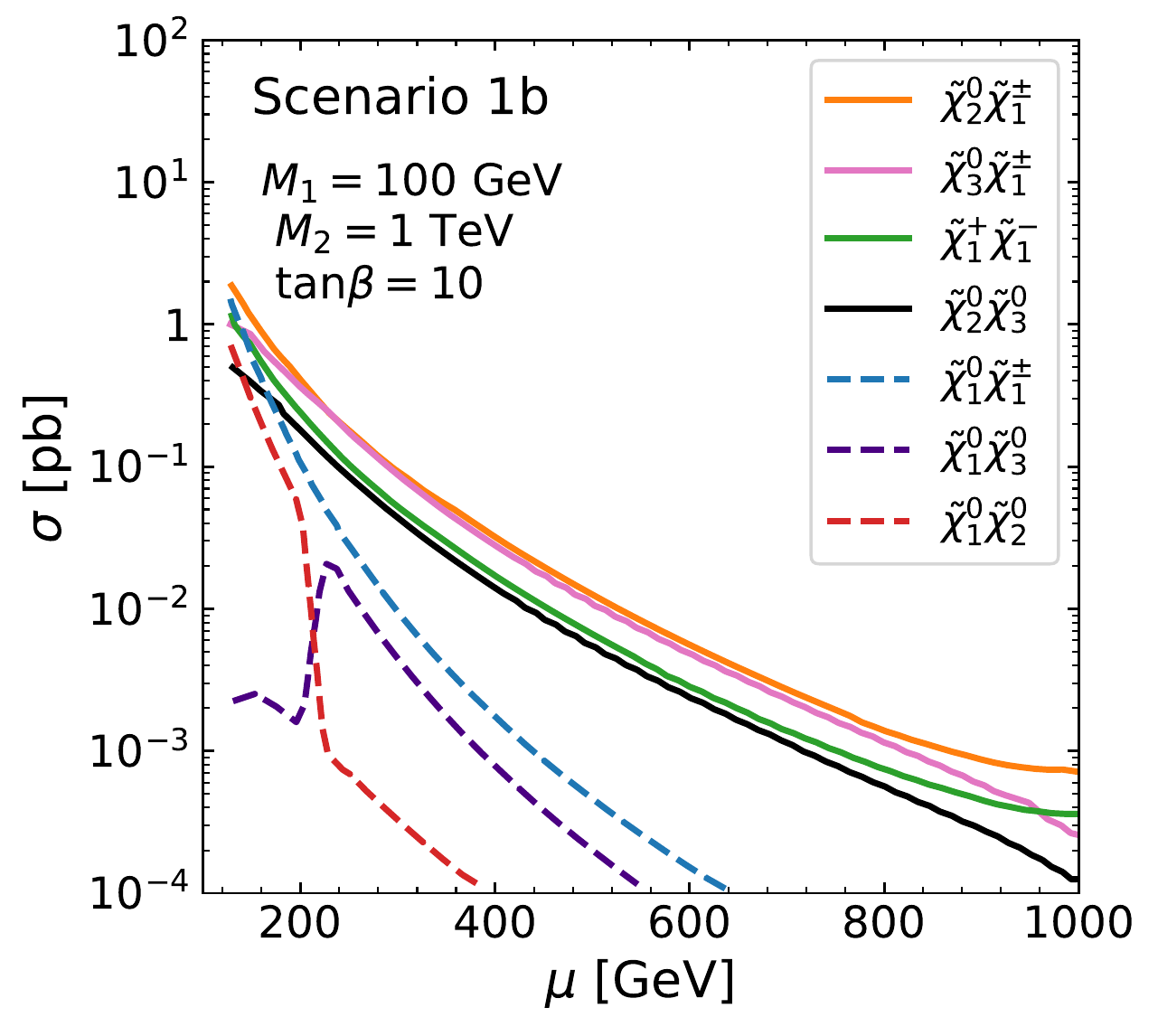}
\vspace{-0.7cm}
\caption{}
\vspace{0.7cm}
\end{subfigure} \\
\begin{subfigure}{.4\textwidth}
\centering
\includegraphics[width=\textwidth]{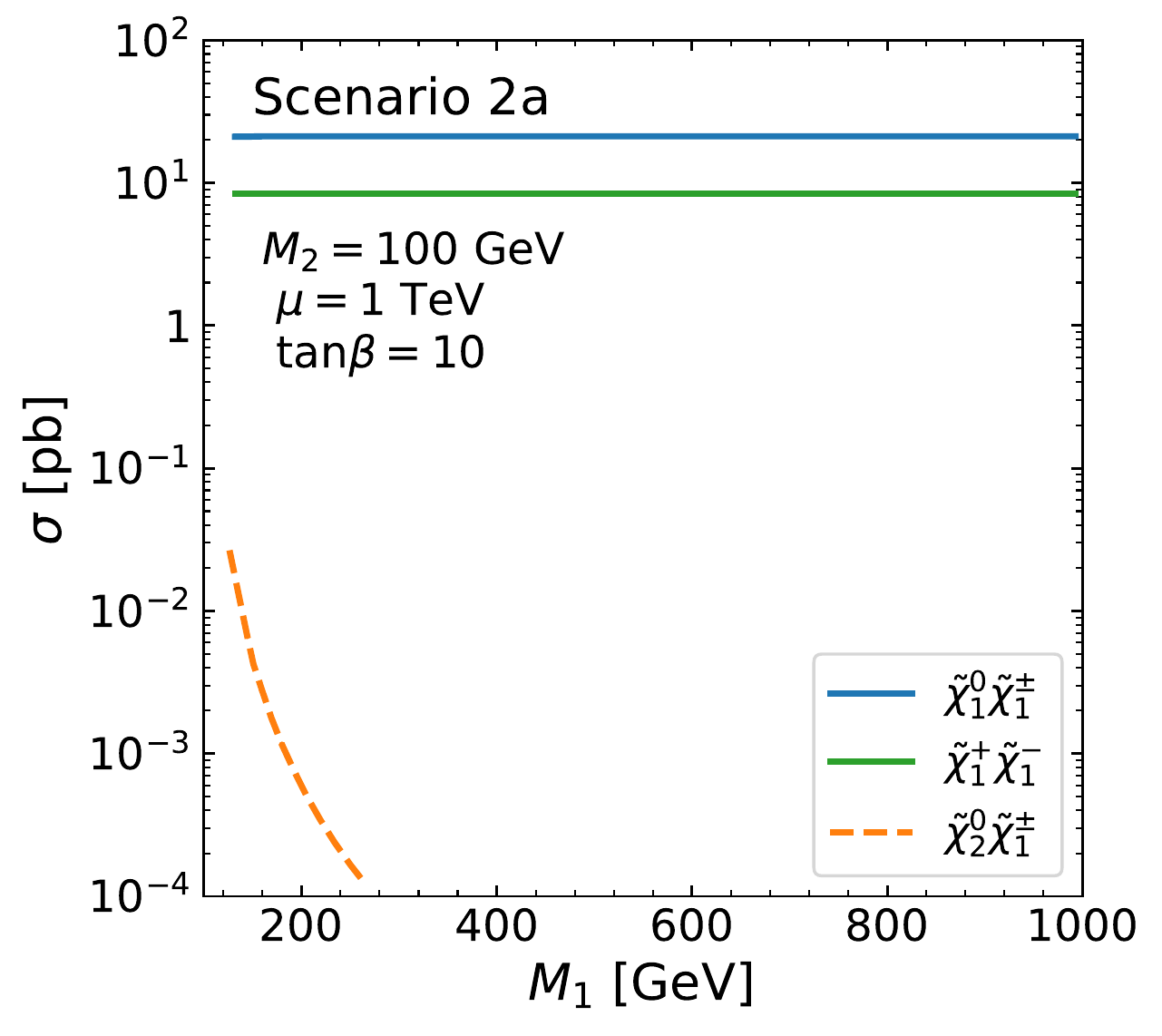}
\vspace{-0.7cm}
\caption{}
\vspace{0.7cm}
\end{subfigure}
\begin{subfigure}{.4\textwidth}
\centering
\includegraphics[width=\textwidth]{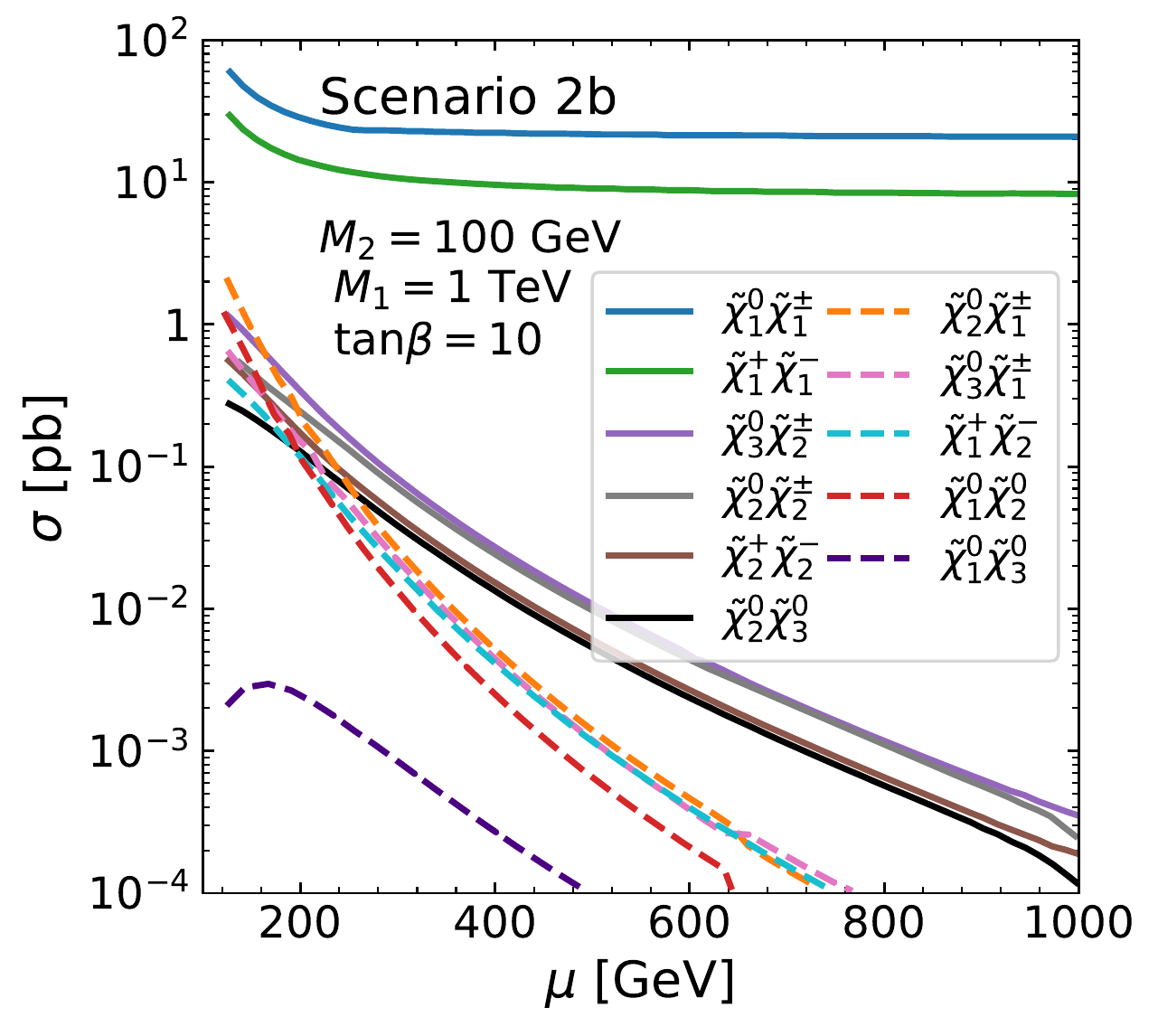}
\vspace{-0.7cm}
\caption{}
\vspace{0.7cm}
\end{subfigure} \\
\begin{subfigure}{.4\textwidth}
\centering
\includegraphics[width=\textwidth]{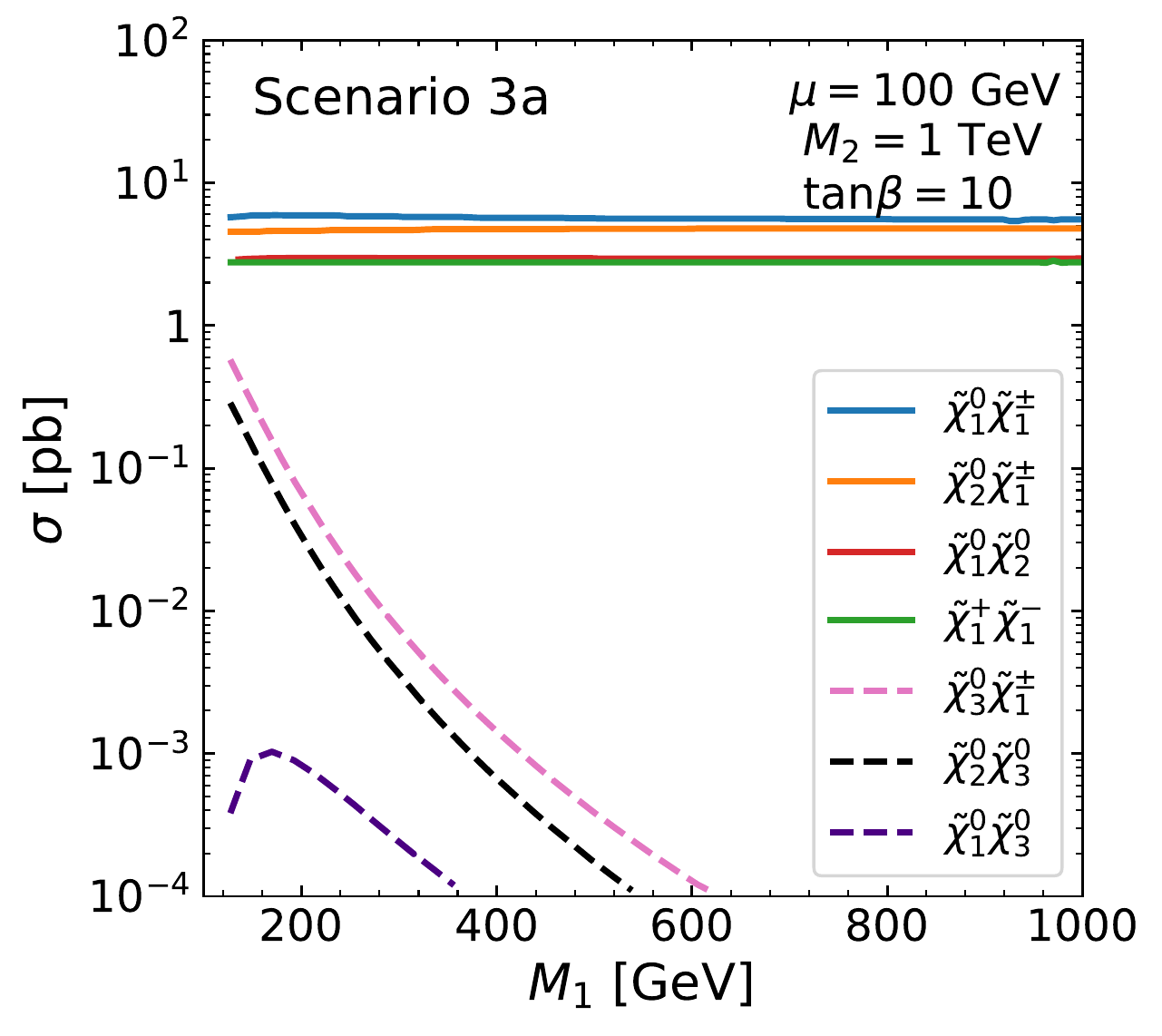}
\vspace{-0.7cm}
\caption{}
\vspace{0.7cm}
\end{subfigure}
\begin{subfigure}{.4\textwidth}
\centering
\includegraphics[width=\textwidth]{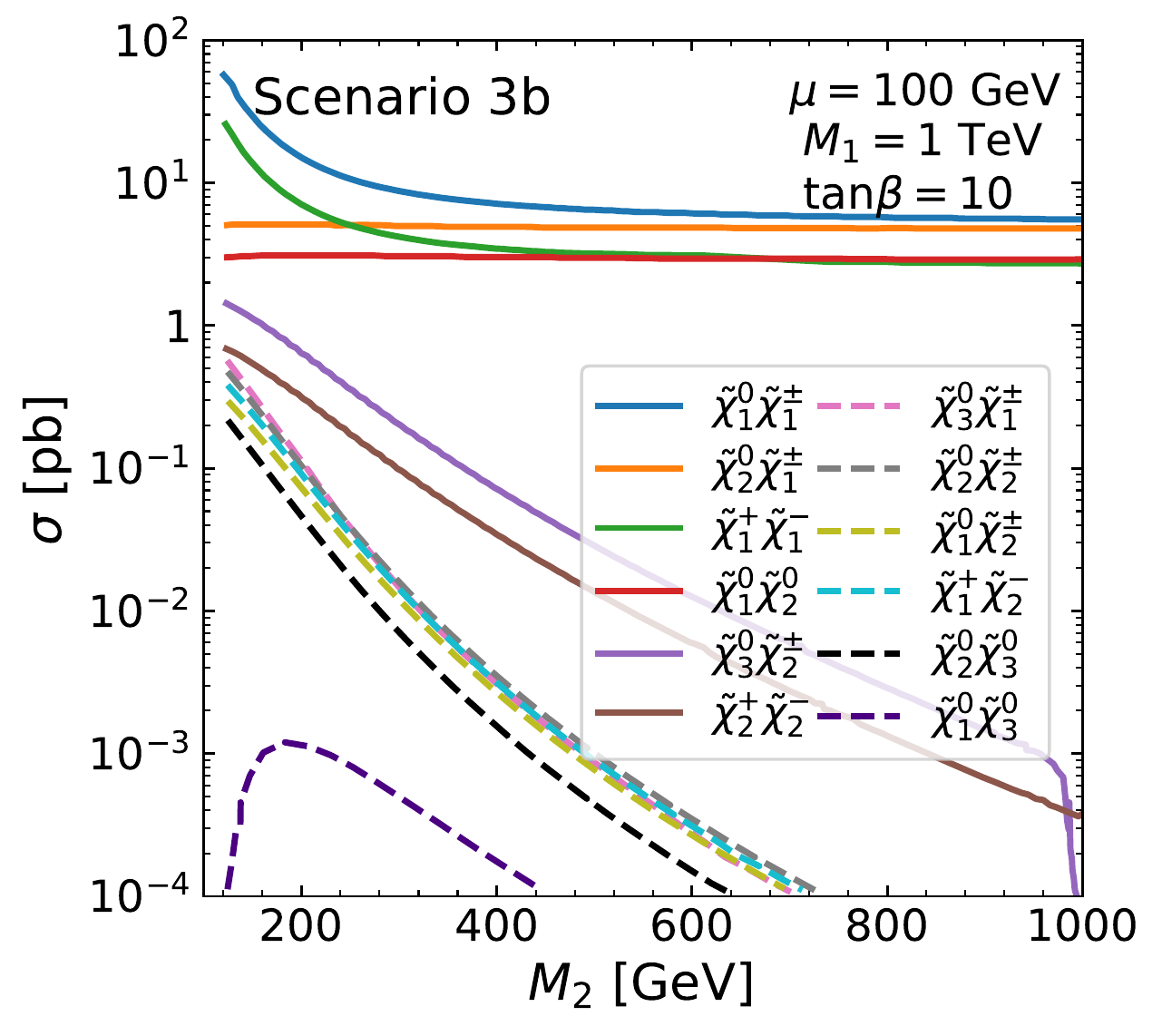}
\vspace{-0.7cm}
\caption{}
\vspace{0.7cm}
\end{subfigure}
\vspace{-0.5cm}
\caption{Electroweakino production cross sections at the LHC $\sqrt s = 14$ TeV \cite{Han:2013kza} versus the NLSP mass parameter for the three scenarios described in Sec.~\ref{sec:model}. The LSP mass parameter is set to be 100 GeV, the heaviest mass parameter is set to be 1 TeV, and $\tan\beta =10$, as stated in the panel legend. }
\label{fig:xsec}
\end{figure}

\begin{figure}
\centering
\begin{subfigure}{.32\textwidth}
\centering
\includegraphics[width=\textwidth]{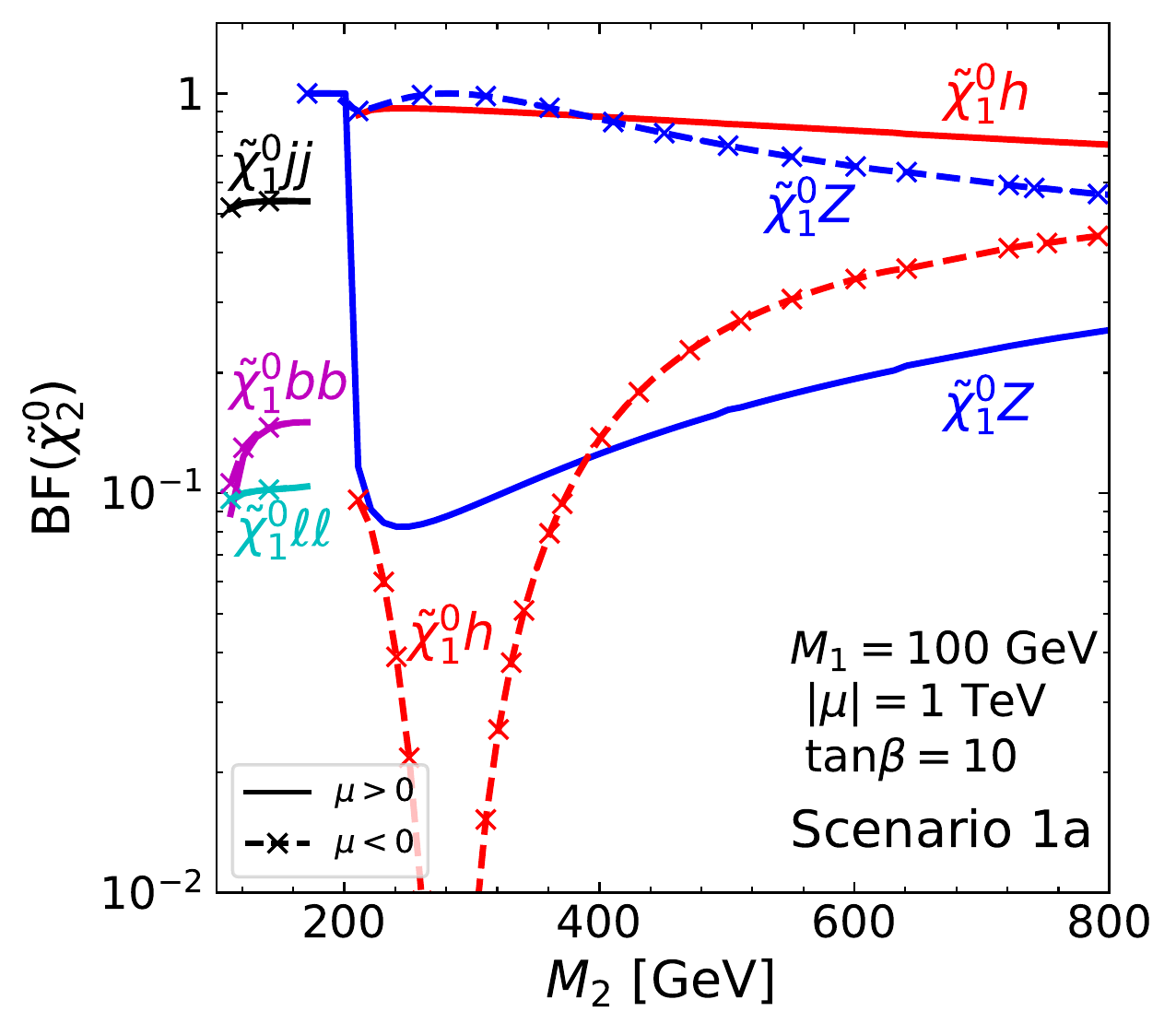}
\vspace{-0.7cm}
\caption{}
\vspace{0.7cm}
\end{subfigure}
\begin{subfigure}{.32\textwidth}
\centering
\includegraphics[width=\textwidth]{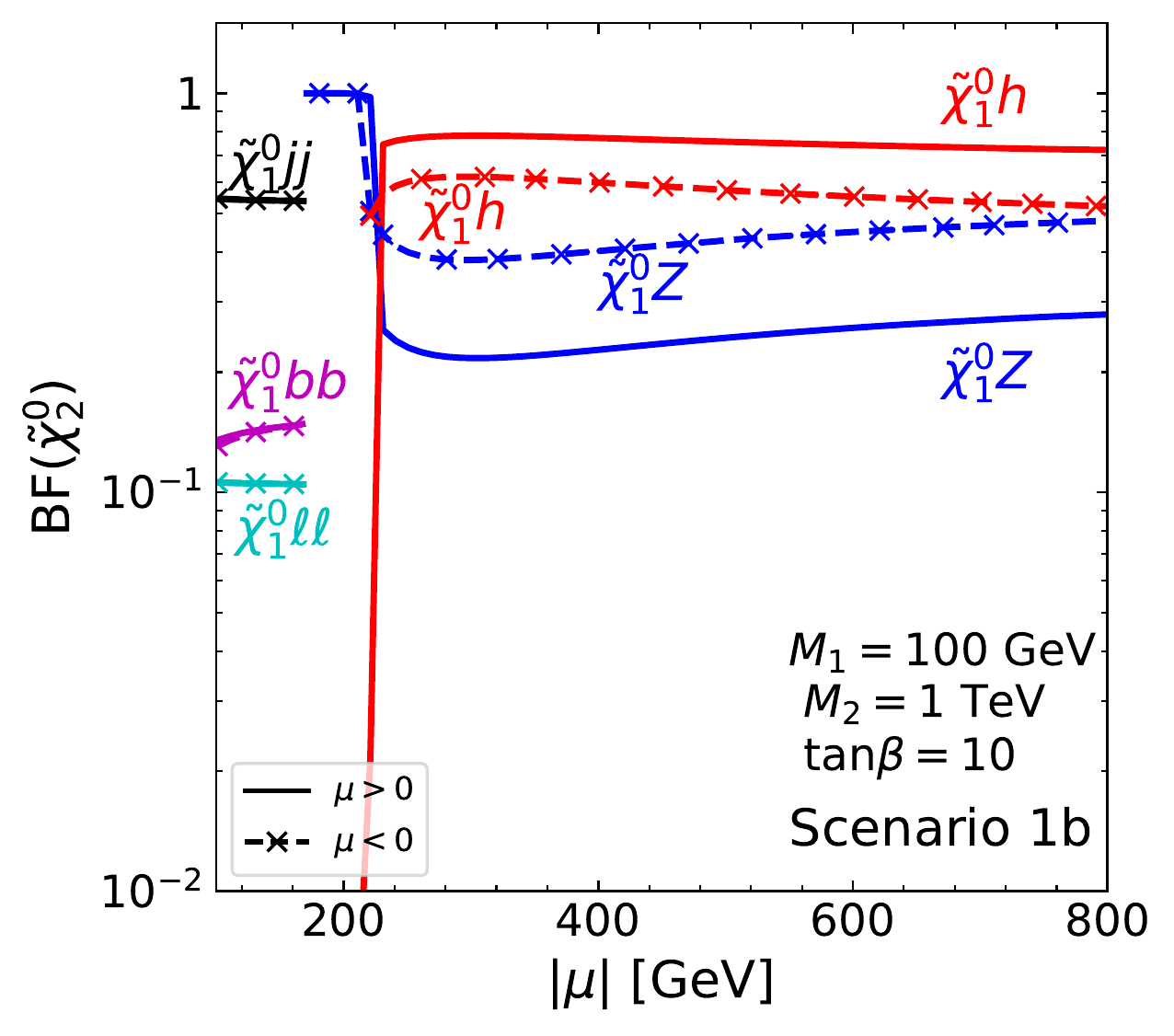}
\vspace{-0.7cm}
\caption{}
\vspace{0.7cm}
\end{subfigure}
\begin{subfigure}{.32\textwidth}
\centering
\includegraphics[width=\textwidth]{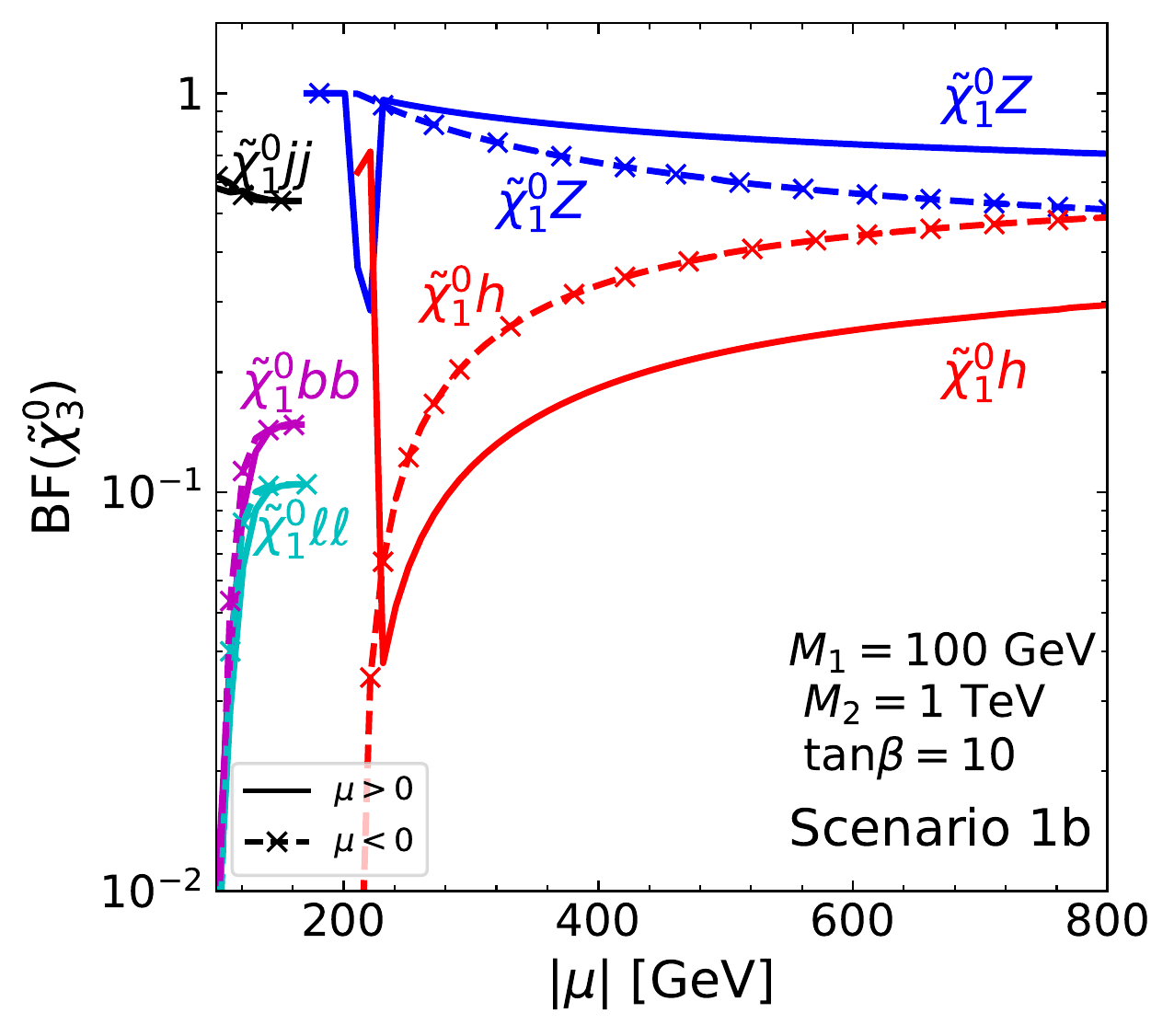}
\vspace{-0.7cm}
\caption{}
\vspace{0.7cm}
\end{subfigure}
\vspace{-0.5cm}
\caption{Decay branching fractions of (a) Wino-like NLSPs in Scenario 1a;  (b), (c): Higgsino-like NLSPs in Scenario 1b.
}
\label{fig:decay_1}
\end{figure}

In Fig.~\ref{fig:xsec}, we plot the pair production cross sections for the EWkinos via the DY processes at the LHC $\sqrt s = 14$ TeV, following the three representative scenarios described in Sec.~\ref{sec:model}.
Scenario 1a is characterized by a Bino-like LSP and three Wino-like NLSPs. 
With the unsuppressed SU(2)$_{\rm L}$ couplings, the leading production channels are the triplet Wino-like NLSPs
\begin{equation}
pp \to \cha\nn\ X,\ \cha\champ X.
\label{eq:winosN}
\end{equation}
As shown in Fig.~\ref{fig:xsec}(a), their cross sections can be the order of 1 pb to 1 fb for $M_2\sim 200$ GeV to 800 GeV. Although kinematically favored, the Bino-like LSP productions of \n\cha\ and \n\nn\ are highly suppressed by the Bino-Wino mixing. 
The Wino NLSPs decay to the LSP \n\ 
plus their SM partners through the mixture of Higgsino states. Therefore, the partial decay widths are scaled with a suppression factor ${\cal O}(M_Z^2/\mu^2)$. 
The branching fraction BF($\cha\to\n W^\pm$) is 100\%. For \nn\ decay, there are two competing channels 
\begin{equation}
\nn \to Z\n, \  h\n, 
\label{eq:AI_chi20}
\end{equation}
once kinematically accessible. Those decay branching fractions are shown in Fig.~\ref{fig:decay_1}(a) versus $M_2$. Solid lines are for $\mu>0$ and dashed lines with crosses are for $\mu<0$. Once ${\chi}_2^0 \rightarrow {\chi}_1^0 h$ channel is open, it quickly dominates for $\mu>0$. In the case of $\mu<0$, the branching fractions of $Z$ and $h$ modes are reversed. In particular, there is a dip in ${\rm BF}(\tilde{\chi}_2^0 \rightarrow \tilde{\chi}_1^0 h)$, as shown in the plot, due to the fact that the partial width is proportional to $(2 \sin(2 \beta)+M_2/\mu)$. Below the threshold for an on-shell $Z$, the branching fractions for various final states through an off-shell $Z$ decays to the SM fermions, about 55\% into light quarks, 15\% into $b\bar b$, 20\% into neutrinos, and 3.3\% into each lepton flavor.  For  $M_2$ slightly above $M_{1}$, the loop-induced radiative decay $\tilde{\chi}_2^0\rightarrow \tilde{\chi}_1^0 \gamma$ becomes appreciable, although the final state photon will be very soft, making its identification difficult. 

Scenario 1b is characterized by a Bino-like LSP and four Higgsino-like NLSPs. The leading production channels are the Higgsino-like NLSPs 
\begin{equation}
pp \to \tilde{\chi}_1^\pm \tilde{\chi}_2^0  X,\ \  \tilde{\chi}_1^\pm \tilde{\chi}_3^0  X,\  \ \tilde{\chi}_1^+ \tilde{\chi}_1^- X,\  \ {\rm and}\ \ \tilde{\chi}_2^0\tilde{\chi}_3^0  X.
\label{eq:higgsinosN}
\end{equation}
As shown in Fig.~\ref{fig:xsec}(b), their cross sections can be the order of 500 fb to 1 fb for $\mu \sim 200$ GeV to 800 GeV. 
Again, the Bino-like LSP production $\tilde{\chi}_1^0 \tilde{\chi}_1^\pm$ etc.~are suppressed except when $M_1\sim \mu$ where the mixing becomes substantial. 
The branching fraction BF$(\tilde{\chi}_1^\pm \to \tilde{\chi}_1^0 W^\pm)$ in Scenario 1b is again 100\%. Figures~\ref{fig:decay_1}(b) and \ref{fig:decay_1}(c) show the decay branching fractions of $\tilde{\chi}^{0}_{2}$ and $\tilde{\chi}_3^0$, respectively, through $Z/h$ bosons, versus $\mu$ for the Higgsino NLSPs. For $\mu \gtrsim 250$ GeV, the decay pattern for $\tilde{\chi}_2^0$ is   qualitatively similar to that of the light wino Scenario 1a with $\mu>0$.  Branching fraction of $\tilde{\chi}_2^0 \rightarrow \tilde{\chi}_1^0 h$ and $\tilde{\chi}_2^0 \rightarrow \tilde{\chi}_1^0 Z$ is about 75\% and 25\% for $\mu=500$ GeV, respectively.   The decays of $\tilde{\chi}_3^0$, however, are more preferable to $\tilde{\chi}_1^0 Z$.  The difference in the decay pattern of $\tilde{\chi}_2^0$ and $\tilde{\chi}_3^0$ is due to the different composition of $\tilde{\chi}_{2,3}^0$ as  $\frac{1}{\sqrt{2}} ({\tilde{H}}_d^0 \mp {\tilde{H}}_u^0)$. 
Note that in Fig.~\ref{fig:decay_1}(c) the branching fraction of  $\tilde{\chi}_3^0 \rightarrow \tilde{\chi}_1^0 h$ shows a sudden drop around 230 GeV, coming from  the level crossing of the two Higgsino-like mass eigenstates.   
For $m_{\tilde{\chi}_{2}^0} - m_{\tilde{\chi}_1^0} < m_Z$, the off-shell decay of $\tilde{\chi}^0_2$ via $Z^*$ again dominates, with the branching fraction of fermion final states similar to that of $\tilde{\chi}_2^0$ in Scenario 1a. The off-shell decays of $\tilde{\chi}^0_3$, on the other hand, occur via both $\tilde{\chi}^{0}_{3}\to \tilde{\chi}^{\pm}_{1} W^{*}$ and $\tilde{\chi}^{0}_{2} Z^{*}$. Even with the phase space suppression comparing to the decay of $\tilde{\chi}_3^0$ directly down to $\tilde{\chi}_1^0$,  the branching fractions for $\tilde{\chi}_3^0 \rightarrow \tilde{\chi}_1^\pm W^*$ could dominate over $\tilde{\chi}_3^0 \rightarrow \tilde{\chi}_1^0 Z^*$, as can be seen from the difference between the black and magenta lines in Fig.~\ref{fig:decay_1}(c), since the coupling $\tilde{\chi}_3^0 \tilde{\chi}_1^\pm W$ is unsuppressed, while $\tilde{\chi}_3^0 \tilde{\chi}_1^0 Z$ suffers from the small Bino-Higgsino mixing.

For Scenario 2a with three Wino-like LSPs and a Bino-like NLSP $\tilde{\chi}_2^0$, the leading production channels are the Wino-like triplet LSPs, similar to Eq.~(\ref{eq:winosN})
\begin{equation}
pp \to \tilde{\chi}_1^\pm \tilde{\chi}_1^0 X ,\ \ \tilde{\chi}_1^+ \tilde{\chi}_1^- X.
\label{eq:winosL}
\end{equation}
The production cross sections at the LHC are shown in Fig.~\ref{fig:xsec}(c) and they are about $(10-20)$ pb for $M_2 = 100$ GeV. Although characterized by a large cross section,
these processes bear a  significant experimental challenge due to the  small mass splitting of $m_{\tilde{\chi}_1^\pm} - m_{\tilde{\chi}_1^0}$, leading to \cha\ decays into \n\ through the
emission of  pions, muons, or electrons. The final states will contain modest missing transverse momentum and very  low transverse momentum tracks, requiring dedicated
reconstruction techniques. We will present the LHC searches in the later sections.

Scenario 2b is characterized by three Wino-like LSPs and four Higgsino-like NLSPs. 
The leading production channels are those Wino-like LSPs like in Eq.~(\ref{eq:winosL}).
The production cross sections at the LHC are shown in Fig.~\ref{fig:xsec}(d) and the rate can be as large as 20 pb for $M_2=100$ GeV. From the observational aspect, similar to the situation of Scenario 2a, the compressed Wino-like LSPs would be challenging to search for, as mentioned earlier, and to be discussed in the next section.
On the other hand, although sub-leading, the Higgsino-like NLSP production is similar to that in  Eq.~(\ref{eq:higgsinosN})
\begin{equation}
pp \to \tilde{\chi}_2^\pm \tilde{\chi}_2^0  X,\ \  \tilde{\chi}_2^\pm \tilde{\chi}_3^0  X,\  \ \tilde{\chi}_2^+ \tilde{\chi}_2^- X,\  \ {\rm and}\ \ \tilde{\chi}_2^0\tilde{\chi}_3^0  X, 
\end{equation}
The cross sections are shown in Fig.~\ref{fig:xsec}(d) and are quite sizable with the 
unsuppressed SU(2)$_{\rm L}$ couplings, reaching the order of 500 fb to 1 fb for $\mu \sim 200$ GeV to 800 GeV, quite similar to the case of Scenario 1b with Higgsino-like NLSPs. 

\begin{figure}[t]
\centering
\begin{subfigure}{.32\textwidth}
\centering
\includegraphics[width=\textwidth]{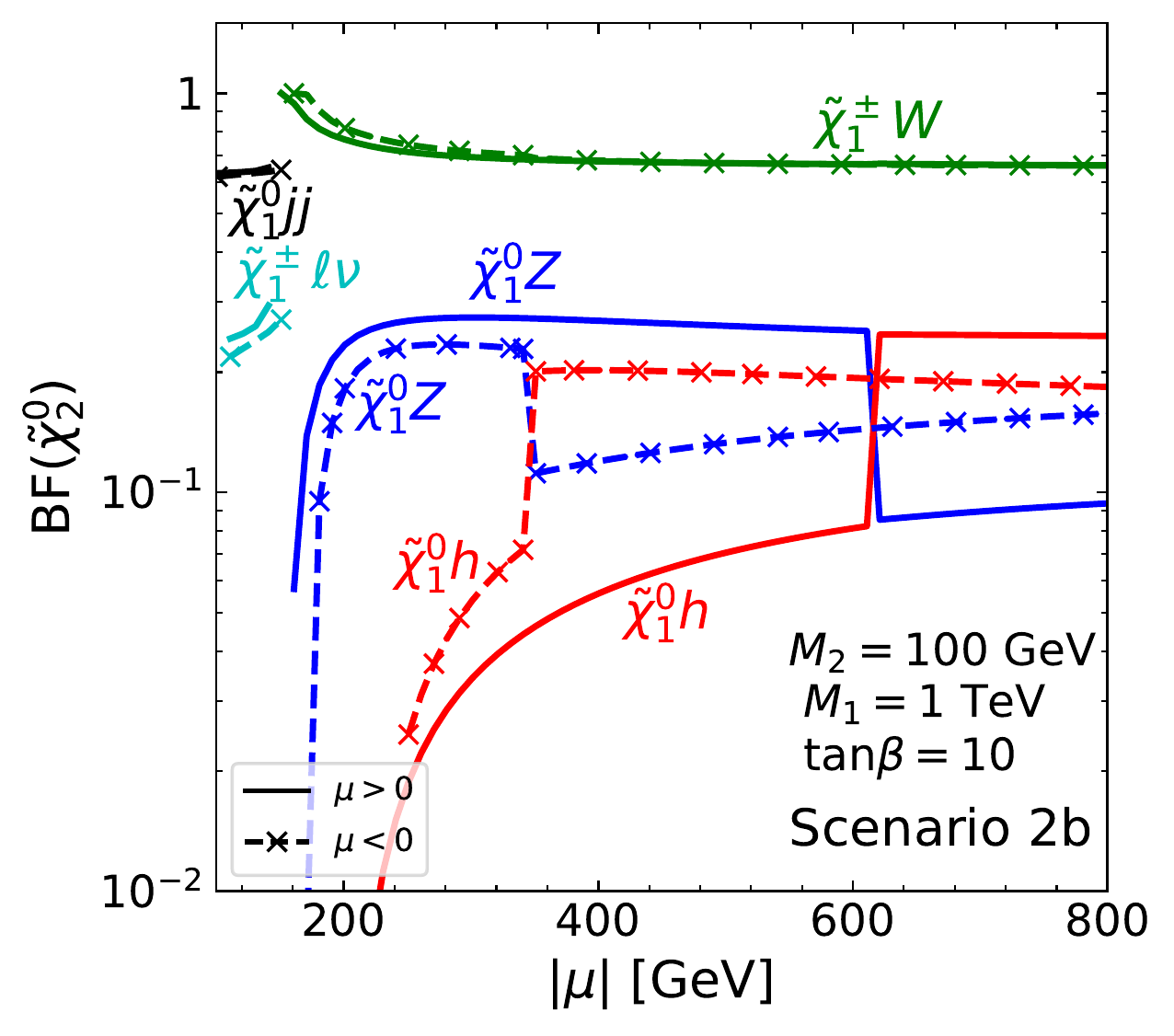}
\vspace{-0.7cm}
\caption{}
\vspace{0.7cm}
\end{subfigure}
\begin{subfigure}{.32\textwidth}
\centering
\includegraphics[width=\textwidth]{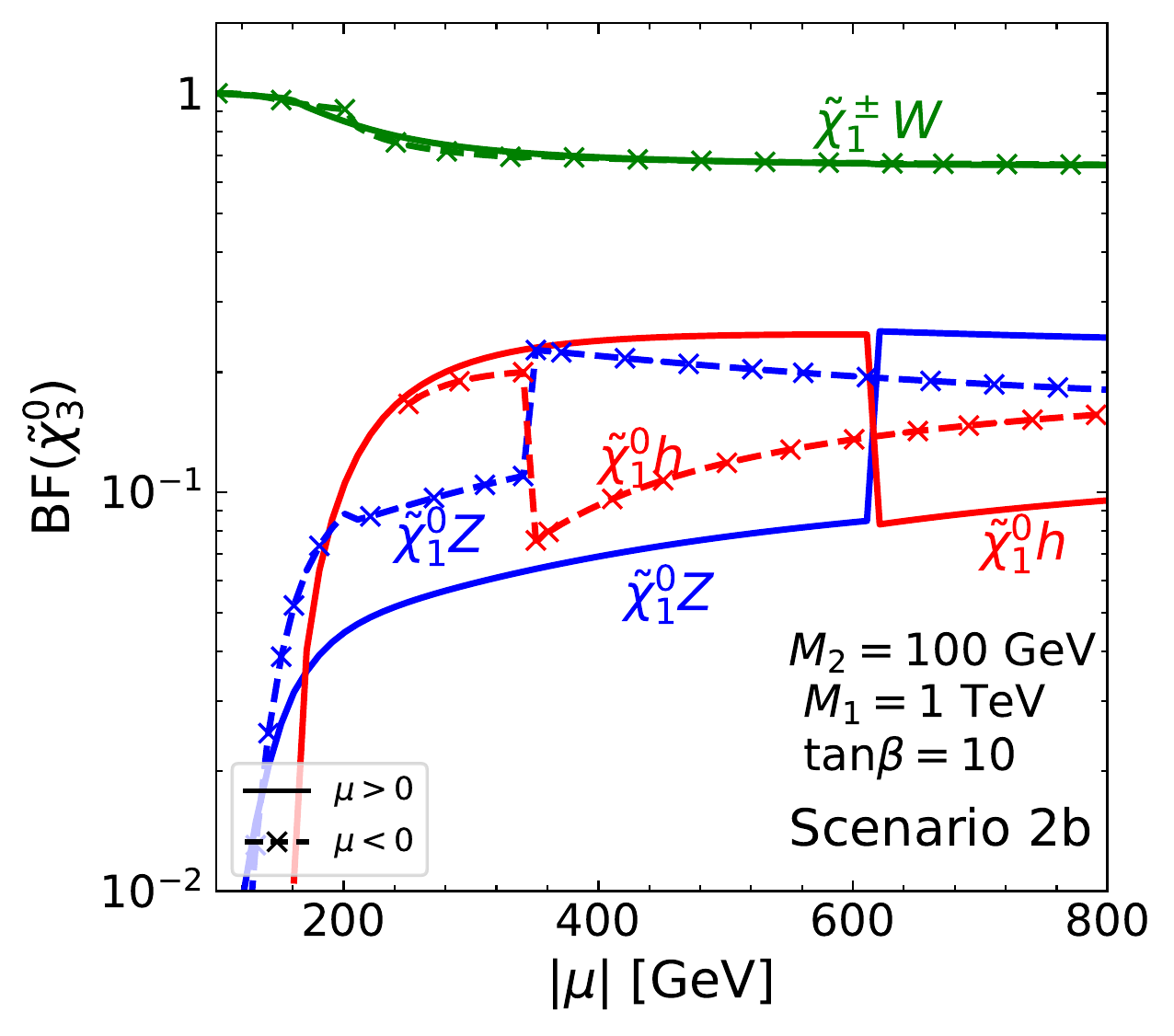}
\vspace{-0.7cm}
\caption{}
\vspace{0.7cm}
\end{subfigure}
\begin{subfigure}{.32\textwidth}
\centering
\includegraphics[width=\textwidth]{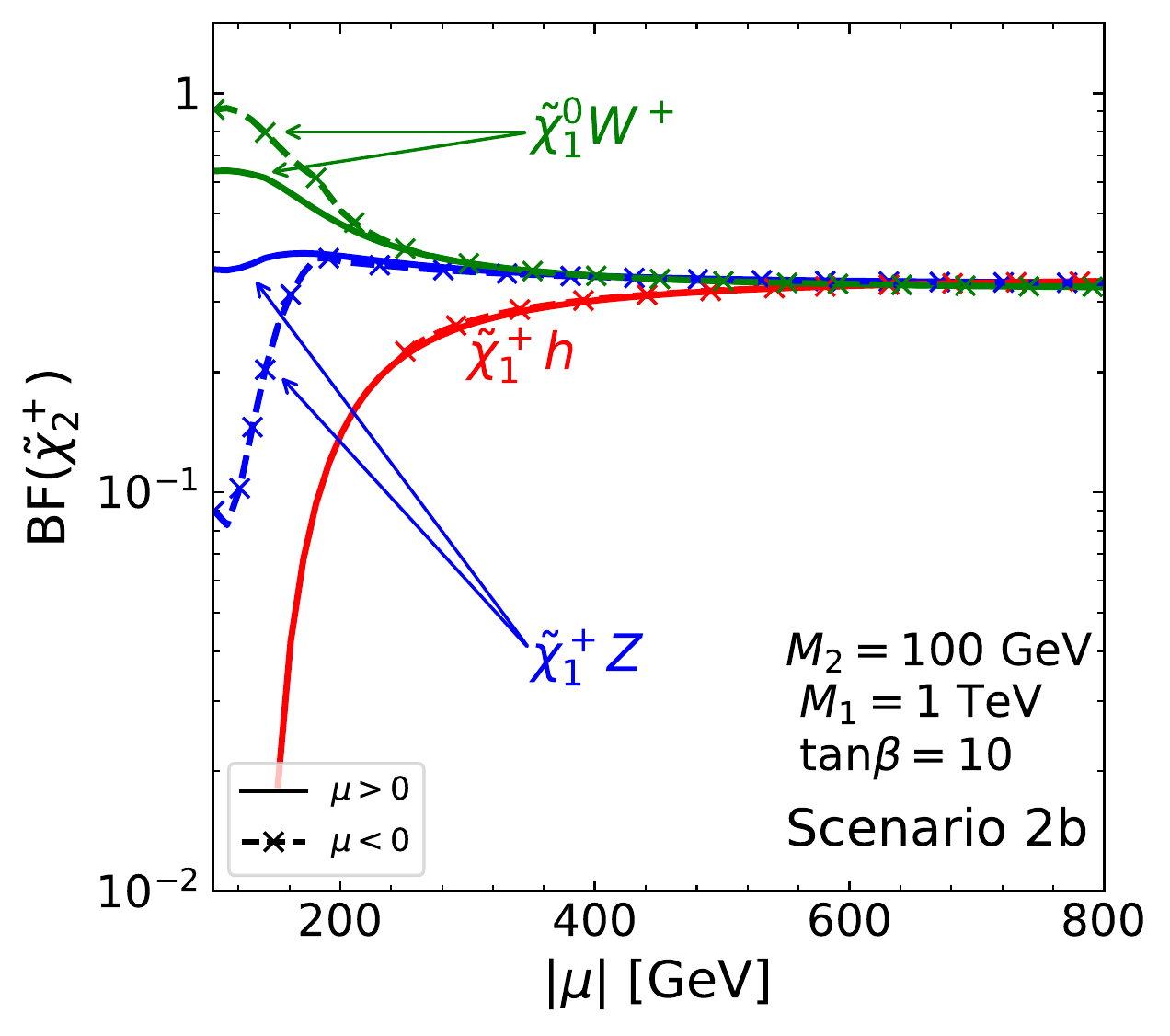}
\vspace{-0.7cm}
\caption{}
\vspace{0.7cm}
\end{subfigure}
\vspace{-0.5cm}
\caption{Decay branching fractions of Higgsino-like NLSPs in Scenario 2b.}
\label{fig:decay_2}
\end{figure}

The decay patterns for the Higgsino-like NLSPs are much richer. Generically, $\tilde{\chi}^{0}_{2,3}$ and $\tilde{\chi}^{\pm}_{2}$ decay to a $W/Z/h$-boson plus its corresponding LSP. 
The decay channels for the two NLSP neutralinos  $\tilde{\chi}_{2,3}^0$ are 
\begin{equation}
\tilde{\chi}_{2,3}^0 \rightarrow \tilde{\chi}_1^\pm W^\mp, \tilde{\chi}_1^0 Z, ~\tilde{\chi}_1^0 h.
\label{eq:BII_chi2030}
\end{equation}
Their decay branching fractions are shown in Figs.~\ref{fig:decay_2}(a) and \ref{fig:decay_2}(b), respectively. They are Majorana fermions and decay to both $\tilde{\chi}_1^+W^-$ and  $\tilde{\chi}_1^-W^+$ equally. 
Under the limit of $ |\mu \pm M_2| \gg m_Z$, the following simplified relation holds for the partial decay widths (and decay branching fractions as well) of $\tilde{\chi}_{2,3}^0$
\begin{equation}
\Gamma_{\tilde{\chi}_1^+W^-}=\Gamma_{\tilde{\chi}_1^-W^+}\approx\Gamma_{\tilde{\chi}_1^0Z}+\Gamma_{\tilde{\chi}_1^0h},
\end{equation}
in accordance to the Goldstone boson equivalence theorem \cite{Lee:1977eg, Chanowitz:1985hj, Bagger:1989fc, He:1992nga}.  The 
$\tilde{\chi}_2^0$ is more likely to decay to $Z$ while $\tilde{\chi}_3^0$ is more likely to decay to $h$ for $\mu>0$. The sudden changes for the $\tilde{\chi}^0_1Z$ and $\tilde{\chi}^0_1h$ channels in Figs.~\ref{fig:decay_2}(a) and (b) are due to level crossing.
For $\tilde{\chi}_2^\pm$, the dominant decay modes are 
\begin{equation}
\tilde{\chi}_2^\pm \rightarrow \tilde{\chi}_1^0 W, \tilde{\chi}_1^\pm Z, ~\tilde{\chi}_1^\pm h.
\label{eq:BII_chi2pm}
\end{equation}
Their decay branching fractions are shown in Figs.~\ref{fig:decay_2}(c). 
Under the limit of $  |\mu \pm M_2| \gg m_Z$, 
the ratios of the partial decay widths are roughly $\Gamma_{\tilde{\chi}_1^0 W}:\Gamma_{\tilde{\chi}_1^\pm Z}:\Gamma_{\tilde{\chi}_1^\pm h} \approx 1:1:1$, with small deviation caused by phase space effects. 

For Scenario 3a with four Higgsino-like LSPs and a Bino-like NLSP $\tilde{\chi}_3^0$, 
the leading production channels are the LSP pairs, similar to that in Eq.~(\ref{eq:higgsinosN})
\begin{equation}
pp \to \tilde{\chi}_1^\pm \tilde{\chi}_{1,2}^0 \ X,\ \  \tilde{\chi}_1^+ \tilde{\chi}_1^- \ X,\ \  {\rm and}\ \  \tilde{\chi}_1^0 \tilde{\chi}_2^0 \ X.
\label{eq:higgsinoL}
\end{equation}
The production cross sections at the LHC are shown in Fig.~\ref{fig:xsec}(e) and the rate is about 5 pb for $\mu=100$ GeV. Similar to Scenario 2, such channels are difficult to probe with conventional searches due to the compressed spectrum for the LSPs.

For Scenario 3b with four Higgsino-like LSPs and three Wino-like NLSPs, the leading production channels are the same as above for the Higgsino-like LSP pairs in Eq.~(\ref{eq:higgsinoL}). The production cross sections at the LHC are shown in Fig.~\ref{fig:xsec}(f) and the rate can be as large as 5 pb for $\mu=100$ GeV, similar to Scenario 3a. 
Again from the observational aspect, it is similar to the situations of Scenarios 2a, 2b, and 3a: The compressed LSPs would be challenging to search for, as mentioned earlier, and to be discussed in the next section. On the other hand, the sub-leading channels for the Wino-like NLSP production as in Eq.~(\ref{eq:winosN}) come to rescue.
The cross sections are shown in Fig.~\ref{fig:xsec}(f) and can be the order of 1 pb to 1 fb for 
$M_2 \sim 200$ GeV to 800 GeV, similar to the case of Scenario 1a. 

\begin{figure}[t]
\centering
\begin{subfigure}{.32\textwidth}
\centering
\includegraphics[width=\textwidth]{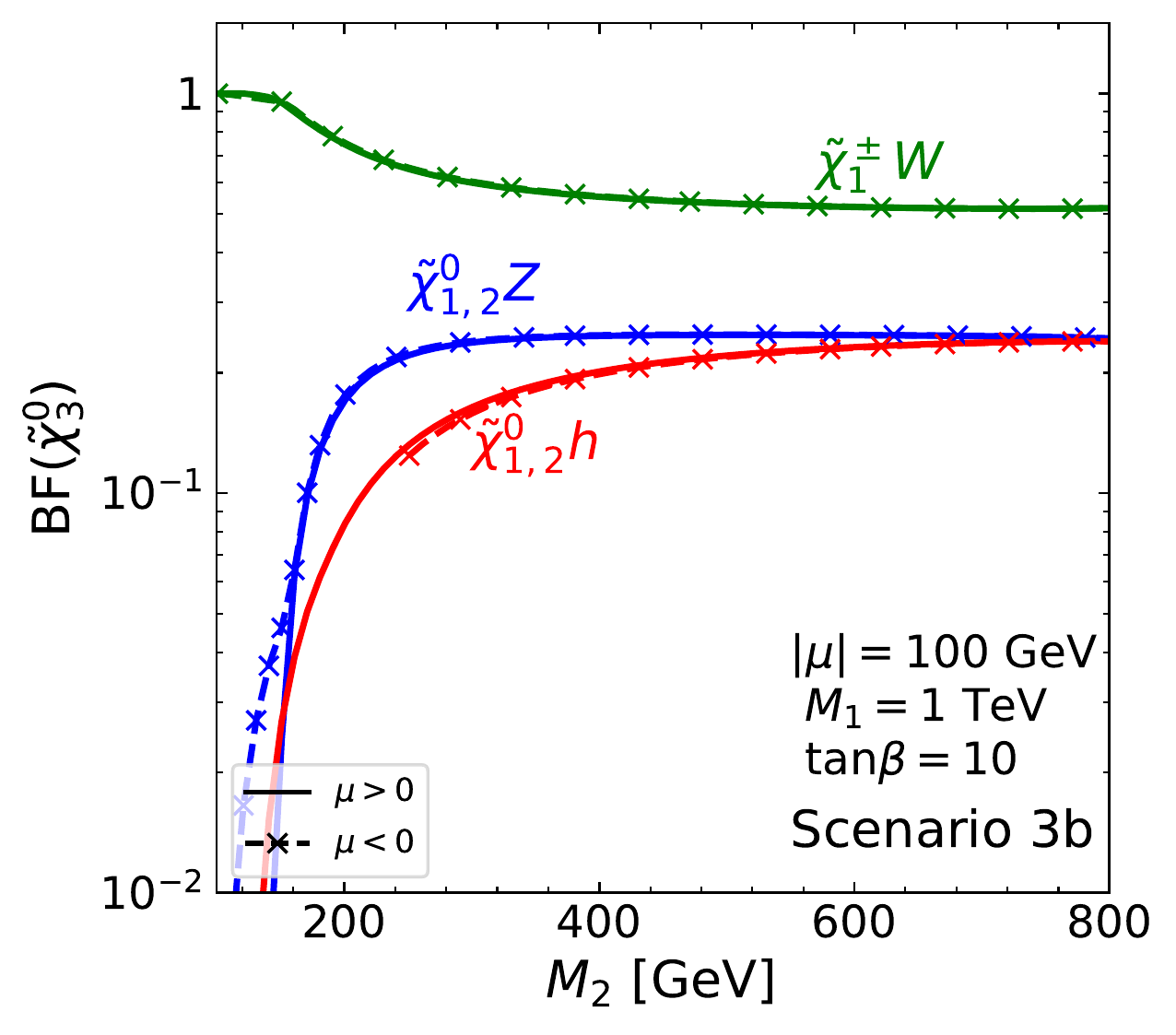}
\vspace{-0.7cm}
\caption{}
\vspace{0.7cm}
\end{subfigure}
\begin{subfigure}{.32\textwidth}
\centering
\includegraphics[width=\textwidth]{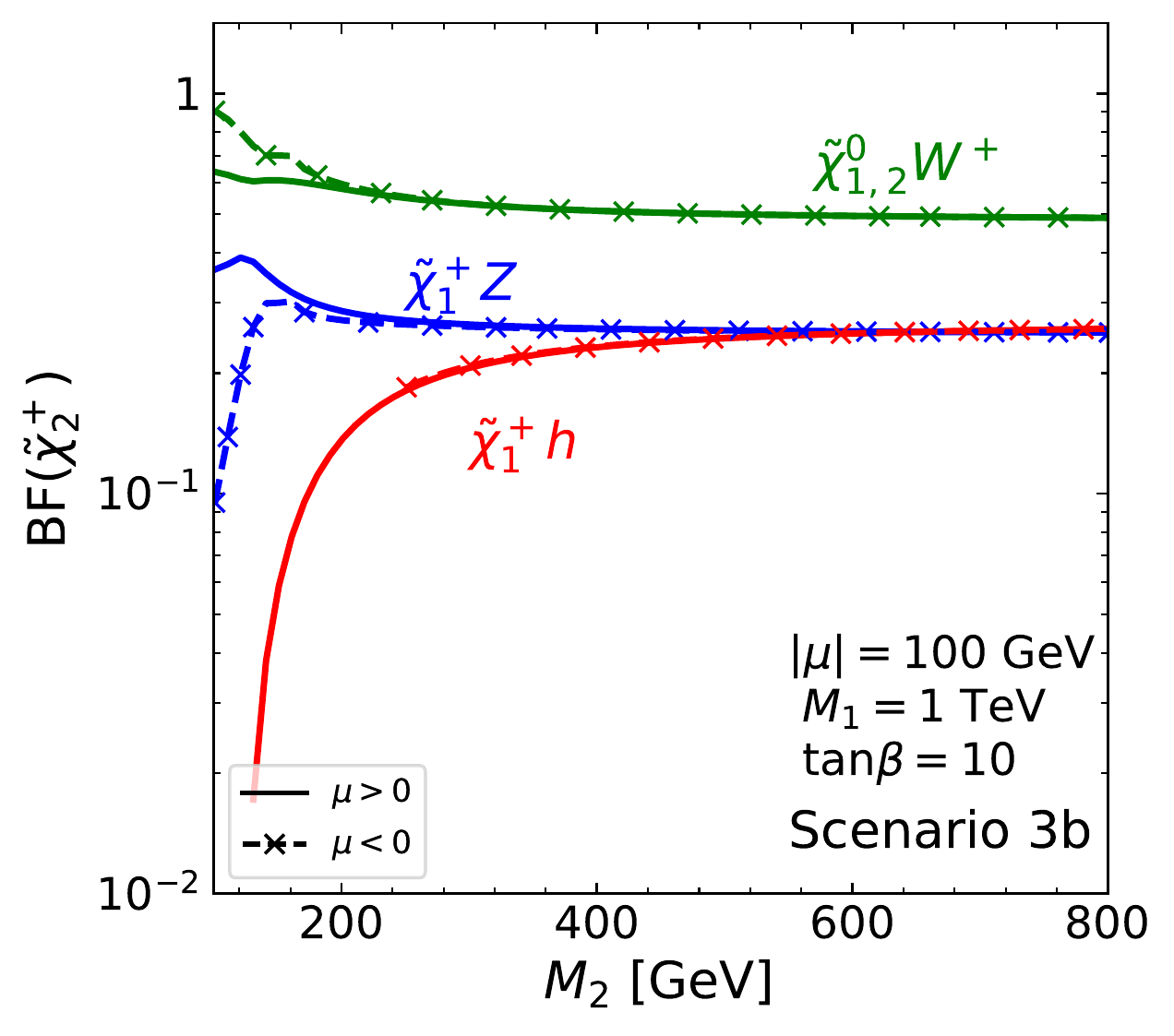}
\vspace{-0.7cm}
\caption{}
\vspace{0.7cm}
\end{subfigure}
\vspace{-0.5cm}
\caption{Decay branching fractions of Wino-like NLSPs in Scenario 3b.}
\label{fig:decay_3}
\end{figure}

The decay branching fractions for the NLSPs $\tilde{\chi}_2^\pm$  and $\tilde{\chi}_3^0$ in Scenario 3b are shown in Figs.~\ref{fig:decay_3}(a) and \ref{fig:decay_3}(b).   
For $\tilde{\chi}_2^\pm$, the dominant decay modes are 
\begin{equation}
\tilde{\chi}_2^\pm \rightarrow \tilde{\chi}_1^0 W,  ~\tilde{\chi}_2^0 W, ~\tilde{\chi}_1^\pm Z, ~\tilde{\chi}_1^\pm h.
\label{eq:CII_chi2pm}
\end{equation}
Under the limit of $|M_2 \pm \mu|\gg m_Z$, the ratios of the partial decay widths are roughly $\Gamma_{\tilde{\chi}_1^0W}:\Gamma_{\tilde{\chi}_2^0W}: \Gamma_{\tilde{\chi}_1^\pm Z}:\Gamma_{\tilde{\chi}_1^\pm h} \approx 1:1:1:1$. Due to the LSP degeneracy of $\tilde{\chi}_1^0$ and $\tilde{\chi}_2^0$, $\tilde{\chi}_1^0W$ and $\tilde{\chi}_2^0W$ final states would be indistinguishable experimentally. Combining these two channels, the branching fractions of $\tilde{\chi}_2^\pm$ to $W$, $Z$ and $h$ channels are roughly 51\%, 26\%, and 23\%, respectively.  In the limit of large $M_2$, the branching fractions approach the asymptotic limit
${\rm BF}(\tilde{\chi}_{2}^\pm \rightarrow \tilde{\chi}_{1,2}^0 W) \approx 2 {\rm BF}(\tilde{\chi}_{2}^\pm \rightarrow \tilde{\chi}_1^\pm h) \approx 2 {\rm BF}(\tilde{\chi}_{2}^\pm \rightarrow \tilde{\chi}_1^\pm Z) \approx   $ 50 \%. Combining $\tilde{\chi}_1^0$ and $\tilde{\chi}_2^0$ final states, the branching fraction of $Z$ channel is almost the same as the $h$ channel at very large $|M_2 \pm \mu|\gg m_Z$, which is about half of the branching fraction of the $W$ final states. 

If kinematically accessible, the heavy Higgs bosons $A^0/H^{0,\pm}$ may decay to a pair of EWkinos with branching fractions of ${\cal O}(10\% - 30\%)$, thereby provide new channels for the search \cite{Gori:2018pmk}. 

The EWkinos could also be produced via weak vector boson fusion processes (VBF) \cite{Datta:2001cy, Datta:2001hv, Cho:2006sx, Dutta:2012xe, Cotta:2012nj, Delannoy:2013ata}
\beq
qq' \to qq'  \tilde{\chi}_i^+ \tilde{\chi}_j^0,  \ \  qq' \tilde{\chi}_i^+ \tilde{\chi}_j^-,  \ \  qq' \tilde{\chi}_i^0 \tilde{\chi}_j^0\ .
\eeq
The production rate for this mechanism is typically smaller than that of the DY processes by about two orders of magnitude depending on their masses.  Thus these channels do not contribute much to the inclusive signal~\cite{Giudice:2010wb}. 
On the other hand, the unique kinematics of the companying forward-backward jets make the signal quite characteristic and the search very promising, as shown in Sec.~\ref{sec:compressed}. 


\subsection{Searches at the ATLAS and CMS experiments}
\label{sec:LHC}

Since the very beginning of the LHC era, direct searches for SUSY have represented one of the
major science drivers of the ATLAS and CMS experiments. 
However,  searches for EWkinos have become the very core of the SUSY program at the LHC after the discovery of
a Higgs boson in 2012 and the collection of  large datasets of proton-proton collisions at 8 and 13~TeV center-of-mass energy.  
Besides the EWkino mass scale that governs the production rate and decay kinematics, the other most characteristic parameter 
for the experimental searches  
is the mass difference between the decaying parent $\tilde\chi_\text{parent}$ and the daughter $\tilde\chi_\text{daughter}$, denoted by
$$
\Delta M = m_{\tilde{\chi}_\text{parent}} - m_{\tilde{\chi}_\text{daughter}},
$$
which determines the average transverse momentum of the daughter particles and thus dictates how candidate events are reconstructed by the experiments. 
For $\Delta M \gtrsim M_Z/M_W/m_h$, we consider it as ``non-compressed'' spectra, 
while $\Delta M  \sim {\cal O}$(1 GeV) and $\Delta M  \sim {\cal O}$(100 MeV) correspond 
to the ``compressed'' and ``nearly-degenerate'' spectra, respectively.

The ATLAS and CMS collaborations have designed a comprehensive
search to target scenarios with non-compressed and compressed spectra signified in Scenario 1 with a Bino-like LSP, or in Scenarios 2b/3b with lower-lying Wino/Higgsino states. The leading search channels address the generic DY pair production of
\begin{itemize}
\item charged and    neutral EWkinos  with subsequent decays into $W\n$ and $Z/h\n$; 
\item two charged EWkinos decaying into $W\n\ W\n$.
\end{itemize}
Results from  these analyses can then be  interpreted  in terms of the theory parameters associated with the scenarios described in Sec.~\ref{sec:model}, and thus can be connected to the underlying theoretical models. Constraints can then be imposed on models predicting decays via  other SUSY states, including {\it e.g.} heavy Higgs bosons, if kinematically allowed.

Nearly-degenerate spectra arise in Scenarios 2 and 3 when the heavier multiplets are decoupled from the lightest one.
As a result, the only accessible decays happen within the lightest Wino-like or Higgsino-like multiplets resulting in low transverse momentum decay products or long-lived EWkinos. These scenarios require dedicated experimental techniques.

Searches for non-compressed scenarios are presented in Sec.~\ref{sec:nonCompressed} and those for compressed  and nearly-degenerate spectra are summarized in Sec.~\ref{sec:compressed} and Sec.~\ref{sec:veryCompressed}, respectively.

\subsubsection{Search methodology}

The ATLAS and CMS collaborations conduct searches for SUSY as  ``blind" analyses in that the signal regions are defined by optimizing the expected sensitivity with respect to a selected model, where a model may be either a realistic framework assuming a specific SUSY breaking and mediation mechanism, or the phenomenological model referred to as ÒpMSSMÓ, or the so-called ÒsimplifiedÓ models. In the simplified models the re-interpretation of the search results is presented in the parameter space defined by the masses of the charginos
and neutralinos, under the assumption of pure states and of 100\% BF into the final state of interest (unless specified). In the pMSSM the space is instead defined by the $\mu$, $M_1$, and $M_2$ parameters governing
the EWkinos masses and composition, and thus their production cross-section and decay branching fractions.

Several SM processes lead to events similar to those expected from the EWkinos' production and subsequent decays. 
The backgrounds due to multijet, bosons plus jet, and top quarks pair production are typically
estimated using data driven methods based on ``control" regions (CR), a  
 subset of events with negligible signal contributions  used to constrain the yield of
 SM processes.   
 Backgrounds 
due to electroweak production of bosons and rare processes ({\it e.g.}~di- and tri-boson or Higgs 
production) are instead
estimated using Monte Carlo simulated data with yields normalized to the
state-of-the-art calculated cross-sections. 
The background predictions obtained from a background-only fit of the CRs can be compared with the
observed data in validation regions to verify the accuracy of the background modeling.

To extend the reach to the largest  possible region of parameter space, candidate
events are   classified depending on the value of selected 
 observables ({\it e.g.}~the  missing transverse momentum or \met): the observable's spectrum
 is ``binned" into multiple (up to hundreds) signal regions or SRs. 
 If the SM background expectations in all SRs are found in agreement with the
observed data within the estimated statistical and systematic 
uncertainties, the results from the search are interpreted as
an upper limit on the SUSY production cross-section. Likelihood
fits are deployed assuming a background-only hypothesis, a
model independent signal plus background hypothesis, and 
a model-dependent signal plus background hypothesis. 
The likelihood incorporates information from  all signal 
and control regions as they are defined in the analysis. This 
approach enables to constrain the expected background to the
yields observed in the data and to reduce the systematic
uncertainties. 
The systematic  uncertainties are considered in the fit 
as nuisance parameters and are constrained by selected distributions
while taking correlations between signal and backgrounds   into account.
The upper limits on the number of SUSY events in each SR and
the  upper limits on the  SUSY cross-sections are  computed at 95\% confidence level 
(CL) using the CL$_s$ method~\cite{asymptotic,confidencelevel,CL}. 
The model independent upper limits are computed using Monte Carlo pseudo-experiments while
the model dependent upper limits using the asymptotic formulae~\cite{1007.1727}.

\input{nonCompressed}

\input{compressed}

\input{veryCompressed}

\input{future}

%% file: nonCompressed.tex
\subsubsection{Searches for non-compressed SUSY spectra\label{sec:nonCompressed}}

Searches for  EWkinos  in non-compressed spectra 
are optimized for the $s$-channel production of mass-degenerate Wino-like states \cha\ and \nn.  Their production cross section at the LHC is shown in Fig.~\ref{fig:xsec}(a, f), discussed as Scenarios 1a and 3b in Sec.~\ref{sec:ProDecay}. 

Searches for $\cha\nn \rightarrow W\n h\n$ are  typically carried out in final states with at least one lepton from the decay of the $W$-boson, to 
 benefit from a reduction of the multi-jet background, while  various decays of the Higgs boson  are  explored to maximize the sensitivity. 
 The ATLAS collaboration has recently completed a search based on 139\ifb\ of $\sqrt{s}$=13~TeV proton-proton
 collisions targeting Higgs boson's decays into a $b\bar{b}$ pairs~\cite{1909.09226}. 
 Signal to background discrimination is achieved by mean of  several mass observables:
\begin{itemize}
\item the invariant mass of the two $b$-jets system~\footnote{Jets containing b-hadrons are referred to as b-tagged or simply b-jets.}, required to be consistent with the Higgs boson mass;
\item  the transverse mass  $m_T= \sqrt{2\met \pt(1-\cos\phi_{\met\ell})}$. When a particle decays into a charged and a neutral  daughter 
the $m_T$ exhibits an end-point at the value of the mother particle mass. The transverse mass therefore helps to  suppress events where  
 a $W$ boson decays leptonically as $W\rightarrow \ell\nu$; 
\item the invariant mass of the lepton and highest \pt\ $b$-jet, which exhibits an end-point  at $\sqrt{m^2(t)-m^2(W)}$ in $t\bar{t}$ and single-top background events;
\item the  cotransverse mass $m_{CT}=\sqrt{2\pt^{b_1}\pt^{b_2}(1+\cos\Delta\phi_{bb})}$ where $b_i$ ($i=1, 2$) are the  selected $b$-jets
and $\Delta\phi_{bb}$ is the azimuthal angle between them. The  $m_{CT}$ is adopted to suppress the $t\bar{t}$ background as well as it shows  
an end-point at $(m^2(t)-m^2(W))/m(t)$.
\end{itemize} 
Degenerate Wino-like \cha\ and \nn\ with mass up to 740 GeV are excluded  for  massless \n. Results are presented in Fig.~\ref{fig:Wh_ATLAS}  (top) along with
those from a novel search in the fully hadronic mode $\cha\nn \rightarrow Wh\rightarrow q\bar{q} b\bar{b}$~\cite{1812.09432} providing good sensitivity 
in the background-free region  at large $\Delta M(\nn,\n)$.  The latest CMS searches  for   $\cha\nn \rightarrow W\n h\n$  are documented in~\cite{1706.09933,1908.08500}.

\begin{figure}
\includegraphics[height=6cm]{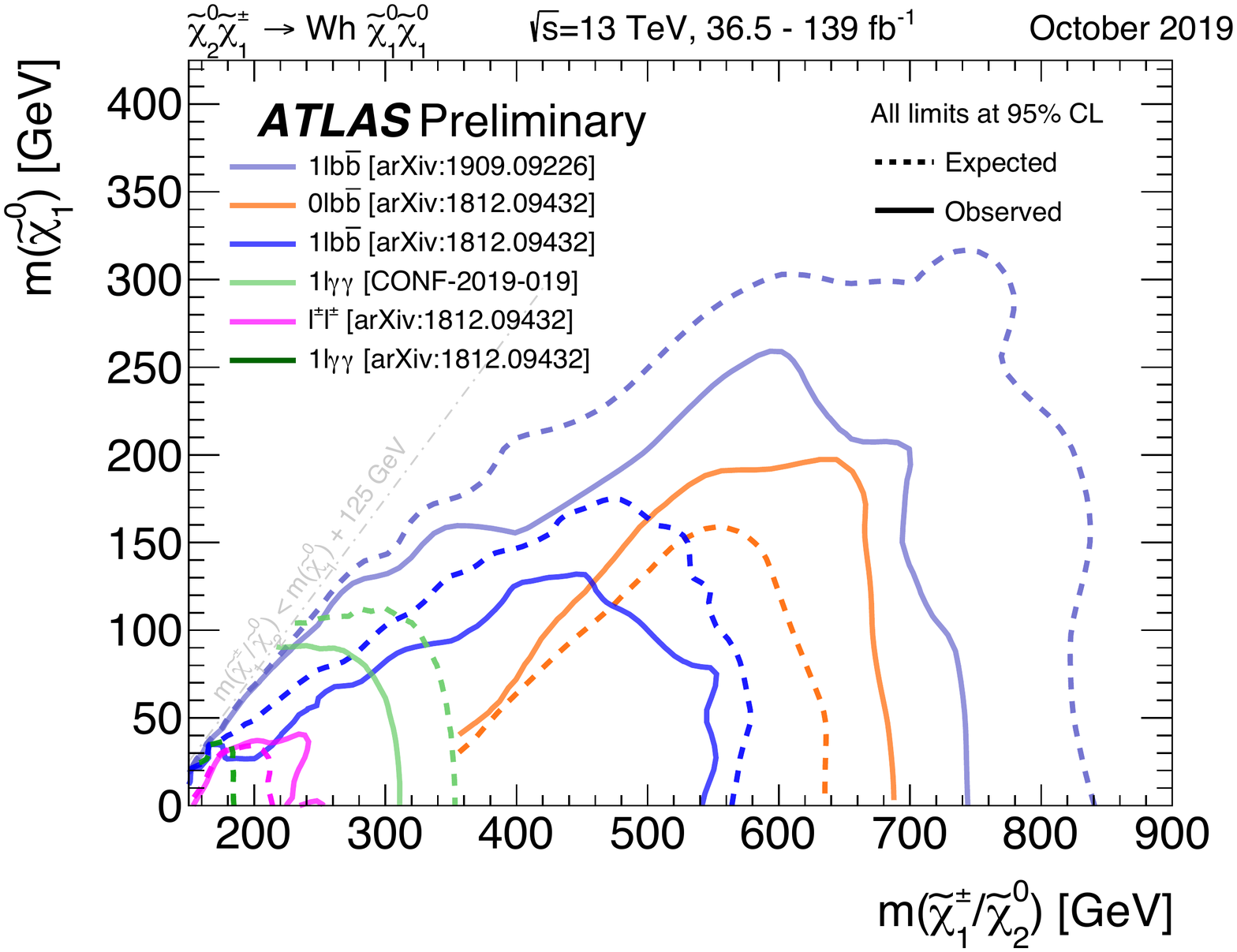}
\includegraphics[height=6cm]{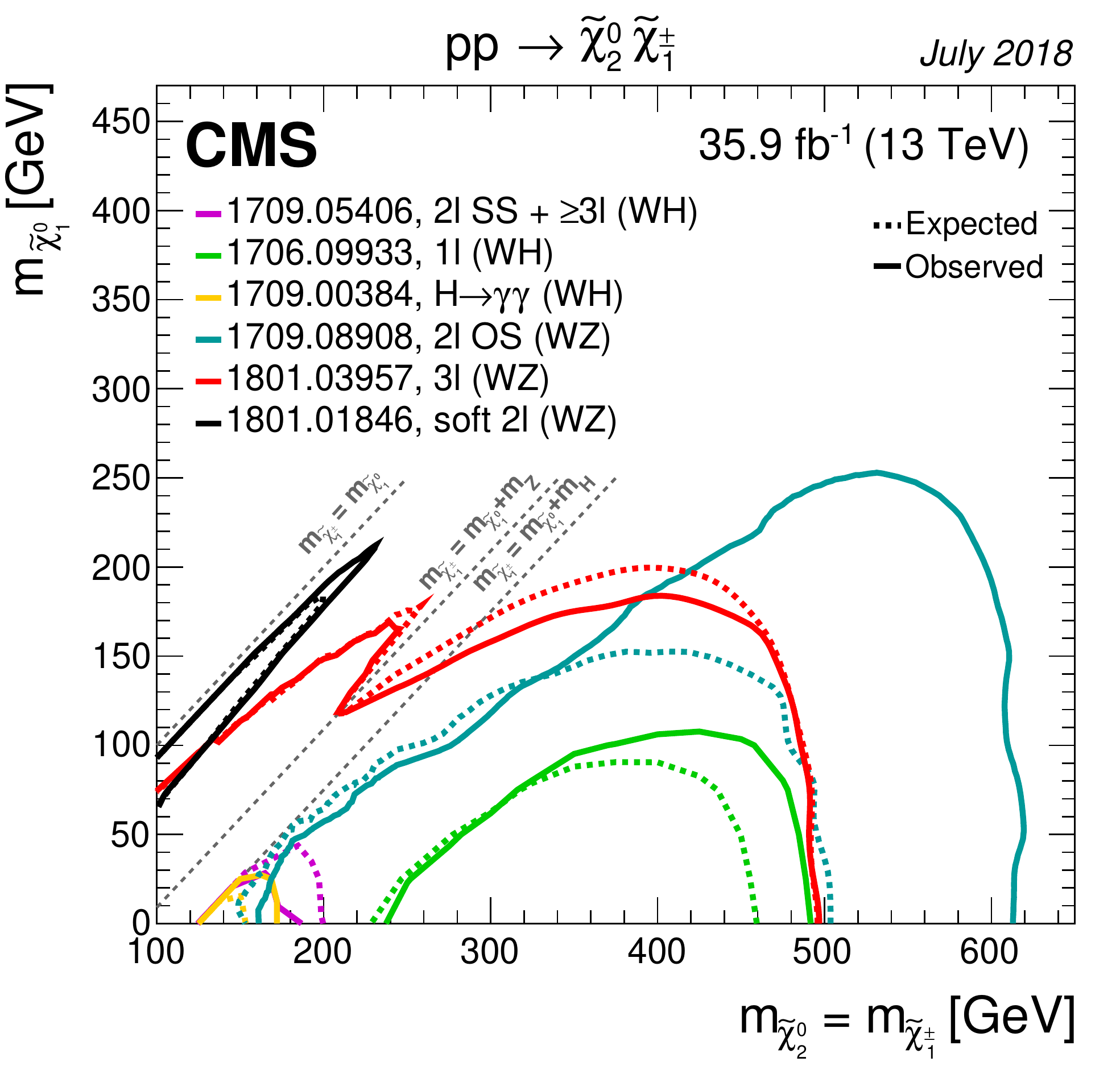}
\caption{
The 95\% CL exclusion limits on \cha\nn\ production as a function of their masses and the \n\ mass.
The \cha ,\nn\ are assumed to decay into \n\ by emitting a $W$   boson and a Higgs boson, respectively, in the top plot.
The bottom plot shows 95\% CL exclusion limits  set assuming various decays of the \nn, including decays via $Z$ bosons. 
In both cases the production cross-section is for   Wino-like \cha\  and \nn. \label{fig:Wh_ATLAS}
}
\end{figure}


In scenarios where the $Z\nn\n$ coupling is significant, 
the search for $\cha\nn \rightarrow W\n Z\n$ can probe a broad area of the $(m_{\nn}, m_{\n})$ 
space thanks to the large $Z$ boson's width. 
If $\Delta M(\nn,\n) \gg m_{Z}$, the  $\cha\nn$ production leads to final states  with high \pt\ leptons or jets from the gauge bosons' decay and  significant  \met. 
Both the ATLAS and CMS collaborations have  developed searches 
 in events with  two leptons from the $Z$ decay and jets from the hadronic decay of the accompanying $W$  ( ``Z+j" search).
 Selecting the leptonic decay of the $Z$ boson enables to suppress the multijet background, while  the exploration of the hadronic decays of the  $W$ maximizes the signal acceptance. 
 In the Z+j CMS search~\cite{1709.08908}, the signal is separated from the remaining $t\bar{t}$ background by 
rejecting events with $b$-jets and by mean of the stransverse mass $m_{T2}$~\cite{stransversemass2}.
The $m_{T2}$ was  originally defined to measure the mass of pair-produced particles, each decaying to a visible and an invisible particle, 
and can  be exploited  to identify the fully leptonic decays of top quarks $t\bar{t} \rightarrow W^+bW^-\bar{b} \rightarrow \ell^+\nu b\ell^-\nu \bar{}b$  as well as those from pair produced $W$s. 
To maximize the  reach, candidate events are   categorized depending on the \met\ and the dijet invariant mass, expected to be consistent with 
the $W$ boson mass in $\cha\nn \rightarrow W\n Z\n$ processes. 
Figure~\ref{fig:Wh_ATLAS} (bottom) shows that the Z+j CMS analysis excludes
mass-degenerate  Wino-like \cha and \nn\  lighter than 
 610~GeV if the \n\ is massless. The Z+j ATLAS search has 
a similar reach~\cite{1806.02293}.

In addition to searches for  $\cha\nn$  production, the exploration of   \cha\ pair production followed by  $W$-mediated decays also represents an  avenue for discovery
of scenarios with  relatively large mass splittings.
Being the \cha\champ\ cross-section comparable to the \cha\nn\ process,  added sensitivity is achieved if   
 the  $W^+W^-$ background is significantly suppressed. In ~\cite{1908.08215}, the ATLAS collaboration targets the challenging dilepton final state from $\cha \rightarrow W\n \rightarrow \ell\nu\n$ 
categorizing  events based on the  
$m_{T2}$,  \met, and  \met\ significance values~\footnote{The \met\  significance is computed on an event-by-event
basis and evaluates the p-value that the observed \met\ be  consistent with the null hypothesis of zero real \met~\cite{ATLASCONF2018038}.
}. The analysis of 139\ifb\ of data  yields sensitivity to Wino-like \cha\ with masses up to 420~GeV if the \n\ is massless.

Scenarios characterized by mass splittings closer to $m_Z$, where the signal kinematics resembles that of the
dominant $WZ$ background, can be probed through the 
fully leptonic decays of the $W, Z$ bosons from  $\cha\nn \rightarrow W\n Z\n$. The analyses, dubbed at ``multilepton" searches, 
typically request events with two leptons of same 
electric charge or $\geq 3$  leptons.
Selecting events with two same-charged leptons increases the
acceptance to scenarios with small  $\Delta M(\nn,\n)$ where one lepton happens to have a transverse momentum below the default threshold. 
The inclusive approach adopted by the multilepton CMS analysis~\cite{1709.05406} relies on splitting events with significant
\met\ into sub-categories based on the number and flavor of leptons (electrons, muons, hadronically decaying taus), topological and kinematical observables including: 
\begin{itemize}
\item the invariant mass of the two oppositely charged same flavor leptons, allowing to identify and suppress the SM $WZ$ background;
\item the transverse momentum of the dilepton system, sensitive to the production of a single resonance and thus further discriminating 
events with and without a $Z$ boson;
\item the minimum transverse mass computed for each lepton in the event, a variable sensitive to the SM production of $W$  bosons decaying into $\ell\nu$;
\item  the stransverse mass, exhibiting an end-point at the $W$ boson mass and therefore helping  to suppress the $W^+W^-$ and 
$t\bar{t}$ SM backgrounds.
\end{itemize} 
The analysis  complements the sensitivity provided by the Z+jets search extending the
reach to the bulk of the ($m_{\nn}, m_{\n}$) space  as shown in Fig.~\ref{fig:Wh_ATLAS}. 
 The CMS collaboration also implemented a statistical combination
of the results from the two searches  and extended the limit on the   \cha and \nn\  mass by approximately 40~GeV
in case of massless \n\ and yielded sensitivity 
to models with intermediate mass values that were not probed by individual analyses~\cite{1801.03957}.  
The multilepton ATLAS analyses are documented in~\cite{1803.02762,1806.02293}.


%% file: compressed.tex
\subsubsection{Searches for compressed SUSY spectra\label{sec:compressed}}

Compressed spectra can emerge in Scenarios 1, as well as Scenarios 2b and 3b. In these cases, the sensitivity of the classical
searches described in Sec.~\ref{sec:nonCompressed} deteriorates significantly. 
These spectra can nevertheless be probed by exploring a subset of  signal events with additional SM objects
 enabling the experiments to efficiently discriminate  the signal from the background:   
 DY events with an  initial-state-radiation (ISR) jet  boosting the sparticle system and increasing the \met\ in the laboratory (``ISR" search), and events where the sparticles are produced via vector boson fusion and are therefore accompanied by two jets from the protons' remnants located in opposite forward-backward regions of the detector (``VBF" search).  
 
 
 In the ISR analyses,   the dominant multi-jet background is typically suppressed
 by reconstructing the two low transverse momentum same-flavor oppositely-charged leptons from the $\nn \rightarrow Z^*\n$ decays and requesting their 
 invariant mass to be compatible with the $Z^*$ mass. To maximize the acceptance for scenarios with very small mass splittings, the ATLAS 
 search~\cite{1911.12606} also includes a signal region based on a lepton and 
 an isolated track with \pt\ in the 1-5~GeV range. This selection targets scenarios with a reconstructed $m_{\ell,tk}$ invariant mass between 0.5 and 5~GeV.
In addition to optimized criteria based on the \met, transverse mass, $b$-jet multiplicity,  subleading lepton \pt, signal to background discrimination in the ATLAS  ISR search is obtained by exploiting:
\begin{itemize}
\item the $m_{\tau\tau}$ observable proposed in \cite{1401.1235,1409.7058,1501.02511},  approximating the invariant mass of a $\tau$ pair where both $\tau$s are boosted and decay leptonically. The  $m_{\tau\tau}$ is deployed to reject events from   $Z/Z^* \rightarrow \tau\tau$;
\item two observables defined using the recursive jigsaw reconstruction technique~\cite{1705.10733}. In the  jigsaw technique, 
the event is split into two hemispheres perpendicular
to the thrust axis  approximating the direction of the recoil of the ISR jets against the sparticles pair: 
one hemisphere is expected to contain the decay products of the \cha\ and \nn\  ($S$ system), while the opposite
hemisphere is associated with the hadronic activity (ISR system). 
 The ratio of the \met\ and the ISR system \pt\ is sensitive to
the sparticles mass splitting, while the transverse mass of the $S$ system can be used to suppress
background events with $W$ bosons thanks to its end-point at the $W$ mass; 
\item the ratio of the \met\ and the scalar sum of the leptons \pt\ expected to be small in SM processes. 
\end{itemize}
The limits in the ($\nn,\Delta M$) plane are obtained by
fitting the  dilepton invariant mass  distribution under the assumption of either Wino-like or Higgsino-like EWkinos. 
The results of the search carried out in 139\ifb\ of data show that Wino-like EWkinos with masses up to 240~GeV are excluded
if $m_{\n} \times m_{\nn} > $ 0 and $\Delta M(\nn,\n)=$7~GeV (Fig.~\ref{fig:sos_atlas_winos}, left). If the $\cha$ mass values are close to the
LEP limit,  mass splittings from 1.5~GeV to 46~GeV are probed. The interpretation of the search results under the assumption of  Higgsinos-like
EWkinos production 
is presented in Fig.~\ref{fig:sos_atlas_winos} (right).  The CMS collaboration published a similar search in~\cite{1801.01846}  
including both an interpretation under the assumption of Wino-like \cha\ and \nn\ (labelled ``soft 2l (WZ)", in Fig.~\ref{fig:Wh_ATLAS}, bottom)
and within a selected region of the pMSSM shown in  Fig.~\ref{fig:sos_atlas_higgsinos} (left). The latter highlights that the LHC
experiments have so far  surpassed the sensitivity achieved at LEP only in few limited regions of parameter space. 
  
\begin{figure}[h!]
\begin{subfigure}{.5\textwidth}
\includegraphics[width=\textwidth]{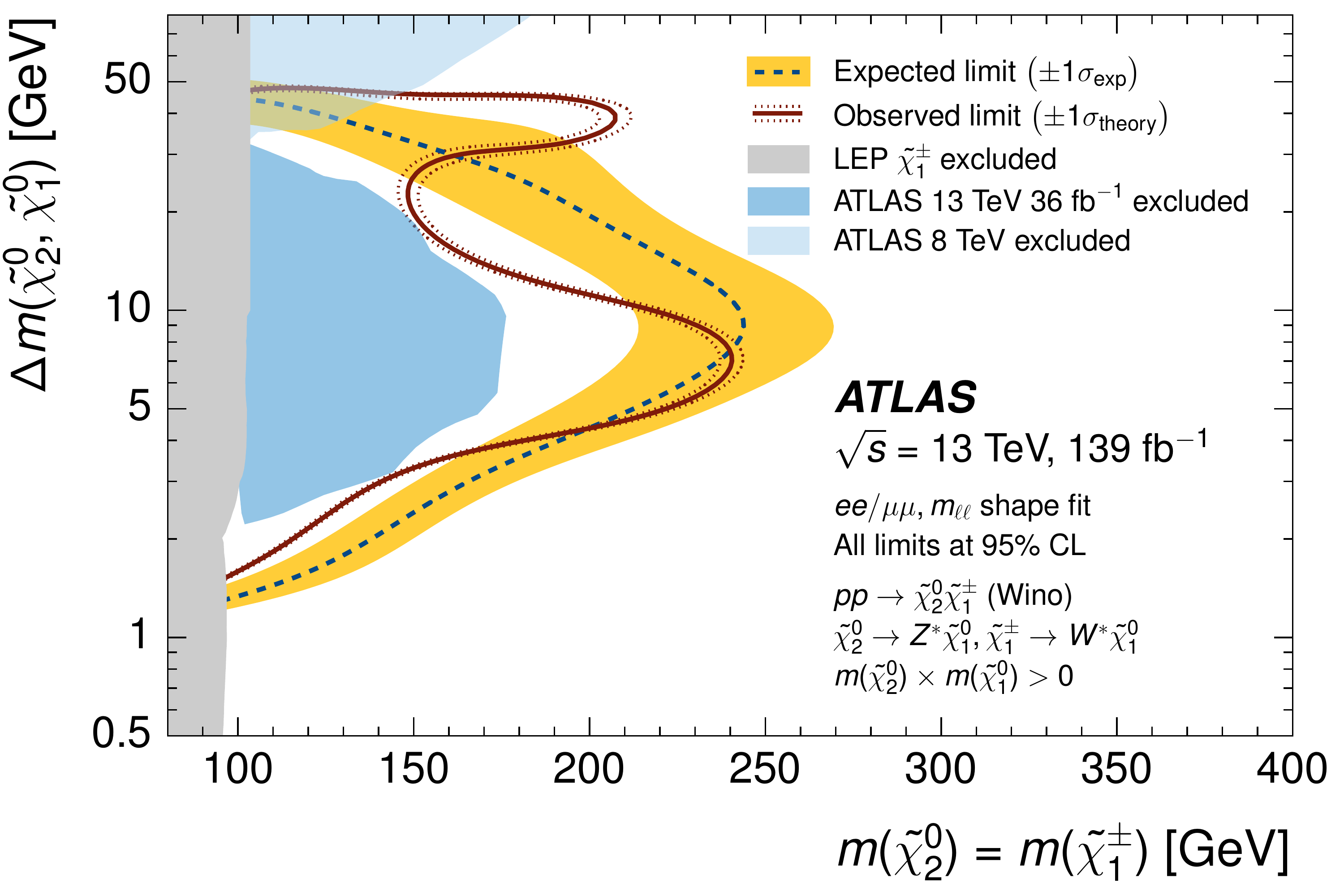}
\end{subfigure}
\begin{subfigure}{.5\textwidth}
\includegraphics[width=\textwidth]{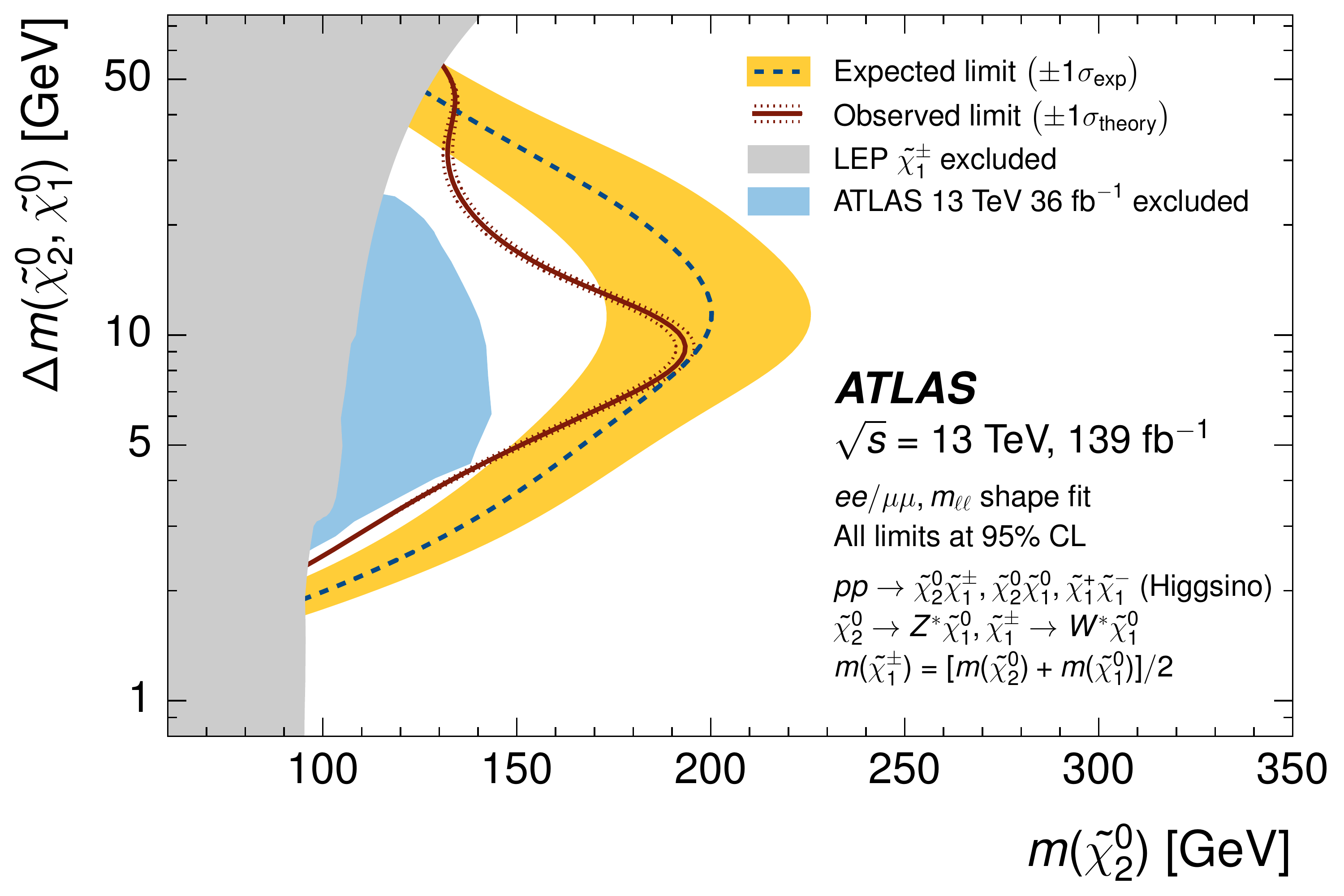}
\end{subfigure}
\caption{(Left) 
Expected 95\% CL exclusion sensitivity  with $\pm 1\sigma_{exp}$  from experimental systematic uncertainties and statistical uncertainties on the data yields, and observed limits   with  $\pm 1\sigma_{theory}$  from signal cross-section uncertainties. 
The Wino-like \cha\ and \nn\ are assumed to be mass degenerate.  In these models, the $m_{\ell\ell}$ shape depends on the relative sign of the $\nn$ and $\n$ mass parameters,  
 $m_{\n} \times m_{\nn} $, assumed to be positive in this case. More details are presented in~\cite{1911.12606}.
 (Right) Expected 95\% CL exclusion sensitivity with $\pm 1\sigma_{exp}$   from experimental systematic uncertainties and statistical uncertainties on the data yields, and observed limits  with  $\pm 1\sigma_{theory}$   from signal cross-section uncertainties. The EWkinos are assumed to be Higgsino-like. 
The chargino $\cha$ mass is assumed to be halfway between the $\nn$ and the $\n$ masses~\cite{1911.12606}. }
\label{fig:sos_atlas_winos} 
\end{figure}

\begin{figure}[h!]
\begin{subfigure}{.5\textwidth}
\includegraphics[width=\textwidth]{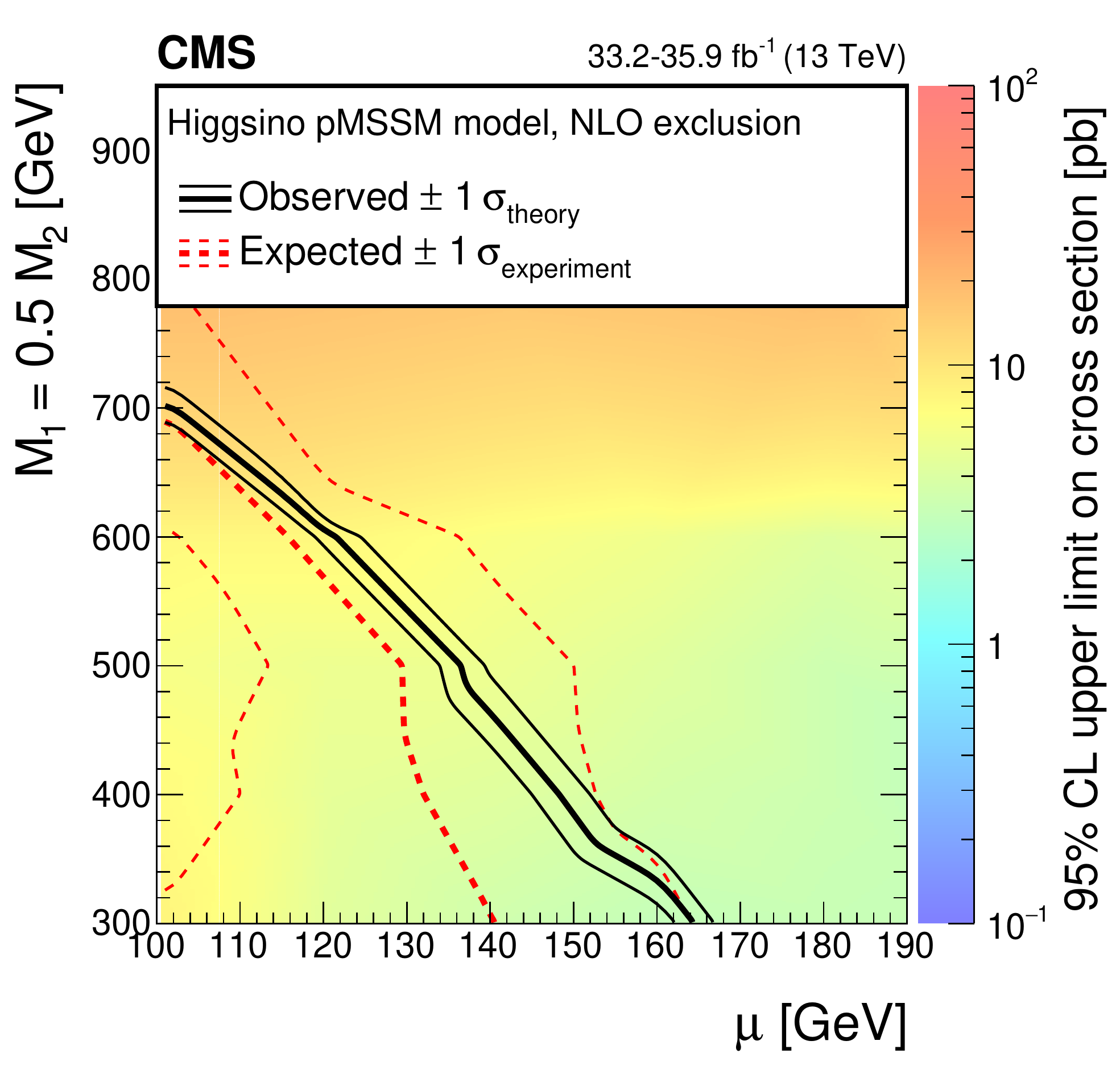}
\end{subfigure}
\begin{subfigure}{.5\textwidth}
\includegraphics[width=\textwidth]{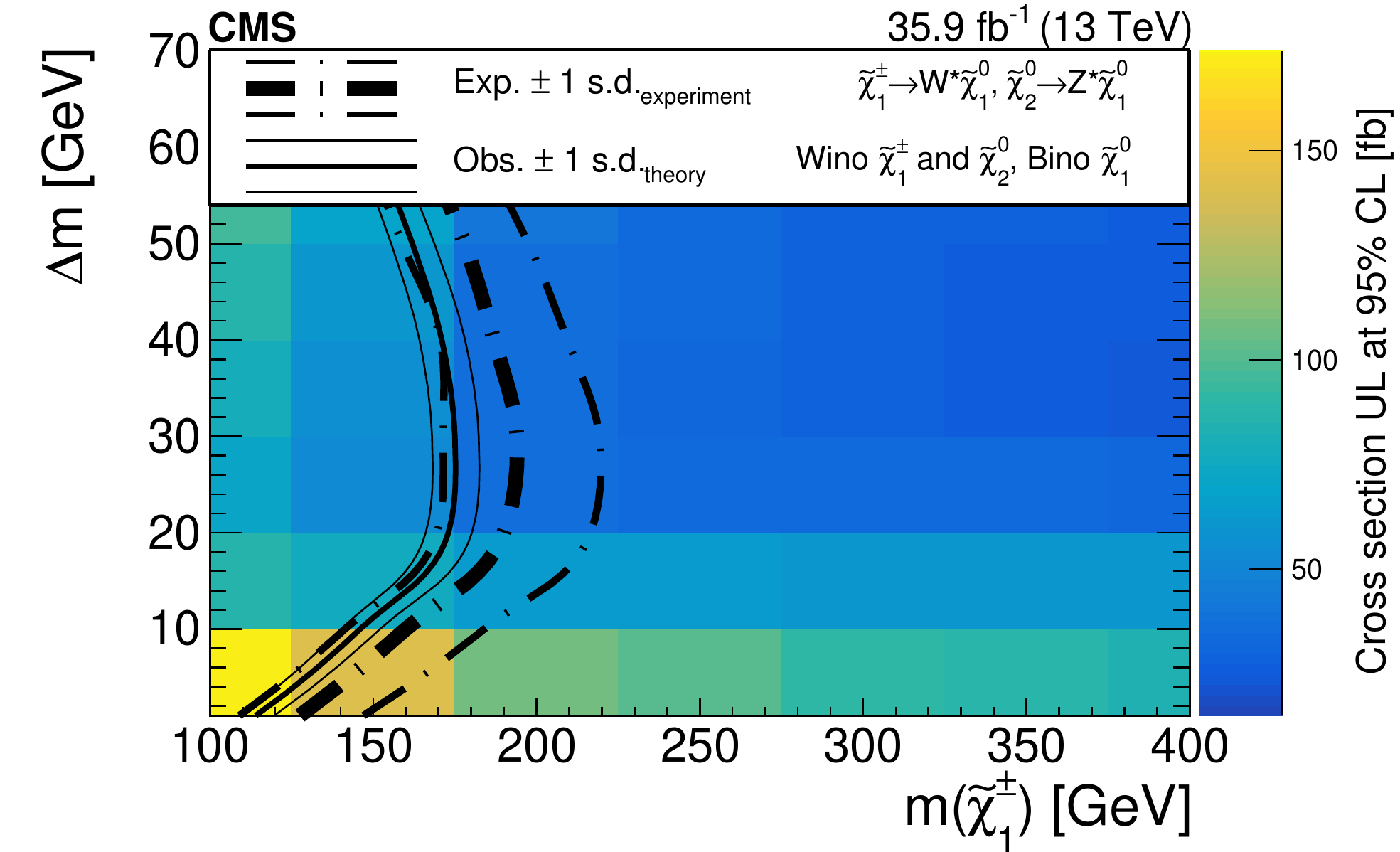}
\end{subfigure}
\caption{ 
(Left) Expected 95\% CL exclusion sensitivity   with $\pm 1\sigma_{exp}$   from experimental systematic uncertainties and statistical uncertainties on the data yields, and observed limits   with  $\pm 1\sigma_{theory}$   from signal cross-section uncertainties in the pMSSM  described in~\cite{1801.01846}. 
(Right) Expected 95\% CL exclusion sensitivity  with $\pm 1\sigma_{exp}$  from experimental systematic uncertainties and statistical uncertainties on the data yields, and observed limits   with  $\pm 1\sigma_{theory}$  from signal cross-section uncertainties.  The colored map reports the 95\% CL upper limits on the cross section.  
The EWkinos are Wino-like and produced via VBF~\cite{1905.13059}. 
}
\label{fig:sos_atlas_higgsinos} 
\end{figure}

Even though the cross section for vector-boson fusion (VBF) production of EWkinos is smaller than for the  $q\bar{q}$ annihilation processes, the striking signature with two forward-backward jets of $\pt \sim M_W$ enables the experiments to efficiently extract the signal from the QCD background. 
The VBF production is usually identified by requesting 
two jets ($j_1, j_2$) with large   invariant mass, large  $\Delta\eta(j_1,j_2)$, and reconstructed in opposite
hemispheres of the detector. VBF signal events are also expected to exhibit large \met\ as the 
 \n\ from the electroweakinos decays receive a boost from the
two forward jets. While adopting a similar baseline event selection,  
the ATLAS and CMS collaborations 
then developed a complementary approach to maximize the reach of their searches. 
 In  \cite{1911.12606}, the ATLAS collaboration focuses on events with two low \pt\ leptons from 
the $\cha\nn \rightarrow W\n Z^*\n$ decays  and fits the dilepton invariant mass to compute the
limits in the ($\nn,\n$) space for both Wino- and Higgsino-like models.  Using 139\ifb\ of data, 
the analysis excludes Wino-like (Higgsino-like) \cha\ for masses up to $\sim$75~GeV ($\sim$55~GeV)  depending on the $\Delta M(\nn, \n)$ mass
splitting.  
 In~\cite{1905.13059}
the CMS collaboration  instead chooses events where
the electroweakinos decay either hadronically or semileptonically and probes Wino-like 
\cha\ with masses  up to 112 GeV for mass splittings  as small as 1~GeV (Fig.~\ref{fig:sos_atlas_higgsinos}, right). This 
analysis assumes 
the production of  \cha\nn\, \cha\cha\, \cha\champ, and \nn\nn.  
Despite targeting a lower production cross-section process,  the VBF search achieves a sensitivity
comparable to that of the ISR analysis exploring a statistically independent set of events.

%% file: veryCompressed.tex
\subsubsection{Searches for nearly-degenerate SUSY spectra\label{sec:veryCompressed}}

\begin{figure}[ht!]
\begin{subfigure}{.5\textwidth}
\includegraphics[width=\textwidth]{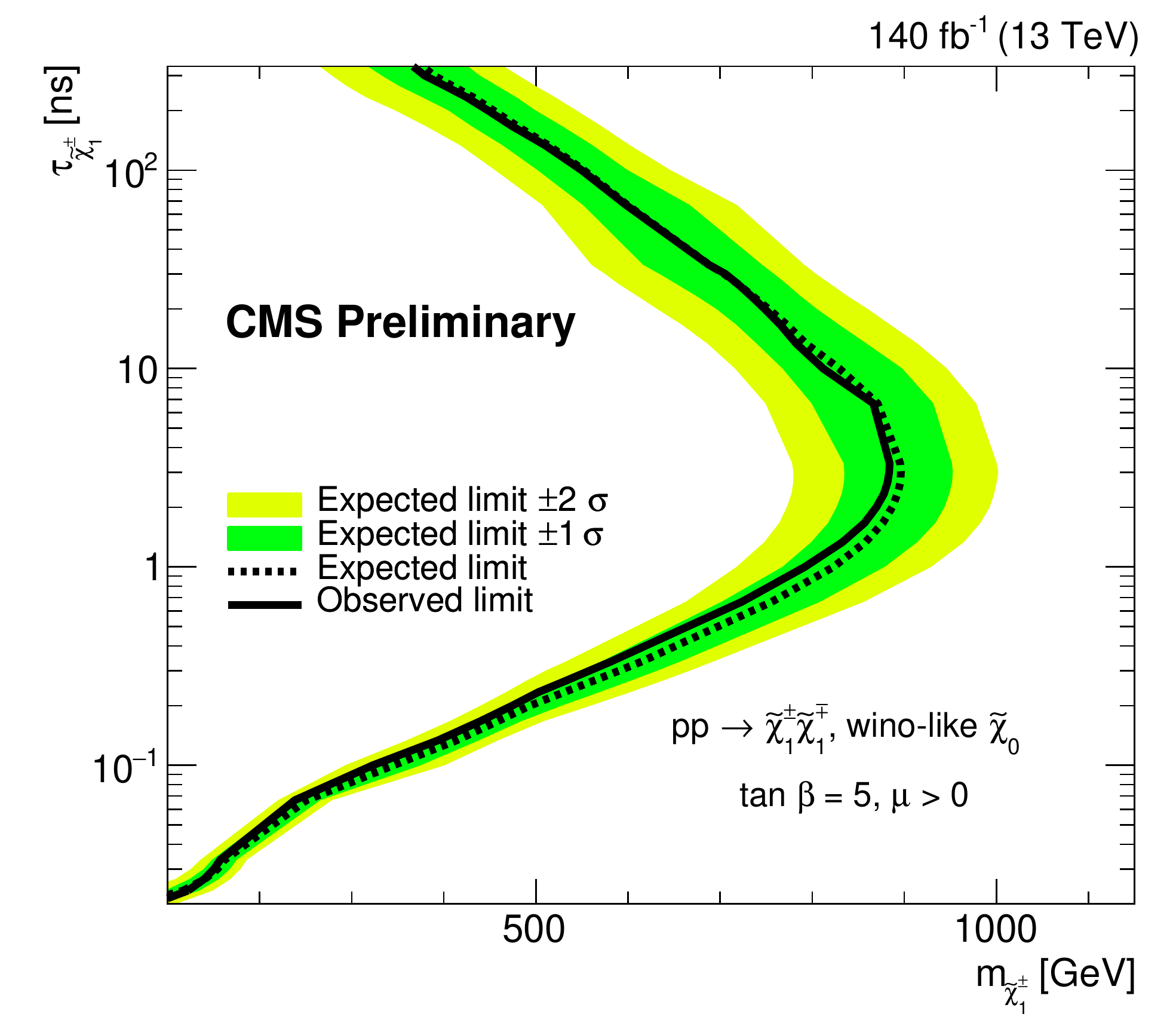}
\end{subfigure}
\begin{subfigure}{.5\textwidth}
\includegraphics[width=\textwidth]{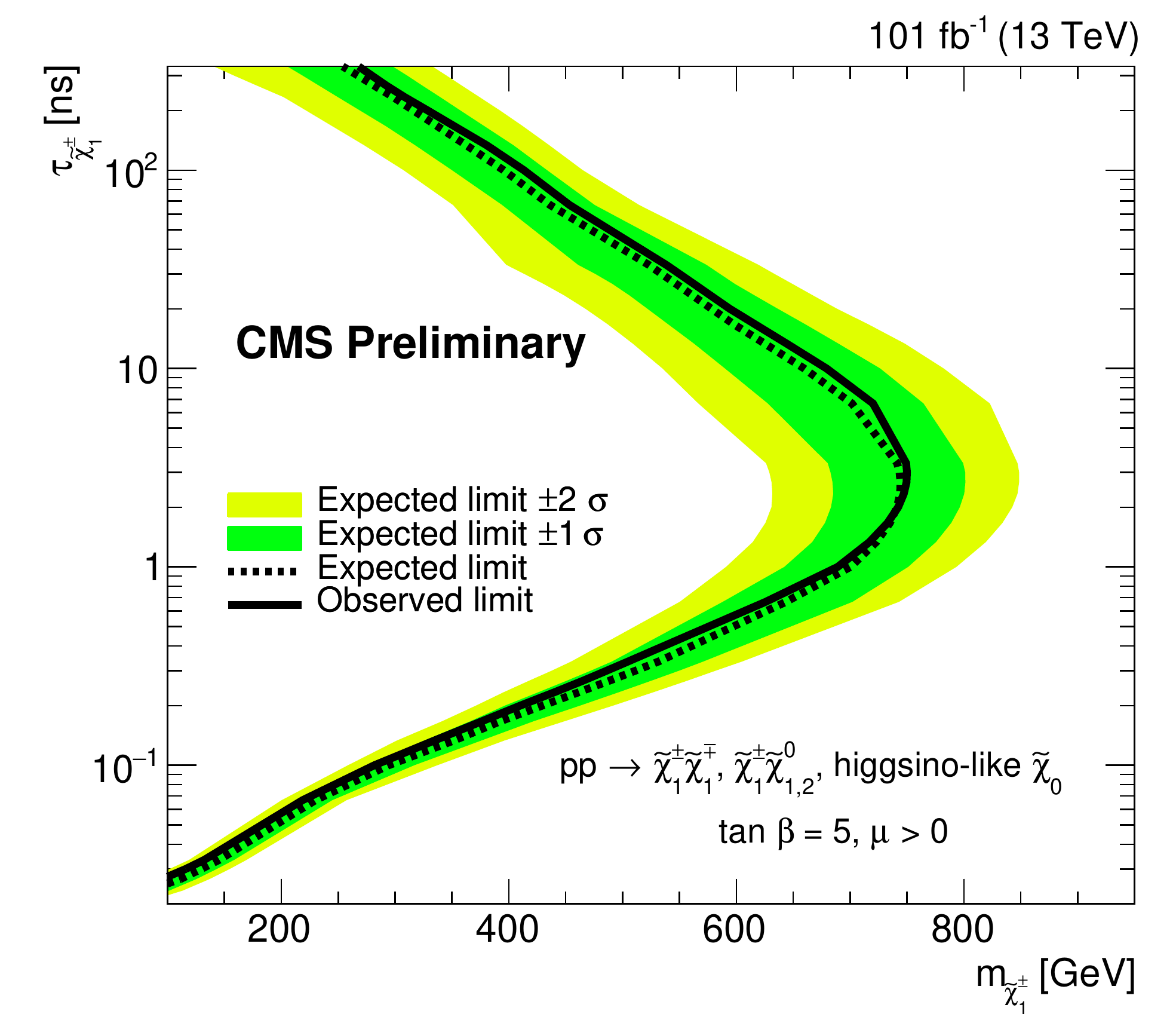}
\end{subfigure}
\caption{ 
 Expected 95\% CL exclusion sensitivity  (left of the curve) with $\pm 1\sigma_{exp}$  from experimental systematic uncertainties and statistical uncertainties on the data yields, and observed limits~\cite{1804.07321}.  The left plot shows the sensitivity to pure Wino \cha, assuming BF($\cha \rightarrow \pi^{\pm}\n$)=100\%), while the right plot assumes the production of Higgsinos.
 The branching fractions of the Higgsino-like \cha\ are reported in the text~\cite{PASEXO19010}.  
\label{fig:DT_CMS} }
\end{figure}

As introduced in Sec.~\ref{sec:modelWinos} and \ref{sec:modelHiggsinos}, the EWkinos' lifetime  is almost uniquely determined by the mass-splitting among states. 
In case of pure Higgsino states, the mass difference of 340~MeV leads to a 
lifetime of 0.05~ns while the lifetime for Wino states, with  $\Delta M$ of 164~MeV, 
 is as large as
0.2~ns.  In Scenarios 2 and 3, if  the heavier multiplets are
 decoupled from the lightest one, the NLSP can  become
long lived and  decay into \n\ at a significant distance with respect to the production point. 
For lifetimes up to a few ns,  the \cha\ from the high cross-section $pp \rightarrow \cha\champ$ and  $pp \rightarrow \cha\n$  
processes decays in the experiment's tracker volume as  $\cha \rightarrow \pi^{\pm}\n$ where the pion has a very low transfer momentum  
and cannot be reconstructed. The branching fraction is close to 100\%. This decay therefore  leads to a peculiar signature of a track with 
 hits only in the innermost layers and no hits in the portions of the tracker at larger radii (``disappearing" track). 
In  the recent ~\cite{PASEXO19010}, the CMS collaboration selects events containing  a disappearing track along with  an ISR jet boosting the sparticles'
system and producing significant \met. The disappearing track candidate 
is required to be compatible with the collision vertex and to have no missing inner
and middle hits~\cite{PASEXO19010}   to reduce the otherwise dominant background from spurious tracks due to pattern recognition errors.
The background from leptons originating in $W$ and $Z$  decays is suppressed by ensuring that the candidate track be 
 spatially separated from  reconstructed leptons. 
The results from this search, 
presented in Fig.~\ref{fig:DT_CMS} (left), indicate that pure Winos  with a lifetime of 3 (0.2) ns are excluded up to a mass of 884
(474)~GeV. The disappearing track search is also sensitive to the production of Higgsinos via $pp \rightarrow \cha\champ$ and
$pp \rightarrow \cha\nn$.  In this case the branching ratio of the \cha\ is modified due to the
presence of the almost mass-degenerate \nn\ as BF($\cha \rightarrow \pi\n$)=95.5\%, BF($\cha \rightarrow e\nu_e \nn\,\n$)=3\%,
and BF($\cha \rightarrow \mu\nu_e \nn\,\n$)=1.5\%. Under these assumptions, 
the analysis probes \cha\ masses as high as 750~GeV (175 GeV) for a lifetime of 3 (0.05) ns
(Fig.~\ref{fig:DT_CMS}, right).

 
\begin{figure}[hb!]
\centering
\includegraphics[width=0.5\textwidth]{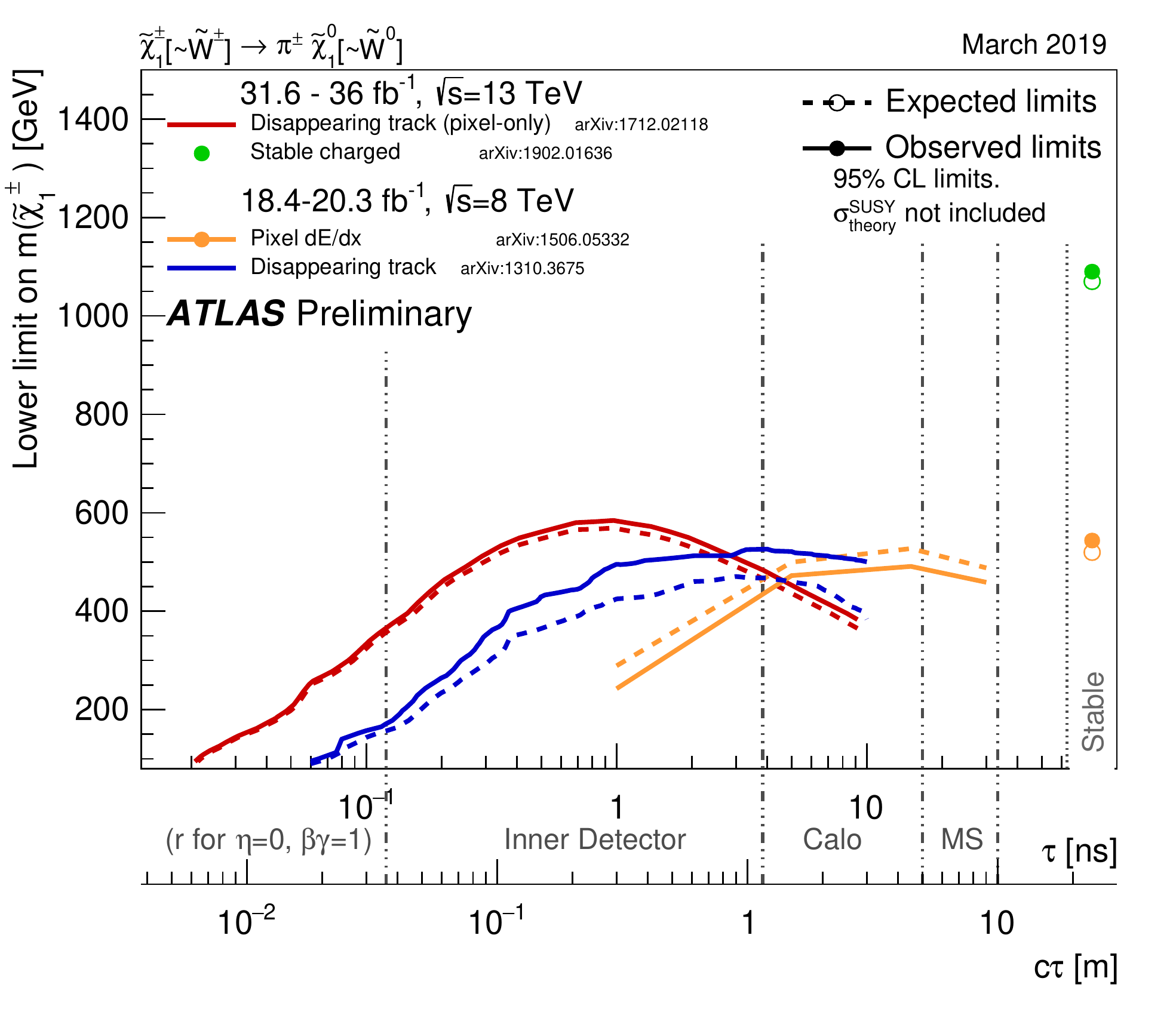}
\caption{
Constraints on the \cha\ mass-vs-lifetime plane for an AMSB model. In this model the Wino-like chargino is pair-produced and decays  as $\cha \rightarrow \pi\n$ into a Wino-like \n. 
It is important to note that  the analyses have sensitivity at lifetimes other than those shown, but only the limits at tested lifetimes are shown. 
}
\label{fig:wino_ATLAS} 
\end{figure}

If the EWkino is stable on the scale of the detector, the sensitivity of the disappearing track searches deteriorates 
since the $\cha$  traverses the entire tracker leaving hits on all layers:  experimental
techniques designed to detect massive charged particles moving at a speed significantly lower than the 
speed of light are adopted. 
In~\cite{1902.01636} ,  the ATLAS collaboration exploits the   ionization energy loss  and time of flight of the candidate particle (identified as a high-quality track)  
to determine the particle's 
mass, which is then used as the main observable to discriminate the signal from the background. The analysis is carried
out in events with   an ISR
jet and significant \met. Sensitivity to stable Winos  with  masses below 1090~GeV is achieved as shown in  Fig.\ref{fig:wino_ATLAS}. 
Results from a previous ATLAS search~\cite{1506.05332} carried out  in 8~TeV data and presented in Fig.\ref{fig:wino_ATLAS} too, and 
indicate that  analyses based on  ionization energy losses offer sensitivity to metastable Winos as well. 
 
%% file: future.tex
\subsection{Expected sensitivity at future colliders}

A significant body of work has been produced in preparation for the 
European Particle Physics Strategy Update ($2018-2020$) as documented
in~\cite{Strategy:2019vxc} and references therein. 
The sensitivity of future colliders  to the SUSY electroweakino sector is determined from projections of results from searches carried out in LHC data as well as from 
dedicated analyses utilizing either a parameterization of the detector
performance tuned to full simulation or a 
 fast multipurpose detector response simulation, Delphes~\cite{1307.6346}.
It is likely that further optimization of these searches may improve the sensitivity demonstrated so far. 

\begin{figure}[h!]
\begin{subfigure}{.5\textwidth}
\includegraphics[width=\textwidth]{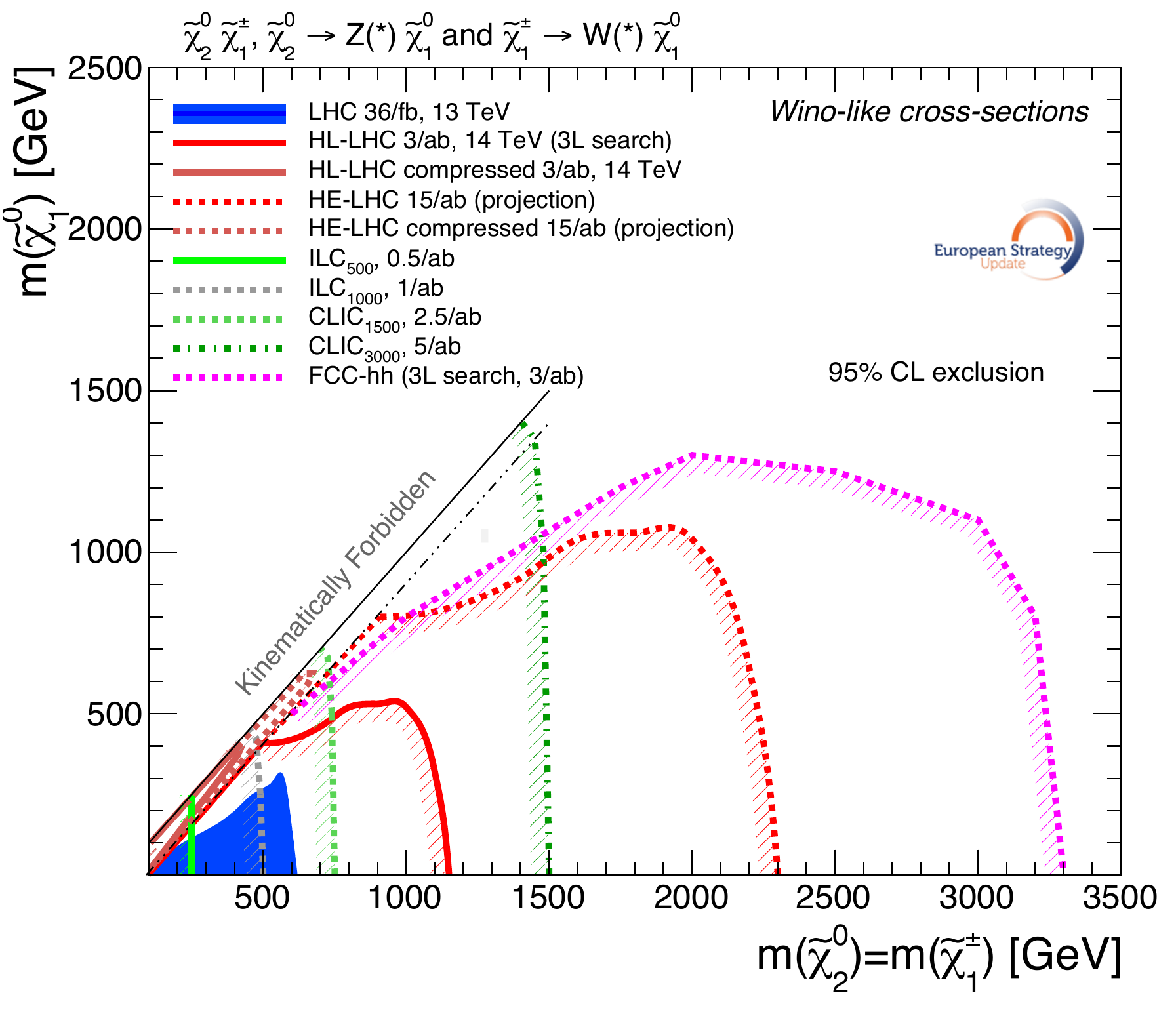}
\end{subfigure}
\begin{subfigure}{.5\textwidth}
\includegraphics[width=\textwidth]{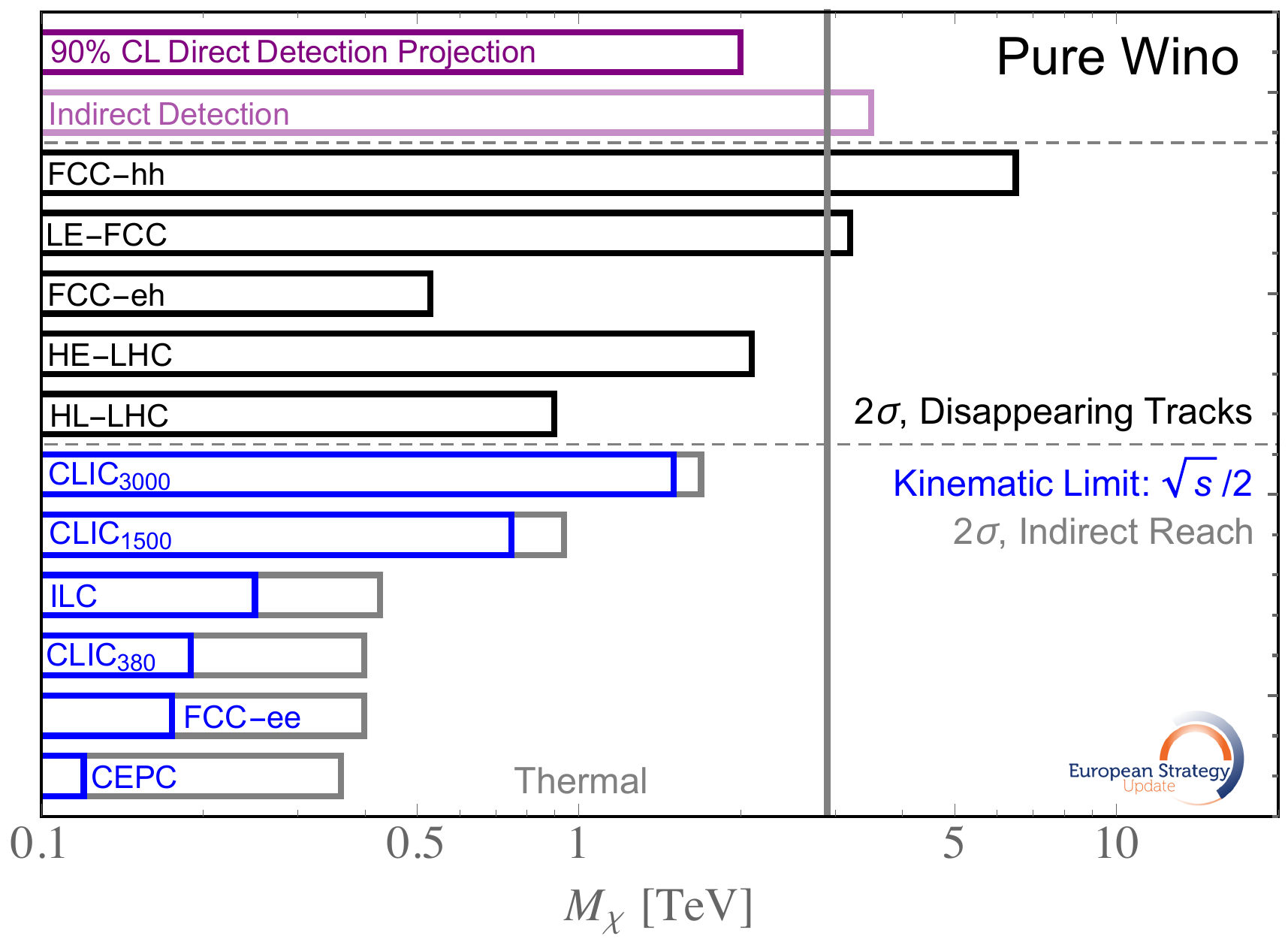}
\end{subfigure}
\caption{(Left) 
 Expected 95\% CL exclusion sensitivity for Wino-like mass degenerate \cha\ and \nn\ as a function of their mass
 and the \n\ mass. 
 (Right) Summary of the 2$\sigma$ sensitivity reach to pure Winos at future colliders. 
 The vertical line indicates the mass corresponding to dark matter thermal relic.
}
\label{fig:futurewinos} 
\end{figure}

Figure~\ref{fig:futurewinos} provides an overview of the reach for Wino-like EWkinos, both in scenarios with a significant mass splitting between the \cha\ and the \n\ (left), and in scenarios with degenerate pure states (right). At hadron colliders, the bulk of the $\Delta M(\cha,\n)$ parameter
space is explored through searches for \cha\nn\ decaying as $\cha\rightarrow W\n$, $\nn \rightarrow Z\n$
in multi-lepton final states. Thanks to the higher center of mass-energy
and larger dataset, the FCC-hh can exclude \cha\ and \nn\
as heavy as 3.3~TeV in scenarios with massless \n. The sensitivity of this 
multilepton search is significantly reduced if the mass gap between states 
becomes of the order of 100~GeV. The HL-LHC  and HE-LHC 
yield sensitivity to heavy electroweakinos with masses of 1~TeV and 2~TeV,
respectively. It is interesting to note that the sensitivity at the HE-LHC is comparable 
to that at the FCC-hh if the \cha,\nn\ masses are smaller than 2~TeV
and the \n\ mass  is close to 1~TeV. 
Future  linear lepton colliders, by scanning the pair production of new particles at the threshold, provide sensitivity to masses as high as $\sqrt{s}/2$. The reach is almost independent of the mass splitting 
among the states under investigation making these machines complementary
to hadron colliders in case of compressed spectra. 
The disappearing track is selected as probe for the production of pure Wino states at 
 hadron machines and yields the sensitivity presented in Fig.~\ref{fig:futurewinos}  (right). 
This search demonstrates that the HL-LHC and HE-LHC can cover the parameter space characterized by pure Winos as heavy as 1 and 2~TeV, respectively, while the FCC-hh extends the sensitivity above 6~TeV
and thus uniquely tests the  hypothesis of thermal dark matter. Linear lepton   colliders offer sensitivity to pure Wino states up $\sqrt{s}/2$. 

\begin{figure}[h!]
\begin{subfigure}{.5\textwidth}
\includegraphics[width=\textwidth]{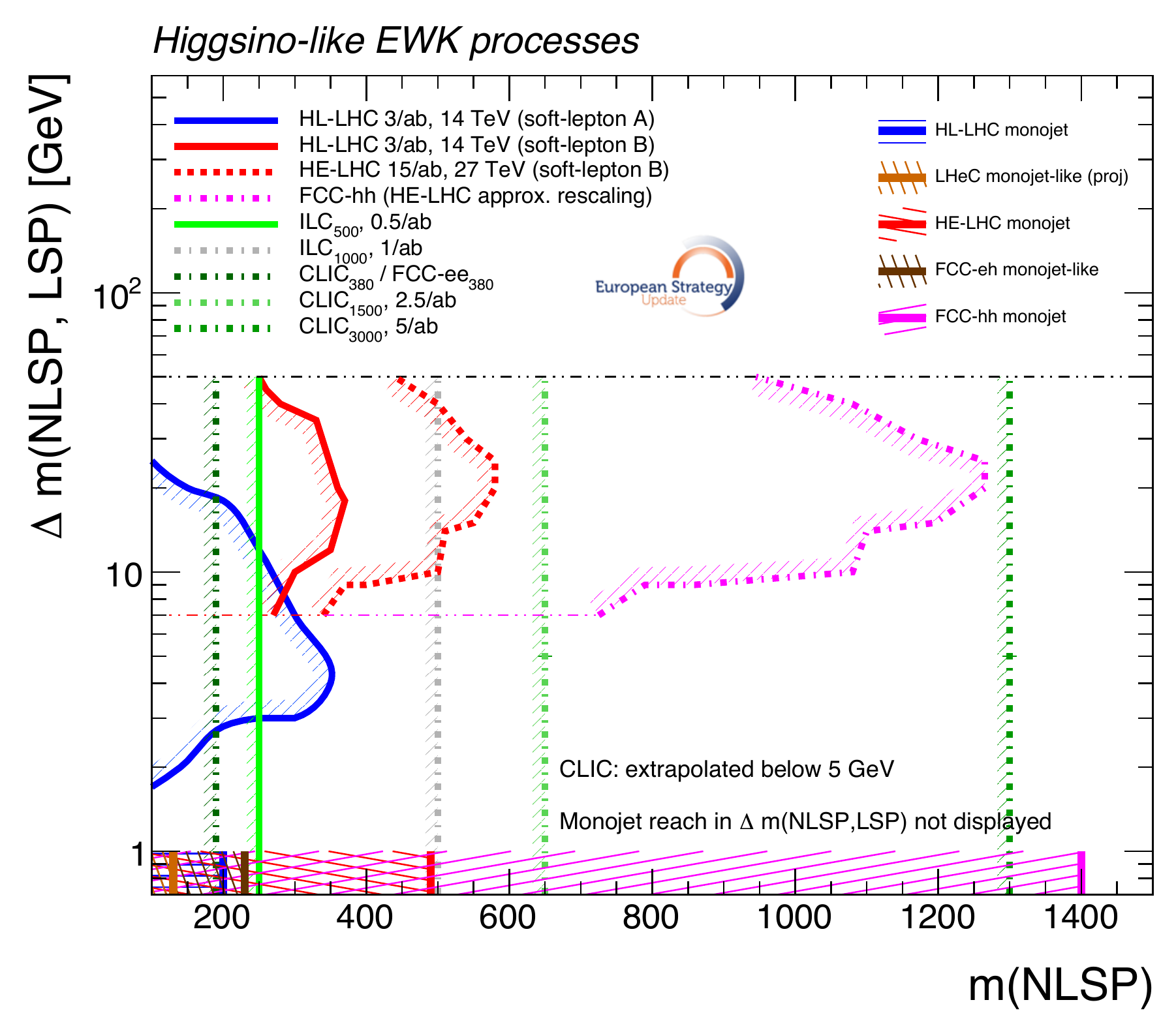}
\end{subfigure}
\begin{subfigure}{.5\textwidth}
\includegraphics[width=\textwidth]{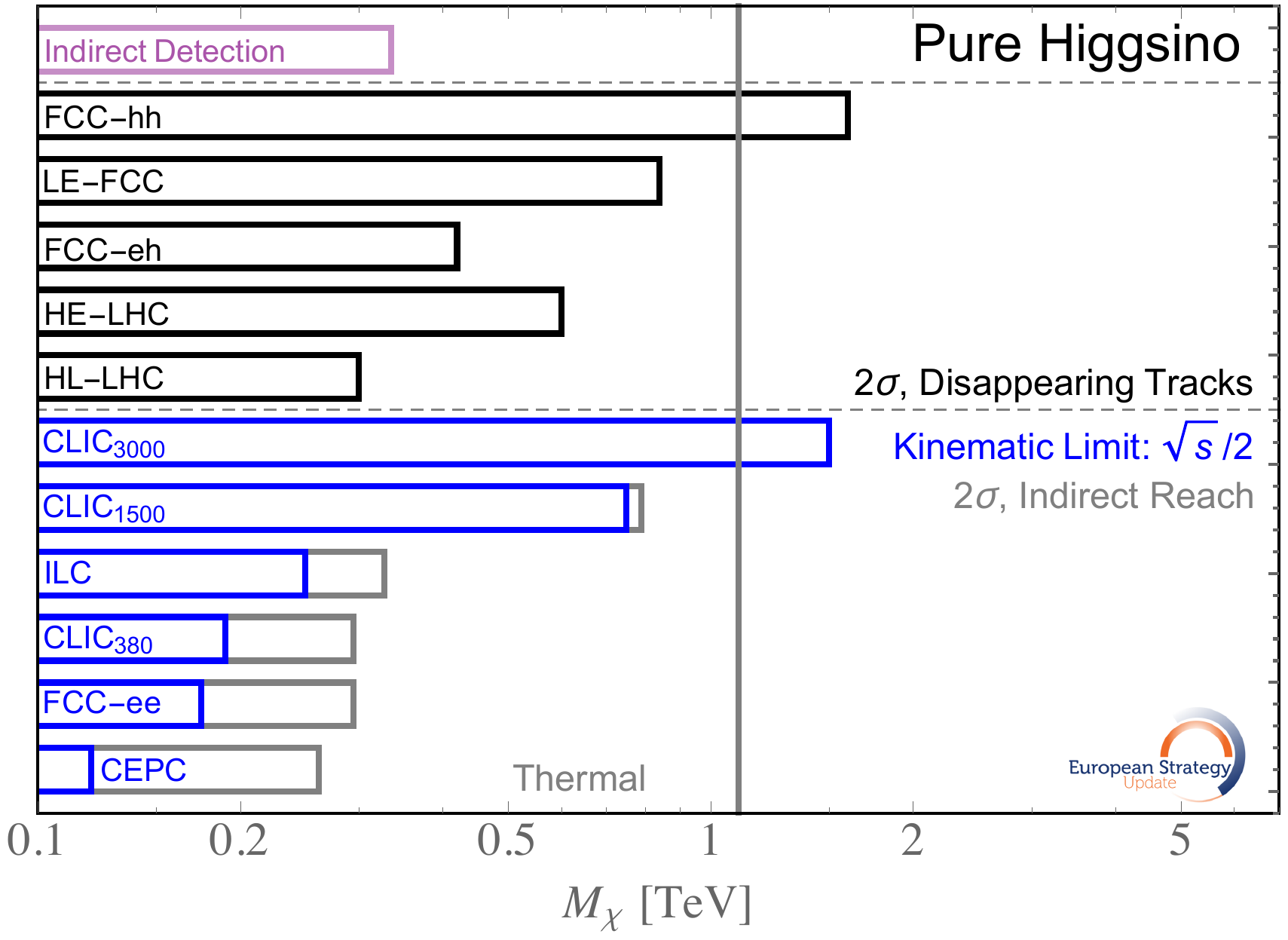}
\end{subfigure}
\caption{(Left) 
Expected 95\% CL exclusion sensitivity  for Higgsino-like NLSP \cha\ and \nn\  
 as a function of  its mass and the mass difference between the NLSP and the \n. The exclusion reach 
of  the mono-jet searches at $pp$ and $ep$ colliders is also superimposed. 
(Right) Summary of the 2$\sigma$ sensitivity reach to pure Higgsinos at future colliders. 
 The vertical line indicates the mass corresponding to dark matter thermal relic.
}
\label{fig:futurehiggsinos} 
\end{figure}
The sensitivity to Higgsino-like electroweakinos is assessed by the 
 ISR search in events with two low \pt\ leptons. Figure~\ref{fig:futurehiggsinos} (left) 
shows that the HL-LHC will probe the parameter space with EWkinos  lighter than
350~GeV and mass splittings larger than a few GeV. The HE-LHC reach is
60\% higher. The FCC-hh instead can yield sensitivity to Higgsino-like electroweakinos as heavy
as 1.3~TeV for mass splittings of 20~TeV\footnote{The reach is determined as 
the extrapolation of the ``soft-lepton'' analysis B corresponding to the 
``ISR'' search developed by the CMS collaboration.}. The ILC shows sensitivities to masses
up to $\sqrt{s}/2$ while CLIC$_{1500}$ (CLIC$_{3000}$) can probe for the existence of sparticles 
as heavy as 650 and 1.3~TeV, respectively. At hadron colliders, 
compressed spectra with $\Delta M$  smaller than 1~GeV are probed by means of the so-called ``monojet'' search, an analysis based on events with one jet and large \met\
and  designed to probe for  electroweakinos decaying into very low \pt\ particles which
cannot be reconstructed by the detectors.  In this case as well the FCC-hh is the machine yielding the
largest sensitivity by probing masses up to 2~TeV while both the FCC-hh and 
CLIC$_{3000}$ would be able to test the dark matter thermal relic hypothesis (Fig.~\ref{fig:futurehiggsinos} (left)). The disappearing track search is instead developed to probe for pure states and allows to test mass hypotheses up to 300~GeV (500~GeV) at the HL-LHC (HE-LHC) and approximately 1.5~TeV at the FCC-hh (Fig.~\ref{fig:futurehiggsinos} (right)).  The linear lepton colliders are sensitive up to $\sqrt{s}/2$ with CLIC$_{3000}$ yielding sensitivity to a large part of the parameter space in the WIMP thermal relic model.

%% file: 5_sum.tex
\section{Summary and Future Prospects}
\label{sec:sum}

With the milestone discovery of a Higgs boson at the LHC, the Standard Model of elementary particle physics is complete. Yet, theoretical considerations and experimental observations strongly suggest the existence of physics beyond the SM, preferably at a scale not far from the EW scale. Weak-scale supersymmetry is one of the top contenders. In this review, we consider a general theoretical framework for fermionic color-singlet states, including a singlet, a doublet and a triplet under the Standard Model SU$_L$(2) gauge symmetry, corresponding to the Bino, Higgsino and Wino in Supersymmetric theories, generically dubbed as ``electroweakinos" (EWkinos) for their mass eigenstates. 

Assuming R-parity conservation, no new sources of CP-violation and decoupling the SUSY scalar and color states, the EWkino sector is simply specified by the three soft SUSY breaking mass parameters $M_1, M_2,\mu$ plus $\tan\beta$. Those parameters govern the phenomenology and the observable signatures: the lighter parameter determines the LSP mass; the heavier one tends to decouple; and those with a similar value will lead to substantial state mixing. R-parity conservation leads to the stability of the LSP state that can be a natural cold dark matter candidate. 

Extensive direct searches for EWkinos have been carried out by experiments at colliders for decades. The ATLAS and CMS experiments have pushed the boundary of knowledge thanks to the outstanding performance of the LHC and the experiments themselves. Breakthrough analyses techniques made it possible to achieve great sensitivity.
\begin{itemize}
\item
Under the assumption of non-compressed scenarios, Wino-like EWkinos
decaying into Higgsino- or Bino-like LSP are excluded at 95\% CL for masses of $600-700$ GeV if the \n\ is massless.
\item
The sensitivity to both Wino- and Higgsino-like 
\cha\ and \nn\ in compressed scenario is challenged by the complexity of reconstructing low \pt\ objects and reaches a few hundreds GeV for $\Delta M$ between 10 and 50~GeV,
but quickly drops for mass splittings between a few GeV and a few hundreds MeV. 
\item
Scenarios with pure Higgsino- and pure Wino EWkinos, characterized by  $\Delta M \sim$ hundreds of MeV, are probed up to a scale of $700-800$~GeV for lifetimes
of a few ns. 
\item
The reach for stable sparticles is of order of 1~TeV. 
\item
Models predicting metastable EWkinos with lifetimes in between a few ns and hundreds ns have not been 
fully explored yet, as well as those leading to short lived sparticles.  
\end{itemize}

Looking forward, innovative ideas and experimental strategies are being devised by both the ATLAS and CMS collaborations to extend the reach to challenging regions of parameter space, {\it e.g.}, by searching for long-lived sparticles as well  as  for promptly decaying EWkinos with $\Delta M$ in the few GeVs to few hundreds of MeV mass range. Furthermore, the fast development of boosted bosons identification ($W$, $Z$, Higgs) is enabling the search for heavier EWkinos in non-compressed spectra. 

It is worth noting that the quoted limits are set at 95\% CL and are valid in the context of simplified models where the EWkinos are typically assumed to be pure states and their  branching fractions in the experimental searches are set to 100\%. 
The re-interpretation of the search results within realistic frameworks, such as those presented in Sec.~\ref{sec:model},  indicates the need for further optimization of analyses to target scenarios
where the EWkinos decay in various modes. This highlights that 
there is still ample room for discovery of EWkinos at the LHC and HL-LHC \cite{Aad:2015eda, Aad:2015iea, Aad:2015pfx, Aad:2015baa, Khachatryan:2016nvf}. 
Furthermore, there are extensions beyond the MSSM in well-motivated theoretical frameworks, such as 
the singlet extension (NMSSM) \cite{Ellis:1988er, Drees:1988fc}, the inclusion of QCD axions \cite{Kim:1983dt,Baer:2014eja}, that would require certain modification and optimization for the search strategies.

Either the search for or the characterization of EWkinos discovered at the LHC experiments will continue at future colliders. 
\begin{itemize}
\item
A future proton-proton collider at $\sqrt{s}=$100~TeV would enable to extend the reach well above the TeV scale, probing non-compressed spectra up to 3~TeV and the very compressed one up to 5~TeV. 
\item
The electron-positron colliders may serve as discovery machines up to a mass as high as $\sqrt{s} / 2$, only limited by the kinematic threshold, essentially model-independent. They especially complement the hadron machines in parameter space with compressed SUSY spectra, where the signal observation would be challenging at hadron colliders.
\end{itemize}

In the underground experiments optimized to observe the nuclear recoil induced by a WIMP-nucleus elastic scattering on a nuclear target, the direct detections of WIMP DM have achieved very impressive sensitivity, reaching SI cross sections of 10$^{-46}$cm$^2$ for the favorable mass region $m_{\n}\sim$10 GeV. At a lower mass,  the sensitivity drops due to the lack of detectable recoil energy, while the collider searches for EWkinos nicely complement this because of the larger missing kinetic energy for a lighter missing particle. The direct detection sensitivity also drops for a TeV mass DM, due to the lower signal rate, while once again the heavy DM searches at future colliders would be further improved due to the accessible phase space at higher energies. Ideally, the two complementary searches should observe consistent signals in order to ultimately confirm the discovery of a WIMP DM particle. 

The search for EWkinos presented in this review provides a well-defined experimental target within a general and well-motivated theoretical framework, and thus holds a great promise for the future discovery. 

%% file: main.bbl
\begin{thebibliography}{160}
\expandafter\ifx\csname natexlab\endcsname\relax\def\natexlab#1{#1}\fi

\bibitem{Aad:2012tfa}
Aad G, et~al.
\newblock \textit{Phys. Lett.} B716:1 (2012)

\bibitem{Chatrchyan:2012xdj}
Chatrchyan S, et~al.
\newblock \textit{Phys. Lett.} B716:30 (2012)

\bibitem{Weinberg:1975gm}
Weinberg S.
\newblock \textit{Phys. Rev.} D13:974 (1976), [Addendum: Phys.
  Rev.D19,1277(1979)]

\bibitem{Gildener:1976ai}
Gildener E.
\newblock \textit{Phys. Rev.} D14:1667 (1976)

\bibitem{Susskind:1978ms}
Susskind L.
\newblock \textit{Phys. Rev.} D20:2619 (1979)

\bibitem{tHooft:1980xss}
't~Hooft G, et~al.
\newblock \textit{NATO Sci. Ser. B} 59:pp.1 (1980)

\bibitem{Ellis:1990wk}
Ellis JR, Kelley S, Nanopoulos DV.
\newblock \textit{Phys. Lett.} B260:131 (1991)

\bibitem{Amaldi:1991cn}
Amaldi U, de~Boer W, Furstenau H.
\newblock \textit{Phys. Lett.} B260:447 (1991)

\bibitem{Langacker:1991an}
Langacker P, Luo Mx.
\newblock \textit{Phys. Rev.} D44:817 (1991)

\bibitem{Giunti:1991ta}
Giunti C, Kim CW, Lee UW.
\newblock \textit{Mod. Phys. Lett.} A6:1745 (1991)

\bibitem{Golfand:1971iw}
Golfand {\relax Yu}A, Likhtman EP.
\newblock \textit{JETP Lett.} 13:323 (1971), [Pisma Zh. Eksp. Teor.
  Fiz.13,452(1971)]

\bibitem{Volkov:1973ix}
Volkov DV, Akulov VP.
\newblock \textit{Phys. Lett.} 46B:109 (1973)

\bibitem{Wess:1974tw}
Wess J, Zumino B.
\newblock \textit{Nucl. Phys.} B70:39 (1974), [,24(1974)]

\bibitem{Wess:1974jb}
Wess J, Zumino B.
\newblock \textit{Nucl. Phys.} B78:1 (1974)

\bibitem{Ferrara:1974pu}
Ferrara S, Zumino B.
\newblock \textit{Nucl. Phys.} B79:413 (1974)

\bibitem{Salam:1974ig}
Salam A, Strathdee JA.
\newblock \textit{Phys. Lett.} 51B:353 (1974)

\bibitem{Kaplan:1983fs}
Kaplan DB, Georgi H.
\newblock \textit{Phys. Lett.} 136B:183 (1984)

\bibitem{Kaplan:1983sm}
Kaplan DB, Georgi H, Dimopoulos S.
\newblock \textit{Phys. Lett.} 136B:187 (1984)

\bibitem{Georgi:1984af}
Georgi H, Kaplan DB.
\newblock \textit{Phys. Lett.} 145B:216 (1984)

\bibitem{Randall:1999ee}
Randall L, Sundrum R.
\newblock \textit{Phys. Rev. Lett.} 83:3370 (1999)

\bibitem{Randall:1999vf}
Randall L, Sundrum R.
\newblock \textit{Phys. Rev. Lett.} 83:4690 (1999)

\bibitem{Nilles:1983ge}
Nilles HP.
\newblock \textit{Phys. Rept.} 110:1 (1984)

\bibitem{Haber:1984rc}
Haber HE, Kane GL.
\newblock \textit{Phys. Rept.} 117:75 (1985)

\bibitem{Martin:1997ns}
Martin SP :1 arXiv:hep-ph/9709356 [hep-ph] (1997), [Adv. Ser. Direct. High
  Energy Phys.18,1(1998)]

\bibitem{Giudice:2008bi}
Giudice GF :155 arXiv:0801.2562 [hep-ph] (2008)

\bibitem{Feng:2013pwa}
Feng JL.
\newblock \textit{Ann. Rev. Nucl. Part. Sci.} 63:351 (2013)

\bibitem{Jungman:1995df}
Jungman G, Kamionkowski M, Griest K.
\newblock \textit{Phys. Rept.} 267:195 (1996)

\bibitem{ATLASWeb}
ATLAS.
\newblock
  \textit{https://twiki.cern.ch/twiki/bin/view/CMSPublic/PhysicsResultsSUS}

\bibitem{CMSWeb}
CMS.
\newblock
  \textit{https://twiki.cern.ch/twiki/bin/view/AtlasPublic/SupersymmetryPublicResults}

\bibitem{Sirunyan:2019ctn}
Sirunyan AM, et~al.
\newblock \textit{JHEP} 10:244 (2019)

\bibitem{ATLAS-CONF-2019-040}
{Search for squarks and gluinos in final states with jets and missing
  transverse momentum using 139 fb$^{-1}$ of $\sqrt{s}$ =13 TeV $pp$ collision
  data with the ATLAS detector}.
\newblock Tech. Rep. ATLAS-CONF-2019-040, CERN, Geneva (2019)

\bibitem{Feng:1999zg}
Feng JL, Matchev KT, Moroi T.
\newblock \textit{Phys. Rev.} D61:075005 (2000)

\bibitem{Hall:2011aa}
Hall LJ, Pinner D, Ruderman JT.
\newblock \textit{JHEP} 04:131 (2012)

\bibitem{Baer:2012up}
Baer H, et~al.
\newblock \textit{Phys. Rev. Lett.} 109:161802 (2012)

\bibitem{Baer:2020kwz}
Baer H, et~al.  arXiv:2002.03013 [hep-ph] (2020)

\bibitem{Baer:1994nr}
Baer H, Chen Ch, Paige F, Tata X.
\newblock \textit{Phys. Rev.} D50:4508 (1994)

\bibitem{Athron:2018vxy}
Athron P, et~al.
\newblock \textit{Eur. Phys. J.} C79:395 (2019)

\bibitem{ArkaniHamed:2006mb}
Arkani-Hamed N, Delgado A, Giudice GF.
\newblock \textit{Nucl. Phys.} B741:108 (2006)

\bibitem{Giudice:2010wb}
Giudice GF, Han T, Wang K, Wang LT.
\newblock \textit{Phys. Rev.} D81:115011 (2010)

\bibitem{Chamseddine:1982jx}
Chamseddine AH, Arnowitt RL, Nath P.
\newblock \textit{Phys. Rev. Lett.} 49:970 (1982)

\bibitem{Barbieri:1982eh}
Barbieri R, Ferrara S, Savoy CA.
\newblock \textit{Phys. Lett.} 119B:343 (1982)

\bibitem{Ibanez:1982ee}
Ibanez LE.
\newblock \textit{Phys. Lett.} 118B:73 (1982)

\bibitem{Hall:1983iz}
Hall LJ, Lykken JD, Weinberg S.
\newblock \textit{Phys. Rev.} D27:2359 (1983)

\bibitem{Ohta:1982wn}
Ohta N.
\newblock \textit{Prog. Theor. Phys.} 70:542 (1983)

\bibitem{Ellis:1982wr}
Ellis JR, Nanopoulos DV, Tamvakis K.
\newblock \textit{Phys. Lett.} 121B:123 (1983)

\bibitem{AlvarezGaume:1983gj}
Alvarez-Gaume L, Polchinski J, Wise MB.
\newblock \textit{Nucl. Phys.} B221:495 (1983)

\bibitem{Randall:1998uk}
Randall L, Sundrum R.
\newblock \textit{Nucl. Phys.} B557:79 (1999)

\bibitem{Giudice:1998xp}
Giudice GF, Luty MA, Murayama H, Rattazzi R.
\newblock \textit{JHEP} 12:027 (1998)

\bibitem{Gherghetta:1999sw}
Gherghetta T, Giudice GF, Wells JD.
\newblock \textit{Nucl. Phys.} B559:27 (1999)

\bibitem{Ibe:2012sx}
Ibe M, Matsumoto S, Sato R.
\newblock \textit{Phys. Lett.} B721:252 (2013)

\bibitem{Fukuda:2017jmk}
Fukuda H, Nagata N, Otono H, Shirai S.
\newblock \textit{Phys. Lett.} B781:306 (2018)

\bibitem{Chung:2003fi}
Chung DJH, et~al.
\newblock \textit{Phys. Rept.} 407:1 (2005)

\bibitem{Dine:1981gu}
Dine M, Fischler W.
\newblock \textit{Phys. Lett.} 110B:227 (1982)

\bibitem{Nappi:1982hm}
Nappi CR, Ovrut BA.
\newblock \textit{Phys. Lett.} 113B:175 (1982)

\bibitem{AlvarezGaume:1981wy}
Alvarez-Gaume L, Claudson M, Wise MB.
\newblock \textit{Nucl. Phys.} B207:96 (1982)

\bibitem{Dine:1993yw}
Dine M, Nelson AE.
\newblock \textit{Phys. Rev.} D48:1277 (1993)

\bibitem{Dine:1994vc}
Dine M, Nelson AE, Shirman Y.
\newblock \textit{Phys. Rev.} D51:1362 (1995)

\bibitem{Dine:1995ag}
Dine M, Nelson AE, Nir Y, Shirman Y.
\newblock \textit{Phys. Rev.} D53:2658 (1996)

\bibitem{Giudice:1998bp}
Giudice GF, Rattazzi R.
\newblock \textit{Phys. Rept.} 322:419 (1999)

\bibitem{Arbey:2011ab}
Arbey A, et~al.
\newblock \textit{Phys. Lett.} B708:162 (2012)

\bibitem{Ajaib:2012vc}
Ajaib MA, Gogoladze I, Nasir F, Shafi Q.
\newblock \textit{Phys. Lett.} B713:462 (2012)

\bibitem{Kang:2012ra}
Kang Z, et~al.
\newblock \textit{Phys. Rev.} D86:095020 (2012)

\bibitem{Craig:2012xp}
Craig N, Knapen S, Shih D, Zhao Y.
\newblock \textit{JHEP} 03:154 (2013)

\bibitem{Albaid:2012qk}
Albaid A, Babu KS.
\newblock \textit{Phys. Rev.} D88:055007 (2013)

\bibitem{Cohen:2013ama}
Cohen T, Lisanti M, Pierce A, Slatyer TR.
\newblock \textit{JCAP} 1310:061 (2013)

\bibitem{0810.3921}
Alwall J, Schuster P, Toro N.
\newblock \textit{Phys. Rev.} D79:075020 (2009)

\bibitem{1105.2838}
Alves D.
\newblock \textit{J. Phys.} G39:105005 (2012)

\bibitem{0812.0980}
Berger CF, Gainer JS, Hewett JL, Rizzo TG.
\newblock \textit{JHEP} 02:023 (2009)

\bibitem{Lee:1977ua}
Lee BW, Weinberg S.
\newblock \textit{Phys. Rev. Lett.} 39:165 (1977), [,183(1977)]

\bibitem{Goldberg:1983nd}
Goldberg H.
\newblock \textit{Phys. Rev. Lett.} 50:1419 (1983), [Erratum: Phys. Rev.
  Lett.103,099905(2009); ,219(1983)]

\bibitem{Steigman:2012nb}
Steigman G, Dasgupta B, Beacom JF.
\newblock \textit{Phys. Rev.} D86:023506 (2012)

\bibitem{Bramante:2015una}
Bramante J, et~al.
\newblock \textit{Phys. Rev.} D93:063525 (2016)

\bibitem{Griest:1990kh}
Griest K, Seckel D.
\newblock \textit{Phys. Rev.} D43:3191 (1991)

\bibitem{Mizuta:1992qp}
Mizuta S, Yamaguchi M.
\newblock \textit{Phys. Lett.} B298:120 (1993)

\bibitem{Edsjo:1997bg}
Edsjo J, Gondolo P.
\newblock \textit{Phys. Rev.} D56:1879 (1997)

\bibitem{Baer:2002fv}
Baer H, Balazs C, Belyaev A.
\newblock \textit{JHEP} 03:042 (2002)

\bibitem{Boehm:1999bj}
Boehm C, Djouadi A, Drees M.
\newblock \textit{Phys. Rev.} D62:035012 (2000)

\bibitem{Ellis:2001nx}
Ellis JR, Olive KA, Santoso Y.
\newblock \textit{Astropart. Phys.} 18:395 (2003)

\bibitem{Arnowitt:2001yh}
Arnowitt RL, Dutta B, Santoso Y.
\newblock \textit{Nucl. Phys.} B606:59 (2001)

\bibitem{Ellis:1998kh}
Ellis JR, Falk T, Olive KA.
\newblock \textit{Phys. Lett.} B444:367 (1998)

\bibitem{Ellis:1999mm}
Ellis JR, Falk T, Olive KA, Srednicki M.
\newblock \textit{Astropart. Phys.} 13:181 (2000), [Erratum: Astropart.
  Phys.15,413(2001)]

\bibitem{Gomez:1999dk}
Gomez ME, Lazarides G, Pallis C.
\newblock \textit{Phys. Rev.} D61:123512 (2000)

\bibitem{Nihei:2002sc}
Nihei T, Roszkowski L, Ruiz~de Austri R.
\newblock \textit{JHEP} 07:024 (2002)

\bibitem{Baer:2005jq}
Baer H, et~al.
\newblock \textit{JHEP} 12:011 (2005)

\bibitem{Berlin:2015wwa}
Berlin A, Gori S, Lin T, Wang LT.
\newblock \textit{Phys. Rev.} D92:015005 (2015)

\bibitem{Freese:2015ysa}
Freese K, Lopez A, Shah NR, Shakya B.
\newblock \textit{JHEP} 04:059 (2016)

\bibitem{Gilmore:2007aq}
Gilmore RC.
\newblock \textit{Phys. Rev.} D76:043520 (2007)

\bibitem{Kar:2019mcq}
Kar A, Mitra S, Mukhopadhyaya B, Choudhury TR.
\newblock \textit{Phys. Rev.} D99:021302 (2019)

\bibitem{Strategy:2019vxc}
Ellis RK, et~al.  arXiv:1910.11775 [hep-ex] (2019)

\bibitem{Cushman:2013zza}
Cushman P, et~al. 2013.
\newblock In \textit{{Proceedings, 2013 Community Summer Study on the Future of
  U.S. Particle Physics: Snowmass on the Mississippi (CSS2013): Minneapolis,
  MN, USA, July 29-August 6, 2013}}

\bibitem{Hisano:2015rsa}
Hisano J, Ishiwata K, Nagata N.
\newblock \textit{JHEP} 06:097 (2015)

\bibitem{Chen:2019gtm}
Chen Q, Hill RJ  arXiv:1912.07795 [hep-ph] (2019)

\bibitem{Cheung:2012qy}
Cheung C, Hall LJ, Pinner D, Ruderman JT.
\newblock \textit{JHEP} 05:100 (2013)

\bibitem{Han:2016qtc}
Han T, Kling F, Su S, Wu Y.
\newblock \textit{JHEP} 02:057 (2017)

\bibitem{Huang:2014xua}
Huang P, Wagner CEM.
\newblock \textit{Phys. Rev.} D90:015018 (2014)

\bibitem{Baum:2017enm}
Baum S, Carena M, Shah NR, Wagner CEM.
\newblock \textit{JHEP} 04:069 (2018)

\bibitem{Han:2018gej}
Han T, Liu H, Mukhopadhyay S, Wang X.
\newblock \textit{JHEP} 03:080 (2019)

\bibitem{Choi:1998ut}
Choi SY, et~al.
\newblock \textit{Eur. Phys. J.} C7:123 (1999)

\bibitem{Choi:2001ww}
Choi SY, Kalinowski J, Moortgat-Pick GA, Zerwas PM.
\newblock \textit{Eur. Phys. J.} C22:563 (2001), [Addendum: Eur. Phys.
  J.C23,769(2002)]

\bibitem{Djouadi:2007ik}
Aarons G, et~al.  arXiv:0709.1893 [hep-ph] (2007)

\bibitem{Moortgat-Picka:2015yla}
Arbey A, et~al.
\newblock \textit{Eur. Phys. J.} C75:371 (2015)

\bibitem{Carena:1986jp}
Carena M, Wagner CEM.
\newblock \textit{Phys. Lett.} B195:599 (1987)

\bibitem{Chen:1995yu}
Chen CH, Drees M, Gunion JF.
\newblock \textit{Phys. Rev. Lett.} 76:2002 (1996)

\bibitem{Hensel:2002bu}
Hensel C. 2002.
\newblock {Search for nearly mass degenerate charginos and neutralinos in e+ e-
  collisions}.
\newblock Ph.D. thesis, Hamburg U.

\bibitem{Berggren:2013vfa}
Berggren M, et~al.
\newblock \textit{Eur. Phys. J.} C73:2660 (2013)

\bibitem{Birkedal:2004xn}
Birkedal A, Matchev K, Perelstein M.
\newblock \textit{Phys. Rev.} D70:077701 (2004)

\bibitem{Dreiner:2012xm}
Dreiner H, et~al.
\newblock \textit{Phys. Rev.} D87:075015 (2013)

\bibitem{Lebrun:2012hj}
Lebrun P, et~al.  arXiv:1209.2543 [physics.ins-det] (2012)

\bibitem{Tanabashi:2018oca}
Tanabashi M, et~al.
\newblock \textit{Phys. Rev.} D98:030001 (2018)

\bibitem{0901.3485}
Dreiner HK, et~al.
\newblock \textit{Eur. Phys. J.} C62:547 (2009)

\bibitem{LEP1}
ALEPH, DELPHI, L3, OPAL.
\newblock
  \textit{http://lepsusy.web.cern.ch/lepsusy/www/inoslowdmsummer02/charginolowdm\_pub.html}

\bibitem{Han:2013kza}
Han T, Padhi S, Su S.
\newblock \textit{Phys. Rev.} D88:115010 (2013)

\bibitem{Lee:1977eg}
Lee BW, Quigg C, Thacker HB.
\newblock \textit{Phys. Rev.} D16:1519 (1977)

\bibitem{Chanowitz:1985hj}
Chanowitz MS, Gaillard MK.
\newblock \textit{Nucl. Phys.} B261:379 (1985)

\bibitem{Bagger:1989fc}
Bagger J, Schmidt C.
\newblock \textit{Phys. Rev.} D41:264 (1990)

\bibitem{He:1992nga}
He HJ, Kuang YP, Li Xy.
\newblock \textit{Phys. Rev. Lett.} 69:2619 (1992)

\bibitem{Gori:2018pmk}
Gori S, Liu Z, Shakya B.
\newblock \textit{JHEP} 04:049 (2019)

\bibitem{Datta:2001cy}
Datta A, Konar P, Mukhopadhyaya B.
\newblock \textit{Phys. Rev.} D65:055008 (2002)

\bibitem{Datta:2001hv}
Datta A, Konar P, Mukhopadhyaya B.
\newblock \textit{Phys. Rev. Lett.} 88:181802 (2002)

\bibitem{Cho:2006sx}
Cho GC, et~al.
\newblock \textit{Phys. Rev.} D73:054002 (2006)

\bibitem{Dutta:2012xe}
Dutta B, et~al.
\newblock \textit{Phys. Rev.} D87:035029 (2013)

\bibitem{Cotta:2012nj}
Cotta RC, Hewett JL, Le MP, Rizzo TG.
\newblock \textit{Phys. Rev.} D88:116009 (2013)

\bibitem{Delannoy:2013ata}
Delannoy AG, et~al.
\newblock \textit{Phys. Rev. Lett.} 111:061801 (2013)

\bibitem{asymptotic}
Cowan G, Cranmer K, Gross E, Vitells O.
\newblock \textit{Eur. Phys. J. C} 73:2501 (2013)

\bibitem{confidencelevel}
Junk T.
\newblock \textit{Nucl. Instr. Meth. A} 434:435 (1999)

\bibitem{CL}
Read AL.
\newblock \textit{J. Phys. G} 28:2693 (2002)

\bibitem{1007.1727}
Cowan G, Cranmer K, Gross E, Vitells O.
\newblock \textit{Eur. Phys. J.} C71:1554 (2011), [Erratum: Eur. Phys.
  J.C73,2501(2013)]

\bibitem{1909.09226}
Aad G, et~al.  arXiv:1909.09226 [hep-ex] (2019)

\bibitem{1812.09432}
Aaboud M, et~al.
\newblock \textit{Phys. Rev.} D100:012006 (2019)

\bibitem{1706.09933}
Sirunyan AM, et~al.
\newblock \textit{JHEP} 11:029 (2017)

\bibitem{1908.08500}
Sirunyan AM, et~al.
\newblock \textit{JHEP} 11:109 (2019)

\bibitem{1709.08908}
Sirunyan AM, et~al.
\newblock \textit{JHEP} 03:076 (2018)

\bibitem{stransversemass2}
Barr A, Lester C, Stephens P.
\newblock \textit{{J. Phys. G}} 29:2343 (2003)

\bibitem{1806.02293}
Aaboud M, et~al.
\newblock \textit{Phys. Rev.} D98:092012 (2018)

\bibitem{1908.08215}
Aad G, et~al.  arXiv:1908.08215 [hep-ex] (2019)

\bibitem{ATLASCONF2018038}
ATLAS.
\newblock \textit{{https://cds.cern.ch/record/2630948}}

\bibitem{1709.05406}
Sirunyan AM, et~al.
\newblock \textit{JHEP} 03:166 (2018)

\bibitem{1801.03957}
Sirunyan AM, et~al.
\newblock \textit{JHEP} 03:160 (2018)

\bibitem{1803.02762}
Aaboud M, et~al.
\newblock \textit{Eur. Phys. J.} C78:995 (2018)

\bibitem{1911.12606}
ATLAS.
\newblock \textit{{https://arxiv.org/pdf/1911.12606.pdf}}

\bibitem{1401.1235}
Han Z, Kribs GD, Martin A, Menon A.
\newblock \textit{Phys. Rev.} D89:075007 (2014)

\bibitem{1409.7058}
Baer H, Mustafayev A, Tata X.
\newblock \textit{Phys. Rev.} D90:115007 (2014)

\bibitem{1501.02511}
Barr A, Scoville J.
\newblock \textit{JHEP} 04:147 (2015)

\bibitem{1705.10733}
Jackson P, Rogan C.
\newblock \textit{Phys. Rev.} D96:112007 (2017)

\bibitem{1801.01846}
Sirunyan AM, et~al.
\newblock \textit{Phys. Lett.} B782:440 (2018)

\bibitem{1905.13059}
Sirunyan AM, et~al.
\newblock \textit{JHEP} 08:150 (2019)

\bibitem{PASEXO19010}
CMS.
\newblock
  \textit{{http://cms-results.web.cern.ch/cms-results/public-results/preliminary-results/EXO-19-010/}}

\bibitem{1804.07321}
Sirunyan AM, et~al.
\newblock \textit{JHEP} 08:016 (2018)

\bibitem{1902.01636}
Aaboud M, et~al.
\newblock \textit{Phys. Rev.} D99:092007 (2019)

\bibitem{1506.05332}
Aad G, et~al.
\newblock \textit{Eur. Phys. J.} C75:407 (2015)

\bibitem{1307.6346}
de~Favereau J, et~al.
\newblock \textit{JHEP} 02:057 (2014)

\bibitem{Aad:2015eda}
Aad G, et~al.
\newblock \textit{Phys. Rev.} D93:052002 (2016)

\bibitem{Aad:2015iea}
Aad G, et~al.
\newblock \textit{JHEP} 10:054 (2015)

\bibitem{Aad:2015pfx}
Aad G, et~al.
\newblock \textit{Eur. Phys. J.} C75:510 (2015), [Erratum: Eur. Phys.
  J.C76,no.3,153(2016)]

\bibitem{Aad:2015baa}
Aad G, et~al.
\newblock \textit{JHEP} 10:134 (2015)

\bibitem{Khachatryan:2016nvf}
Khachatryan V, et~al.
\newblock \textit{JHEP} 10:129 (2016)

\bibitem{Ellis:1988er}
Ellis JR, et~al.
\newblock \textit{Phys. Rev.} D39:844 (1989)

\bibitem{Drees:1988fc}
Drees M.
\newblock \textit{Int. J. Mod. Phys.} A4:3635 (1989)

\bibitem{Kim:1983dt}
Kim JE, Nilles HP.
\newblock \textit{Phys. Lett.} 138B:150 (1984)

\bibitem{Baer:2014eja}
Baer H, Choi KY, Kim JE, Roszkowski L.
\newblock \textit{Phys. Rept.} 555:1 (2015)

\end{thebibliography}
